\documentclass{JFM-FLM_Au}
\usepackage{bm}
\usepackage{framed}
\usepackage{caption}
\usepackage{subcaption}
\usepackage{array}

\usepackage[normalem]{ulem}

\usepackage{comment}

\lefttitle{I. J. G. Lewis and  C. R. Constante-Amores}
\righttitle{Journal of Fluid Mechanics}

\title{
Exact coherent states underlying chaotic falling-film dynamics
}

\author{Isaac J. G. Lewis \and C. Ricardo Constante-Amores }
\affiliation{Department of Mechanical Science and Engineering, University of Illinois, Urbana Champaign}

\corresau{C. R. Constante-Amores, \email{crconsta@illinois.edu}}

\begin{document}
\maketitle

\begin{abstract}

Dynamical-systems approaches to spatiotemporal chaos have been developed primarily for single-phase flows, where the system state is  defined by bulk velocity fields. Extending these ideas to two-phase flows remains challenging because the dynamics are intrinsically coupled to the evolution of a deforming interface. Here, we address this challenge for a two-dimensional vertical falling film by formulating the dynamics in terms of the interface evolution. Starting from the Navier--Stokes equations, we recover a classical long-wave interface evolution equation, originally derived by \citet{topper_kawahara},
whose coefficients are determined  by the Reynolds, Weber, and Froude numbers. Using this formulation, we perform an extensive parametric study to construct a regime map in the space of domain size and dispersion parameter. The resulting map reveals a rich range of interfacial behaviors, including travelling waves,  bursting travelling waves, and fully chaotic regimes, providing a unified picture of the interfacial phenomenology.
In the chaotic falling film regime, we exploit the dissipative nature of the governing equation, which suggests that the long-time dynamics evolve onto an inertial manifold. Using a data-driven approach, we parameterize this inertial manifold and estimate its intrinsic dimension, suggesting approximately linear growth with domain size. We then construct low-dimensional models in manifold coordinates to facilitate the search for exact coherent states of the full system. 
Using this approach, we identify travelling waves, relative periodic orbits and equilibria embedded within the chaotic attractor. Chaotic trajectories repeatedly approach the neighbourhoods of these invariant solutions, indicating that the recurrent interfacial patterns observed in the dynamics correspond to visits to these coherent states. To the best of our knowledge, this constitutes the first identification of exact coherent structures embedded in chaotic falling-film dynamics, demonstrating that modern state-space approaches can be extended to interfacial chaos.

\end{abstract}


\section{Introduction}

Falling liquid films provide a canonical setting where the interplay of inertia, viscosity, and surface tension gives rise to a hierarchy of instabilities culminating in spatiotemporal chaos \citep{omar,Joo_Davis_1992,chang1994wave,kalliadasis2011falling}.  Beyond their fundamental significance for understanding pattern formation and low-dimensional chaos,  falling films play a vital role in natural and industrial processes, such as coating and reactor design \citep{russo2019falling,Batchvarov,holroyd2024linear,thompson2016falling,Dietze_2014}.

Experiments on falling films confirm the richness of the resulting dynamics, displaying periodic solitary waves with capillary ripples \citep{coatings10060599} and a wide range of nonlinear behaviors across different substrates and Reynolds numbers \citep{polym13081205,pierson_waves,DIETZE_AL-SIBAI_KNEER_2009,Jones_film}. 
The theoretical description of falling films has largely relied on the long wave approximation, derived as an asymptotic reduction of the Navier-Stokes equations.
Classical analyses within this framework revealed
soliton solutions \citep{Korteweg01051895}, nonlinear interactions between capillary and gravity waves, the development of undular bores \citep{johnson_undular_bore}, and hydraulic jumps \citep{Bohr_1993}.
Introducing the parameter $\delta=h_0/l$, defined as the ratio of the equilibrium film thickness $h_0$ to a characteristic wavelength $l$,  enables systematic asymptotic expansions 
in the limit $\delta\rightarrow0$.
In the context of thin-film flows,
this limit corresponds physically to vanishing film thickness rather than infinitely long waves.

In long-wave analyses, the governing dynamics depend on the relative magnitudes of the 
relevant parameters, which are commonly the Reynolds number $Re$ (ratio of inertia to viscous forces), the Froude number $Fr$ (ratio of inertial to gravitational forces)
and the Weber number $We$ (inertia to  surface tension forces). The assumed order of 
these parameters with respect to $\delta$ determines the complexity of the resulting PDEs. A classical example is the work by
\citet{benney_long_waves} 
who expanded the
2D Navier-Stokes equations  to $\mathcal{O}(\delta^2)$
while taking $Re$, $Fr$, and $We$ to be 
$\mathcal{O}(1)$. In this regime,  
 inertial effects dominate, whereas surface tension contributions appear only at
$\mathcal{O}(\delta^3)$.
Benney's expansion yields a critical Reynolds number, $\Rey_c=(5/4)\cot\theta$; for $\Rey>\Rey_c$ nonlinear wave interactions develop, leading to finite-amplitude travelling waves and, eventually, more complex spatiotemporal patterns.

\citet{johnson_undular_bore} modified the analysis of \citet{benney_long_waves}, reducing the Navier-Stokes equations to a variant of the Korteweg-de Vries Burgers (KdV-B) equation, which incorporates a second-order term:
\begin{equation}
\phi_t+\text{Fr}\hspace{.25mm}\phi\phi_x+\text{Fr}\kappa\phi_{xxx}=\frac{1}{3}\text{Re}\hspace{.25mm}\kappa\left(1-\frac{2}{5}\text{Fr}^2\right)\phi_{xx}.\label{kdvb}
\end{equation}

\noindent Here,  $\phi(x,t)$ represents  the height of the film, subscripts denote partial derivatives, and $\kappa$ is the long-wave parameter arising from the asymptotic expansion. The third-order derivative represents dispersive effects, while the second-order term accounts for viscous dissipation. Equation \ref{kdvb} is only valid for values $\text{Fr}<\sqrt{5/2},$ ensuring positive effective viscosity.
Subsequently, the three-dimensional case was derived by \citet{topper_kawahara}, in which the
asymptotic expansion up to the fourth  term,
under the assumptions that $\text{We}=\mathcal{O}(\delta^{-2}),$ \text{Fr}$=\mathcal{O}(1)$ and Re $= \mathcal{O}(\delta)$ yields the equation
\begin{equation}
\phi_t+\phi\phi_x+\alpha\phi_{xx}+\beta\phi_{xxx}+\gamma\phi_{xxxx}=0.\label{kawahara}
\end{equation}
This is a form of the Kawahara equation, in which a fourth-order hyperdiffusion term replaces the fifth-order term common in this family of models.  A full derivation of this equation starting from the three-dimensional Navier-Stokes equations is provided in  Appendix A.
Topper \& Kawahara's equation  represents a generic one-dimensional normal form with free coefficients. To obtain a physically relevant model for falling films, we reinterpret the field $\phi$ as the local film thickness $H$ and extend the formulation to two spatial dimensions \citep{Akrivis2011Linearly}. After appropriate rescaling of space and time, the second-order diffusive coefficient is normalized, and the remaining higher-order contributions are combined into a single dimensionless parameter (dispersion parameter). This procedure yields the two-dimensional dispersive thin-film equation
\begin{equation}
H_t + H H_x + H_{xx} + \delta \nabla^2 H_x + \nabla^4 H = 0,
\label{film_eq}
\end{equation}
which governs the spatiotemporal evolution of the film thickness. Here, $H(x,y,t)$ denotes the dimensionless local film thickness, with $x$ and $y$ the streamwise and spanwise coordinates and $t$ time. The dimensionless parameter $\delta$ controls the relative importance of dispersive effects. More details are provided in Appendix~A.

Equation \ref{film_eq} may be viewed as a reduced, long-wave approximation of the Navier-Stokes equations that preserves their dissipative character through a fourth-order diffusive term. As with many partial differential equations, despite its infinite-dimensional phase space, the long-time dynamics are expected to evolve on a finite-dimensional invariant (or inertial) manifold  \citep{Hopf,Titi}.  From this perspective, the complex spatiotemporal behaviour observed in falling film flows can be interpreted as low-dimensional chaotic dynamics evolving on a finite-dimensional invariant manifold.    Within the dynamical systems description of such chaotic flows, the long-time dynamics are often organized around unstable invariant solutions, commonly referred to as exact coherent states (ECS) \citep{topper_kawahara,Graham2021-om,WALEFFE_2001}.  These include equilibria, travelling waves, periodic orbits, and relative periodic orbits. As a trajectory evolves on the inertial manifold, it recurrently approaches neighborhoods of these unstable ECSs, temporarily adopting their structure before being repelled along unstable manifolds.

The computation of such invariant solutions requires accurate initial conditions in order for Newton-Krylov method to converge to  ECSs. Traditionally, these initial guesses are obtained  by  recurrence analysis \citep{Chandler_Kerswell_2013}. While successful, this approach can be computationally expensive, particularly in high-dimensional systems. An alternative and increasingly effective strategy is to exploit reduced-order representations of the system. For example, the Data-driven Manifold Dynamics (DManD) framework \citep{linot2023dynamics} builds low-dimensional models from high dimension simulation data. Near-recurrent states identified in the reduced space provide high-quality initial guesses for subsequent Newton-Krylov refinement in the full system. The DManD approach has been shown to successfully 
reproduce the dynamics of several canonical dissipative systems, including the 1D Kuramoto-Sivashinsky equation \citep{Linot_deep_learning}, and it has been used to find new ECS in turbulent pipe flow \citep{CRCA_pipe}, and planar Couette flow  \citep{linot2023dynamics}.

To our knowledge, no prior studies have identified exact coherent states embedded in chaotic falling films. Although numerous works have analyzed the spatiotemporal chaos of falling films using reduced-order models \citep{Akrivis,tomlin_ks},
these studies have primarily focused on  instability thresholds, wave statistics, or energy spectra rather than isolating the invariant solutions that might organize the underlying  dynamics in the chaotic regime.
Here, we show that invariant solutions exist  within the spatiotemporally chaotic dynamics of falling liquid films.
Identifying these invariant solutions in the context of thin film flows provides a complementarity dynamical systems perspective on nonlinear wave analyses, and offers a mechanistic framework for interpreting how coherent interfacial patterns emerge, interact, and decay.
Section \ref{framework} introduces the numerical framework used to simulate chaotic films, extract manifold coordinates, and compute new ECS using a Newton-Krylov approach applied to both the DNS and the reduced-order model. Section \ref{results_section} presents the results, including a regime map as a function of the domain size and the dispersion parameter,
as well as the classification of different states, with a focus on the chaotic regime.
Finally, concluding remarks are summarized in Section \ref{conclusion_Section}.

\section{Numerical methods and framework \label{framework}}

\subsection{Numerical simulations of chaotic films}

\begin{figure}
\centering
\begin{tabular}{cc}
\includegraphics[width=0.5\textwidth]{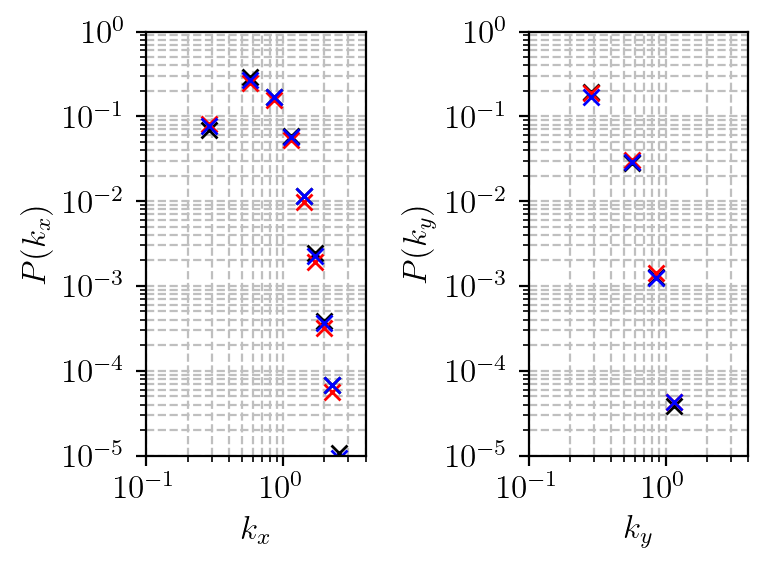}&
\includegraphics[width=0.5\textwidth]{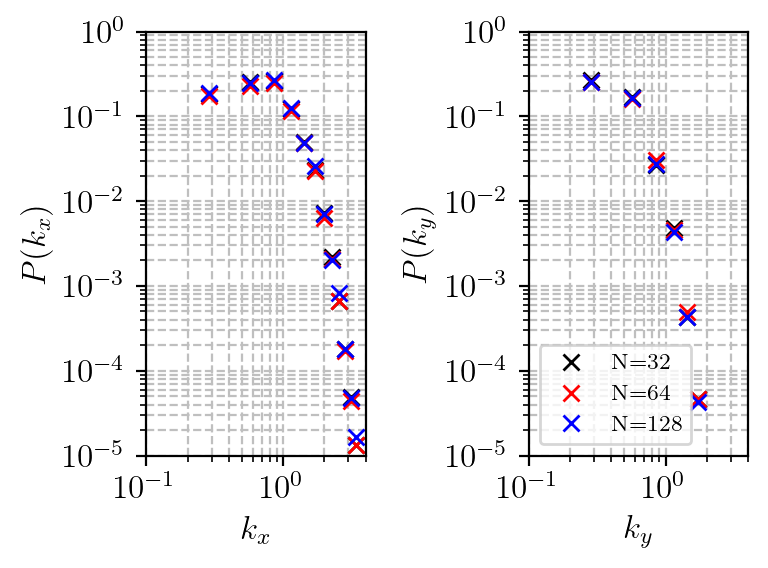}\\
(a) & (b)\\
\end{tabular}
\caption{Streamwise and spanwise power spectra $P(k_x)$ and $P(k_y)$ for  cases $(L=22,\delta=0.002)$ and  $(L=30,\delta=0.65)$ corresponding to panels (a) and (b), respectively. Spectra are shown for Fourier discretizations with $N=32,64,$ and $128$. 
 } \label{fig:spectra_dns}
\end{figure}

To simulate the dynamics of falling films, we employ the long-wave evolution equation \ref{film_eq}, which governs the spatiotemporal evolution of the film thickness $H(x,y,t)$ under the combined effects of inertia, viscosity, and surface tension.
\textcolor{black}{Solutions are computed in a doubly periodic domain $[0,L_x]\times [0,L_y]$;  in this work, we restrict attention to square domains with $L_x=L_y=L$.}
The equations are solved  using the open-source solver
Dedalus 
\citep{Burns_dedalus}. 
Truncating the Fourier representation  to
$N_x$ and $N_y$ modes in the $x$ (streamwise) and $y$ (spanwise)  directions,   reduces the PDE to a finite set of  ODEs. 
{The resulting Fourier coefficients are collected into a finite-dimensional state vector $\boldsymbol{x}(t)\in \mathbb{R}^{N}$, with
$N=N_xN_y$, so that discretized dynamics can be written as $d\boldsymbol{x}/dt=\boldsymbol{f}(\boldsymbol{x})$.}

To assess numerical convergence of solutions, we examine the spatial power spectra in the streamwise and spanwise directions, defined as

\begin{equation}
    P(k_{x})=\frac{1}{N_TN_x}\int_0^{N_T}\left|\langle\hat{H}(k_x)\rangle_y\right|^2dt
\end{equation}
\begin{equation}
    P(k_{y})=\frac{1}{N_TN_x}\int_0^{N_T}\left|\langle\hat{H}(k_y)\rangle_x\right|^2dt
\end{equation}
where $|\cdot|$ is the magnitude of a complex number, $\langle\cdot\rangle_x$ and $\langle\cdot\rangle_y$ are the spatial averages over $x$ and $y$,
respectively, $\hat{H}$ denotes the $n$th Fourier coefficient,
$N_T$ is the number of snapshots used in the average,
and $\mathbf{k_{nm}}=(k_{x,n},k_{y,m})=(2\pi n/L_x,2\pi m/L_y)$ denotes the Fourier wavenumber corresponding to the $(n,m)$ mode.

Figures~\ref{fig:spectra_dns}(a,b) show the  spectra as a function of wavenumbers $k_x$ and $k_y$  for  two characteristic cases used in the ECS search, e.g., $L=22$ with $\delta=0.002$  and $L=30$ with $\delta=0.65373$.
For  $L=22$, the spectral energy distribution is essentially independent of the spectral resolution, while  slight deviations exist for $L=30$. Tests with resolutions up to
$N_x=N_y=128$ show only minor changes. Unless otherwise stated, all simulations use $N_x=N_y=64$.
Linear stability analysis confines exponential growth to the bounded wavenumber band $|\boldsymbol{k}|<1$ with a preferred scale $k^*=1/\sqrt{2}$ \citep{tomlin_ks}; since the maximum resolved wavenumber satisfies $k_{\max}\gg 1$, all dynamically active linear scales lie well within the resolved spectral range.


\subsection{Relevant experimental parameters}

In this section we relate the   dispersion parameter $\delta$ to experimentally relevant conditions. Experimental measurements by \citet{Jones_film} reported film thicknesses in the range $0.12$--$0.29~\mathrm{mm}$ for air--water systems.
Falling liquid films with thicknesses as large as $O(1\,\mathrm{mm})$ have been documented by \citet{Dietze_2014}.
We express $\delta$ in terms of the Reynolds number by assuming an equilibrium film thickness $h_0 \sim 0.2~\mathrm{mm}$ and near-vertical inclination. Under these assumptions, the Reynolds number is given by

\begin{equation}
\text{Re}=\frac{3\sqrt{5}}{\delta|\text{Fr}^2-5/2|^{1/2}\sqrt{2\text{We}}}.\label{eq:Reynolds}
\end{equation}

\begin{figure}
\begin{center}
\begin{tabular}{cc}
\includegraphics[width=0.5\textwidth]{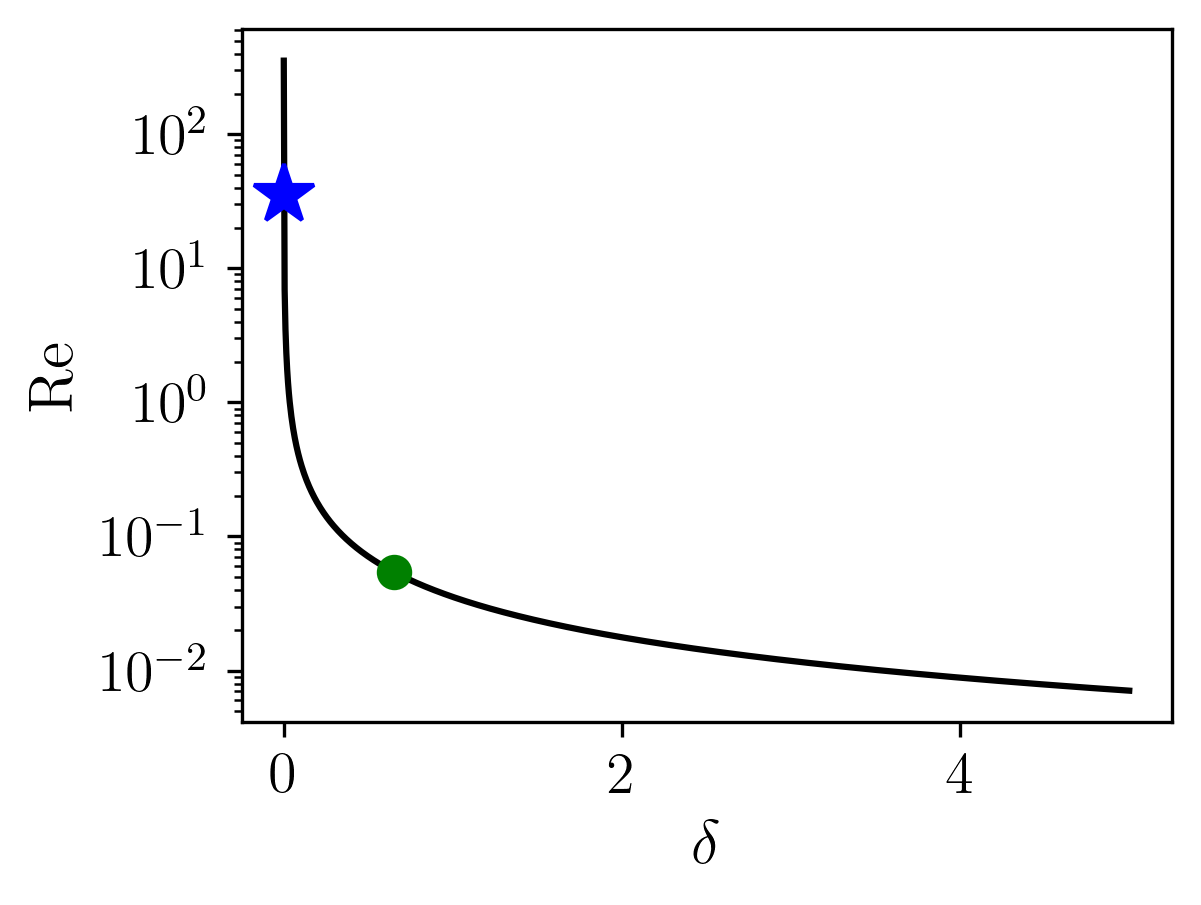} &
\includegraphics[width=0.5\textwidth]{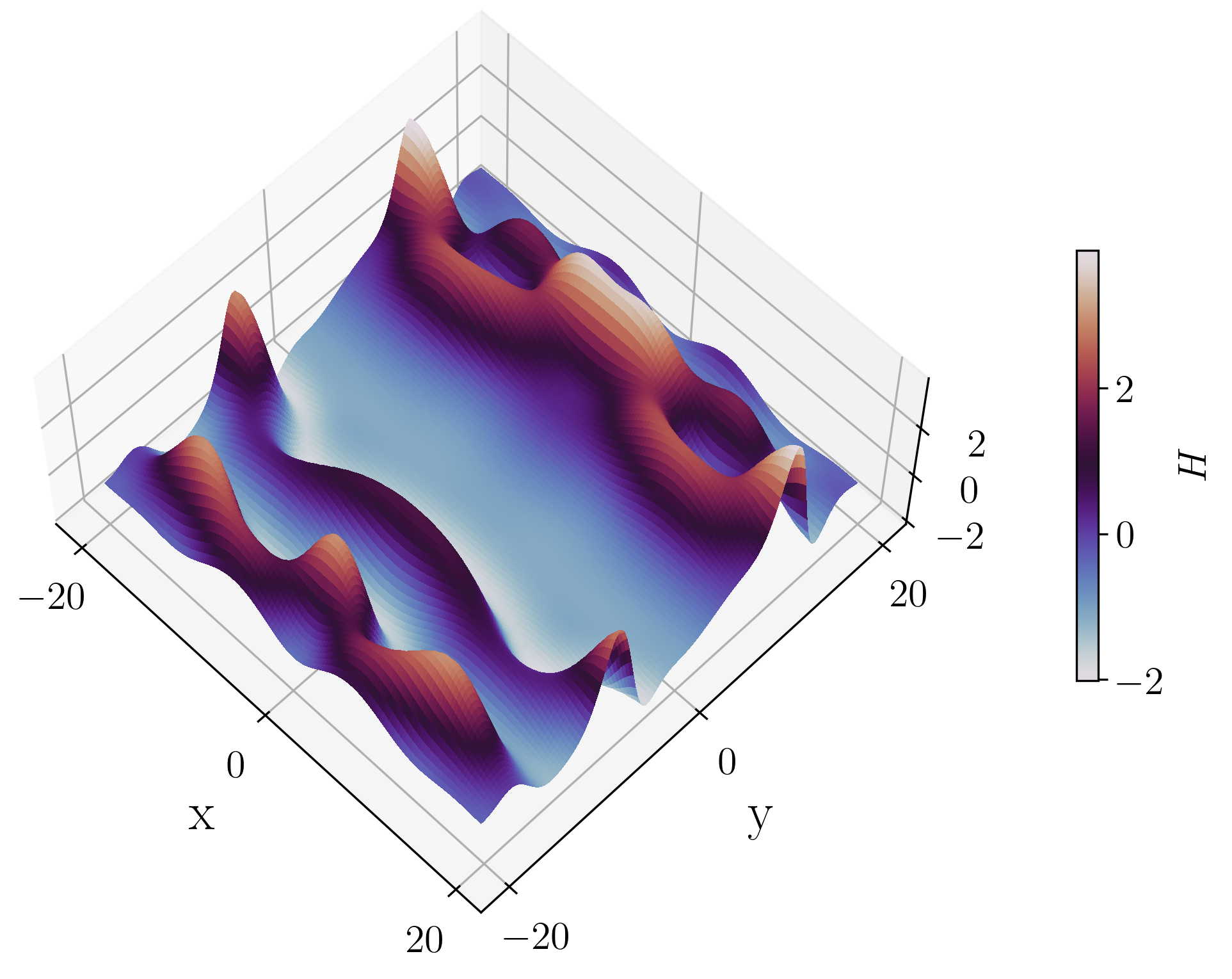}\\
(a) & (b)\\
\end{tabular}
\caption{ (a) 
Dependence of the Reynolds number as a function of the dispersion parameter 
in the range $\delta\in\left[10^{-3},5\right]$ for a falling film with $h_0$=0.2 mm and  $\theta=1.5608$. The blue star and green point correspond to $\delta$ values of 0.002 and 0.65373, respectively, where $\text{Re}(0.002)=35.38$ and $\text{Re}(0.65373)=0.0541.$ (b) Instantaneous snapshot of the film-height field $H(x,y)$ for $\delta=0.002$ in a domain of size $L=40$, shown as an illustrative example of interfacial deformation.
}\label{fig:Re_delta}
\end{center}
\end{figure}

Since the expression for $\delta$ becomes singular at $\theta=\pi/2$, corresponding to a perfectly vertical plate, we consider a near-vertical inclination $\theta=\pi/2-0.01$, yielding $\mathrm{Fr}=98.65$. Figure~\ref{fig:Re_delta}a shows the resulting relationship between $\mathrm{Re}$ and $\delta$ over the range $\delta\in[10^{-3},5]$. In this work we assume $\delta\in[10^{-3},1.1]$, as  this range spans both moderate and low Reynolds number regimes while avoiding the asymptotic limit of weak dispersion. These limits are also shown in figure \ref{fig:Re_delta}a with the green point and blue star (i.e., Re $\in[0.0541,35.38]$).
Figure~\ref{fig:Re_delta}b shows a representative snapshot of the film height field for a domain of size $L=40$ at $\delta=0.002$ and time $t=400$. The interface exhibits a strongly modulated,  wave pattern composed of elongated streamwise ridges and localized crests. These waves propagate predominantly in the streamwise direction and are accompanied by spanwise modulations of comparable wavelength. This snapshot is representative of the dispersive, nonlinear regime explored in this work.

\subsection{Symmetries}
When searching for ECS, we will take advantage of the symmetries of the system. Equation \eqref{film_eq} posed in a doubly periodic domain
$[0,L_x]\times[0,L_y]$ is invariant under continuous spatial translations in the streamwise and spanwise directions. 
Accordingly, if $H(x,y,t)$ is a solution, then
\begin{equation}
H(x+s_x,y+s_y,t)=H(x,y,t),
\end{equation}
for any shift $\boldsymbol{s}=(s_x,s_y)$ with $s_x\in[0,L_x)$ and $s_y\in[0,L_y)$.  These symmetries form the two-parameter translation group $\mathbb{T}^2$ associated with the periodic boundary conditions.

The field is represented by a finite-dimensional state vector $\boldsymbol{x}$, and the action of a spatial translation is represented by a linear operator $\mathbf{G}(\boldsymbol{s})$ acting on $\boldsymbol{x}$. Denoting by $\boldsymbol{\Phi}^t(\boldsymbol{x}_0)$ the flow map of the
discretized dynamical system, the generic invariant solutions are  RPOs, which satisfy
\begin{equation}
\boldsymbol{\Phi}^{T}(\boldsymbol{x}_0)
=
\mathbf{G}(\boldsymbol{s})\,\boldsymbol{x}_0,
\label{eq:RPO}
\end{equation}
for some period $T>0$ and shift $\boldsymbol{s}$. Because of the additional phase shift,  $\boldsymbol{\Phi}^{T}(\boldsymbol{x}_0)\neq \boldsymbol{x}_0$ in general, and RPOs cannot be    detected using the residual for strictly periodic solutions. True periodic orbits, for which $\boldsymbol{s}=\boldsymbol{0}$, are therefore non-generic in the presence of continuous symmetries.

In the numerical implementation, symmetry reduction is done using the first Fourier mode slice outlined in \citet{BudanurFFM}.

\subsection{Mapping to manifold coordinates \label{AEs}}

To identify the manifold coordinates of equation~\eqref{film_eq}, we adopt a two-stage reduction procedure. First, we perform a linear dimension reduction using proper orthogonal decomposition (POD; \citealp{pod_book,Berkooz1993-mt}) to obtain an orthogonal basis that captures the dominant energetic structures of the flow. We then apply a nonlinear reduction using autoencoders to further compress the system onto a low-dimensional manifold.

Each DNS snapshot of the film thickness field is represented by the truncated spectral coefficient vector
$\boldsymbol{x}(t)\in\Re^{N}$, which defines the state of the finite-dimensional dynamical system obtained from \eqref{film_eq}.
The fluctuating component is defined as
\begin{equation}
\boldsymbol{x}'(t)=\boldsymbol{x}(t)-\bar{\boldsymbol{x}},
\end{equation}
where $ \bar{\boldsymbol{x}}=\langle \boldsymbol{x}(t)\rangle$ denotes the temporal mean. 
The aim of POD is to find a function $\boldsymbol{\phi}$ that maximizes
\begin{equation}
\frac{\left\langle
\left|(\boldsymbol{x}',\boldsymbol{\phi})\right|^2
\right\rangle}
{\|\boldsymbol{\phi}\|^2},
\label{eq:pod}
\end{equation}
where $\left \langle \cdot \right \rangle$ denotes the ensemble average, and the inner product is defined to be
\begin{equation}\label{eigenvalue1}
(\boldsymbol{x}_1,\boldsymbol{x}_2) = \int_\Omega \boldsymbol{x}_1 \cdot \boldsymbol{x}_2 \, dx\,dy
\end{equation}
We apply a standard (direct) POD to the full dataset without any prior assumption of symmetries. We reiterate that symmetries are factored out later to search for ECS.

Next, we focus on identifying the intrinsic manifold dimension and the coordinates of the low-dimensional representation from the POD coefficients. For this purpose, we employ a recently developed framework known as 
implicit rank minimizing autoencoder with weight decay (IRMAE-WD), detailed in \citet{zeng2023autoencodersdiscoveringmanifolddimension}. This framework has been shown to recover accurate low-dimensional embeddings and the corresponding inertial manifolds for data lying on the attractor in several canonical flow systems. In particular, it successfully identifies the correct latent representations for the 1D Kuramoto-Sivashinsky equation \citep{zeng2023autoencodersdiscoveringmanifolddimension}, as well as for travelling waves and periodic orbits in 2D Kolmogorov flow and pressure-driven pipe flow \citep{constante2024data}.

The  IRMAE autoencoder is formed by a standard nonlinear encoder and decoder networks with $n$ additional linear layers $\mathcal{W}_n$ (of size $d_z \times d_z$) between them. The encoder finds a compact representation ${\bf z} \in \Re^{d_z}$, and the decoder performs the inverse operation.  The additional linear layers promote minimization of  the rank of the data covariance in the latent representation, precisely aligning  with the dimension of the underlying manifold.  Post-training, a singular value decomposition (SVD) is applied to the covariance matrix of the latent data matrix  $\textbf{z}$ yielding matrices of singular vectors $\mathsfbi{U}$, and  singular values $\mathsfbi{S}$.  Then, we can project $\textbf{z}$ onto $\mathsfbi{U}^T$ to obtain $\mathsfbi{U}^T \textbf{z}= \textbf{h}^+ \in\Re^{d_z}$ in which each coordinate of $\textbf{h}^+$ is orthogonal and ordered by contribution (here, $\textbf{h}^+$ refers to the projection of latent variables onto manifold coordinates). This framework reveals the manifold dimension $d_\mathcal{M}$ as the number of significant singular values, indicating that a coordinate representation exists in which the data spans $d_\mathcal{M}$ directions. Thus, the encoded data avoids spanning directions  associated with nearly zero singular values (i.e., $\mathsfbi{U} \mathsfbi{U}^T {\bf z} \approx  \mathsfbi{\hat{U}} \mathsfbi{\hat{U}}^T {\bf z} $, where $\mathsfbi{\hat{U}}$ are the singular vectors truncated corresponding to singular values that are not nearly zero). Leveraging this insight, we  extract a minimal, orthogonal coordinate system by projecting   ${\bf z}$ onto $\mathsfbi{\hat{U}}$, resulting in a minimal representation $\mathsfbi{\hat{U}}^T {\bf z} = {\bf h} \in \Re^{d_{\mathcal{M}}}$.
The problem then becomes a minimization of the reconstruction loss due to the encoder and decoder functions, i.e\begin{equation}
\mathcal{L}=||\tilde{\textbf{a}}-\mathcal{D}(\mathcal{W}(\mathcal{E}(\textbf{a},\theta_{\mathcal{E}})),\theta_{\mathcal{D}})||_2^2+\lambda||\bm{\theta}||_2^2.\label{loss_irmae}
\end{equation}
Here, $\theta_{\mathcal{E}}$ and $\theta_{\mathcal{E}}$ are the hyperparameters corresponding to the encoder and decoder functions, respectively, and $||\cdot||_2$ is the usual $L^2$ norm.

Here, $\tilde{\textbf{a}}$  denotes the projected POD data.
Determination of the manifold dimension then simply follows by examining magnitudes of the singular values of the reduced space $\boldsymbol{h}$ and comparing them with that of the latent space. A drop in magnitudes of these singular values occur at $\sigma_{d_\mathcal{M}},$ signifying identification of the dimension of the manifold.

To conclude, in this work, there are four distinct representations of the system. Let $\mathcal{H}$ denote the infinite-dimensional solution space of the equation \ref{film_eq}. The direct numerical simulation (DNS) produces trajectories in a finite-dimensional subspace $\Re^{N} \subset \mathcal{H}$, which we refer to as the `full state'. This full state is mapped  to a  linear representation using POD:  $\Re^{N} \to \Re^{d_{POD}}$. Then, this  POD space is projected to a  $d_\mathcal{M}$-dimensional coordinate system via a nonlinear mapping $\mathcal{E} : \Re^{POD} \to \Re^{d_\mathcal{M}}$, obtained from a trained autoencoder,
and a decoder function $\mathcal{D}:\Re^{d_\mathcal{M}}\rightarrow\Re^{d_{POD}}$.

\subsection{Evolution equation in manifold coordinates}

Once we have found an accurate coordinate transformation from the full state to the inertial manifold,
we seek to learn the dynamics in this low-dimensional representation.
We use a neural ODE framework for   continuous-time modelling   of the dynamics   \citep{chen2019neural}.
Then, the problem becomes
\begin{equation}
\frac{d\boldsymbol{h}}{dt}=\boldsymbol{g}(\boldsymbol{h},\theta_f),
\label{node1}
\end{equation}
representing the vector field $\boldsymbol{g}$ on the manifold as a neural network with weights $\theta_f$. We can time-integrate equation \ref{node1}  between $t$ and $t+\delta t$ to yield a prediction $\tilde{\boldsymbol{h}}(t+\delta t)$, such as
\begin{equation}
\tilde{{\boldsymbol{h}}}(t+\delta t)=\boldsymbol{h}(t)+\int_{t}^{t+\delta t}\left(\boldsymbol{g}(\boldsymbol{h}(t');\theta_f) + \mathsfbi{A}\boldsymbol{h}(t')\right)dt'.   
\end{equation}
Here, $\mathsfbi{A}$ is a damping term,
which is introduced to stabilize the learning of the vector field by preventing trajectories from drifting away \citep{linot_stabilised}. 
Given data for $\boldsymbol{h}(t)$ and $\boldsymbol{h}(t +\delta t)$ for a long time series we can train $\boldsymbol{g}$  to minimize the $L_2$ difference between the prediction $\tilde{\boldsymbol{h}}(t+\delta t)$   and the known data  $\boldsymbol{h}(t+\delta t)$. Then the loss function is defined by
\begin{equation}
\mathcal{L}_g=||\boldsymbol{h}(t_i+\delta t)-\tilde{\boldsymbol{h}}(t_i+\delta t)||_2^2.
\label{loss_node}
\end{equation}

We use automatic differentiation to determine the derivatives of $\boldsymbol{g}$ with respect to  $\theta_f$.    To optimize the loss function described in equation \ref{loss_node}, we use an Adam optimiser in PyTorch \citep{pythorch}.

\subsection{Calculation of exact coherent states in DNS and  model in manifold coordinates}
\label{newton_krylov_method}

To identify  invariant solutions in the system, we search for exact coherent states of the associated dynamical system $d\boldsymbol{x}/dt = \boldsymbol{f}(\boldsymbol{x})$. Let $\boldsymbol{\Phi}^{T}$ denote the associated flow map. Time-periodic solutions satisfy
\begin{equation}
\boldsymbol{R}(\boldsymbol{x},T)
=
\boldsymbol{\Phi}^{T}(\boldsymbol{x})-\boldsymbol{x}
=\boldsymbol{0}.
\label{eq:residual_full}
\end{equation}

In manifold coordinates, 
with $\boldsymbol{h}(t)\in\mathbb{R}^{d_{\mathcal{M}}}$ and flow map
$\boldsymbol{\Psi}^{T}$.
Periodic solutions of the reduced system satisfy
\begin{equation}
\boldsymbol{R}_h(\boldsymbol{h},T)
=
\boldsymbol{\Psi}^{T}(\boldsymbol{h})-\boldsymbol{h}
=\boldsymbol{0}.
\label{eq:residual_rom}
\end{equation}

Both \eqref{eq:residual_full} and \eqref{eq:residual_rom}
are solved using a Jacobian-free Newton-Krylov (JFNK) method \citep{willis2019equilibriaperiodicorbitscomputing}. Introducing the combined unknown
$\boldsymbol{q}=(\boldsymbol{x},T)$ (or $\boldsymbol{q}_h=(\boldsymbol{h},T)$ in the reduced space), each Newton step requires the solution of
\begin{equation}
\mathbf{J}\,\delta\boldsymbol{q}
=
-\boldsymbol{R},
\end{equation}
which is carried out using GMRES.

The Jacobian-vector product is evaluated in matrix-free form as
\begin{equation}
\mathbf{J}\,\delta\boldsymbol{q}
\approx
\frac{
\boldsymbol{R}(\boldsymbol{q}+\varepsilon\,\delta\boldsymbol{q})
-
\boldsymbol{R}(\boldsymbol{q})
}{\varepsilon},
\end{equation}
with $\varepsilon\sim10^{-6}$-$10^{-8}\|\boldsymbol{q}\|_2$. An identical formulation is used in the reduced space with $\boldsymbol{R}$ replaced by $\boldsymbol{R}_h$.

In practice, the reduced-order model is used to generate near-recurrent trajectories at a significantly lower computational cost. These are mapped back to the full state space and employed as initial guesses for the Newton-Krylov search in DNS, where final convergence to machine-accurate invariant solutions is obtained.

Convergence to a periodic orbit in the full system is declared when
\begin{equation}
r= \frac{ \|\boldsymbol{\Phi}^{T}(\boldsymbol{x}_0)-\boldsymbol{x}_0\|_2 }{ \|\boldsymbol{x}_0\|_2 } < r_{\mathrm{tol}} .
\end{equation}

All reduced-order computations are performed in the symmetry-reduced coordinates, so that relative periodic orbits of the full system appear as strictly periodic solutions.

We note that Newton--Krylov searches for periodic solutions are dominated by the cost of Krylov subspace construction, which requires repeated evaluations of the flow map at each Newton iteration. For long-period or dynamically complex orbits, GMRES may require $\mathcal{O}(30$--$60)$ iterations per step in the full state space of dimension $N$, and convergence often depends on numerous initial trajectory guesses.
The computational burden therefore grows rapidly with system dimension and domain size $L$. The low-dimensional representation restricts the dynamics to a latent space of dimension $d_\mathcal{M} \ll N$, thereby bounding the Krylov subspace dimension and substantially reducing the cost per Newton iteration. Although neural ODEs trained on single trajectories do not reproduce the full DNS state space and cannot satisfy the strict residual tolerances required for final convergence, they efficiently generate approximate periodic trajectories with residuals of order $\mathcal{O}(10^{-3})$.
These serve as dynamically consistent initial guesses for refinement in DNS. Thus, the ROM acts as a computational preconditioner for the periodic-orbit search rather than a replacement for the full Newton solve, yielding the greatest benefit in chaotic regimes where strongly contracting solutions are absent.

The numerical implementation for carrying out the Newton-Krylov periodic orbit search was adapted from \citet{willis2019equilibriaperiodicorbitscomputing}. The set of equations (2.14-2.18) is implemented in Dedalus, enabling direct coupling of the Newton-Krylov-GMRES algorithm with the fully resolved solutions of the chaotic falling-film equation. This implementation allows the refinement of near-recurrent states into invariant solutions within the same spectral framework used for the direct numerical simulations.

\section{Results \label{results_section}}

\subsection{Phenomenological regime map in $L-\delta$ space}

\begin{figure}
    \centering
    \begin{subfigure}[]{0.32\textwidth}
    \centering
        \includegraphics[width=\textwidth]{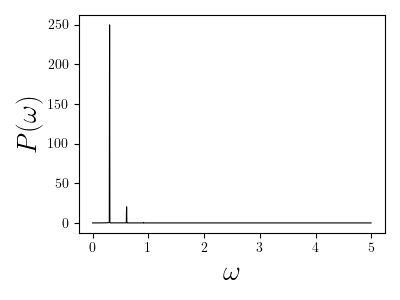}
        \caption{}
    \end{subfigure}
    \hfill
    \begin{subfigure}[]{0.32\textwidth}
    \centering
        \includegraphics[width=\textwidth]{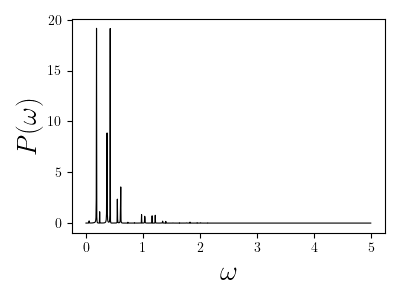}
        \caption{}
    \end{subfigure}
    \hfill
   \begin{subfigure}[]{0.32\textwidth}
       \includegraphics[width=\textwidth]{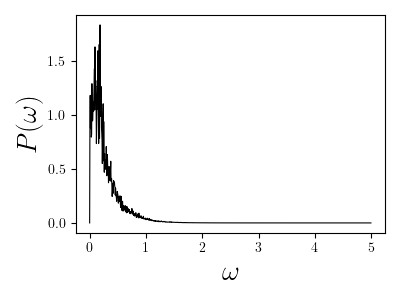}
       \caption{}
   \end{subfigure}
   \hfill
    \begin{subfigure}[]{0.32\textwidth}
        \includegraphics[width=\textwidth]{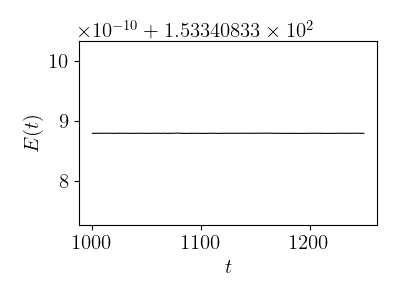}
        \caption{}
    \end{subfigure}
    \hfill
    \begin{subfigure}[]{0.32\textwidth}
        \includegraphics[width=\textwidth]{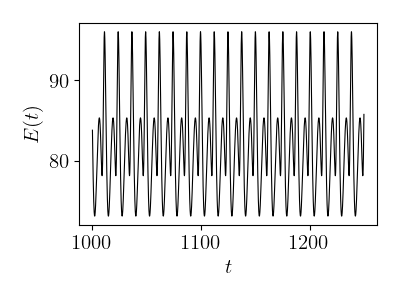}
        \caption{}
    \end{subfigure}
    \hfill
    \begin{subfigure}[]{0.32\textwidth}
        \includegraphics[width=\textwidth]{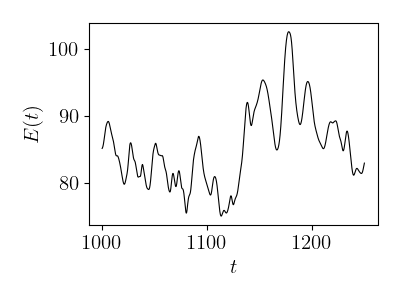}
        \caption{}
    \end{subfigure}
    \hfill
    \caption{Power spectra of solutions representative of the different dynamical regimes identified in the regime map shown in figure \ref{fig:regime_map}. (a): Travelling wave solution at $L=8.57,\delta=1.14329$. (b) 
    Bursting travelling wave solution at $L=15.71,\delta=0.53135$.
    (c) Chaotic solution at $L=31.43,\delta=0.85771$. 
    Panels (d)–(f) show the temporal evolution of the energy for the cases in (a)-(c), respectively.} \label{PowerSpectra}
\end{figure}

\begin{figure}
\begin{center}
\begin{tabular}{cccc}
\multicolumn{4}{c}{\small Travelling wave: $L=8.57$, $\delta=1.14329$} \\
\includegraphics[width=0.2\textwidth]{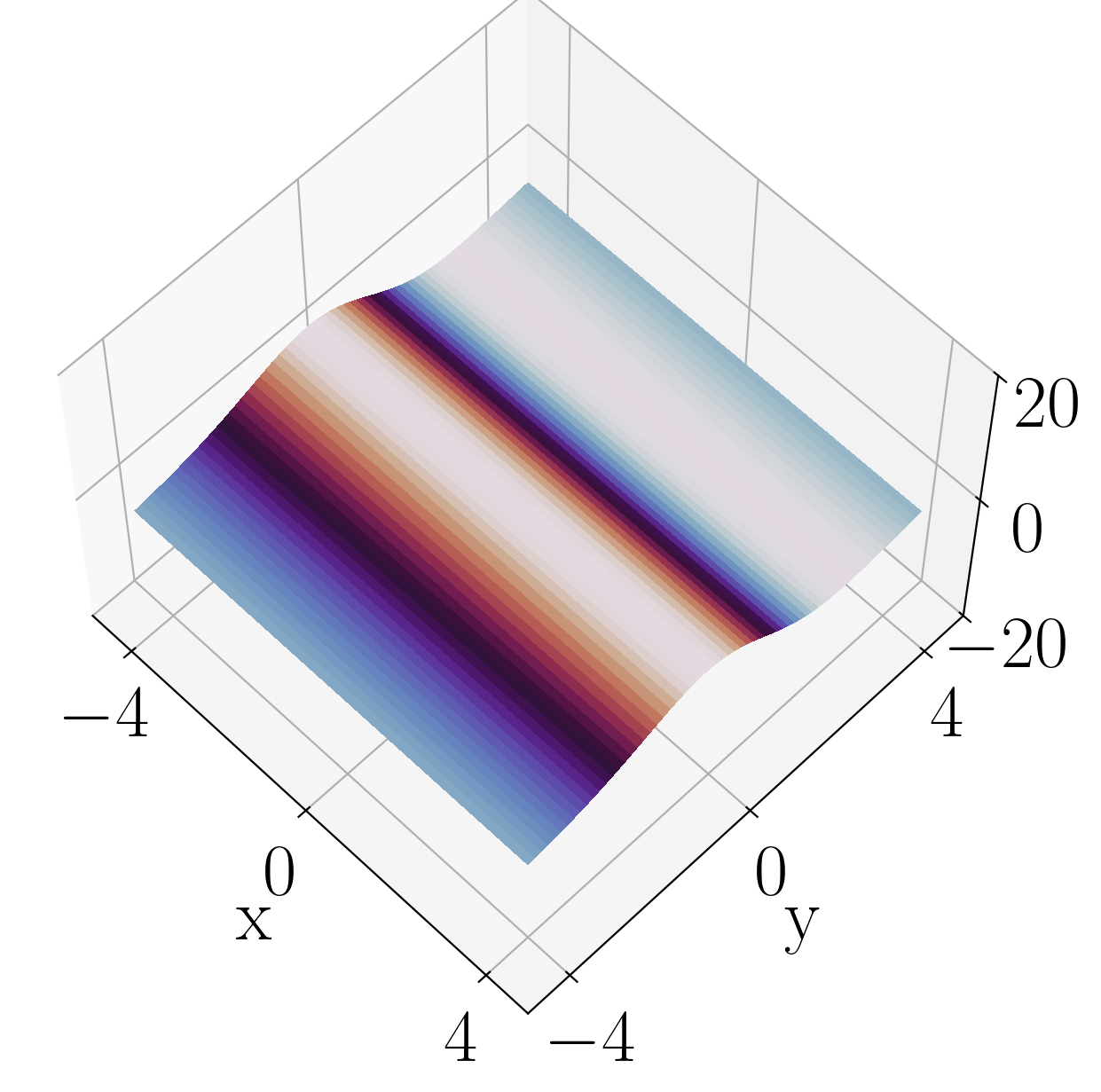} & 
\includegraphics[width=0.2\textwidth]{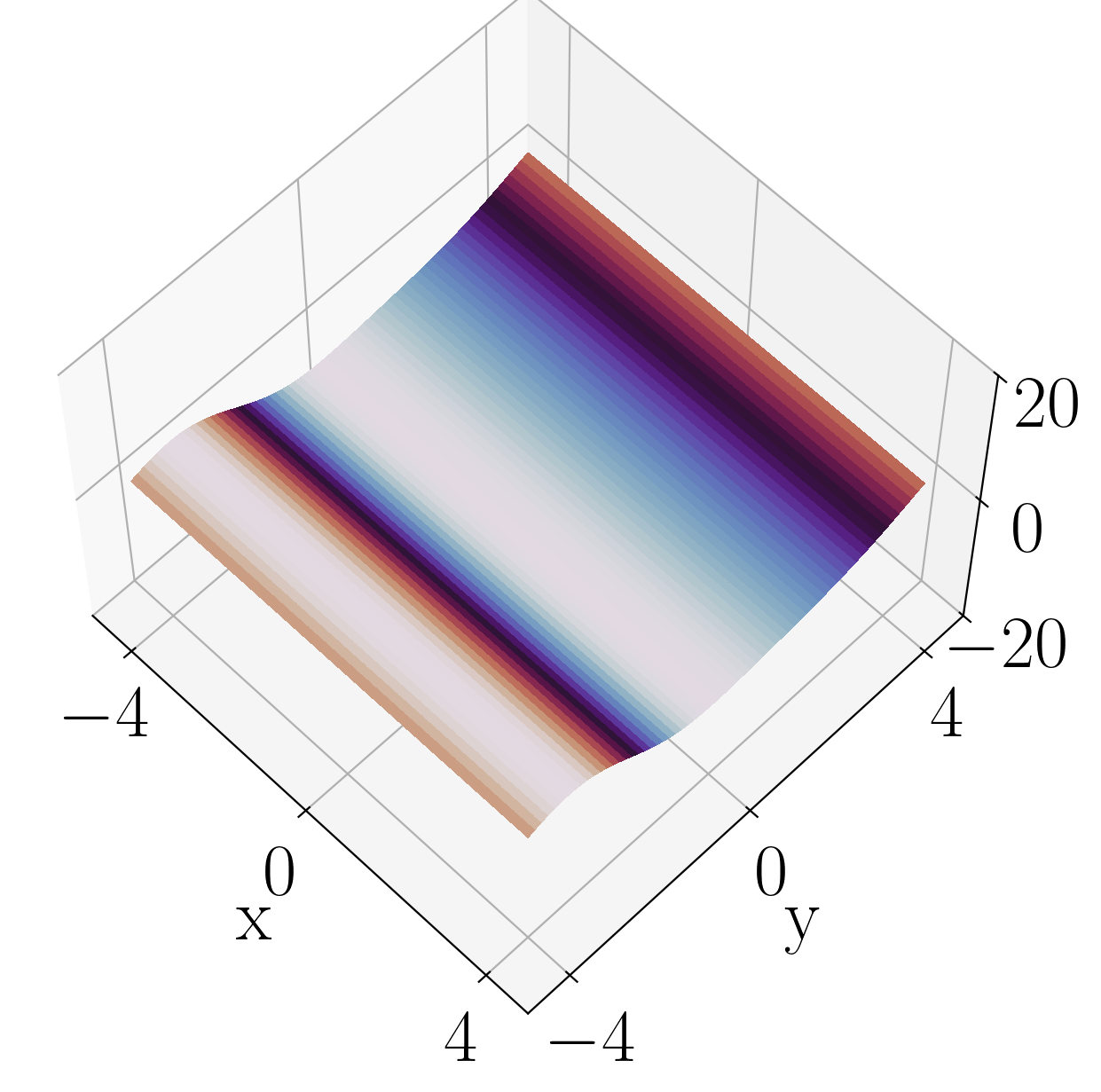} & 
\includegraphics[width=0.2\textwidth]{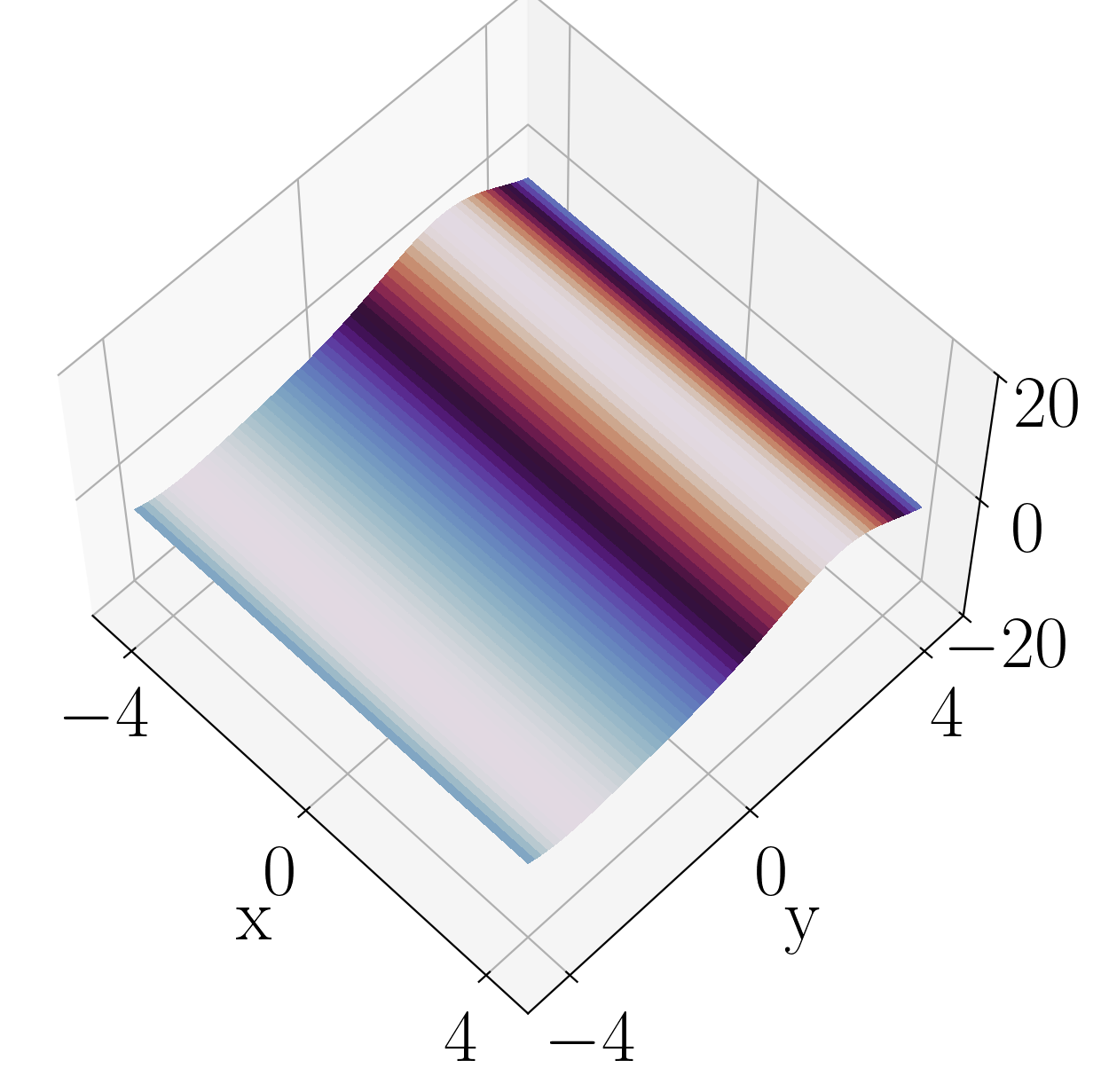} & 
\includegraphics[width=0.26\textwidth]{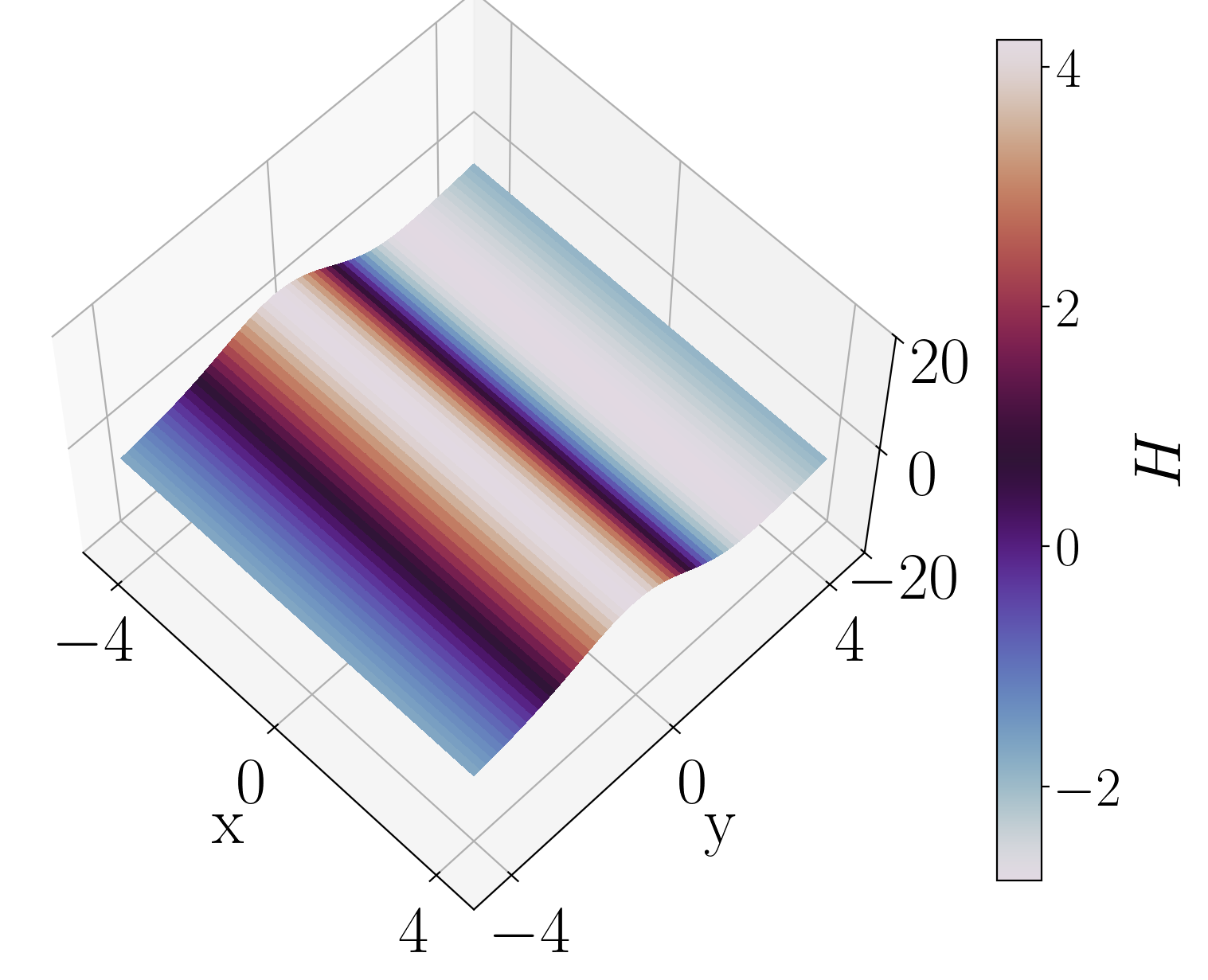} \\
{\footnotesize (a)} & {\footnotesize (b)} & {\footnotesize (c)} & {\footnotesize (d)} \\[8pt]


\multicolumn{4}{c}{\small Bursting travelling wave: $L=15.71$, $\delta=0.53135$} \\
\includegraphics[width=0.2\textwidth]{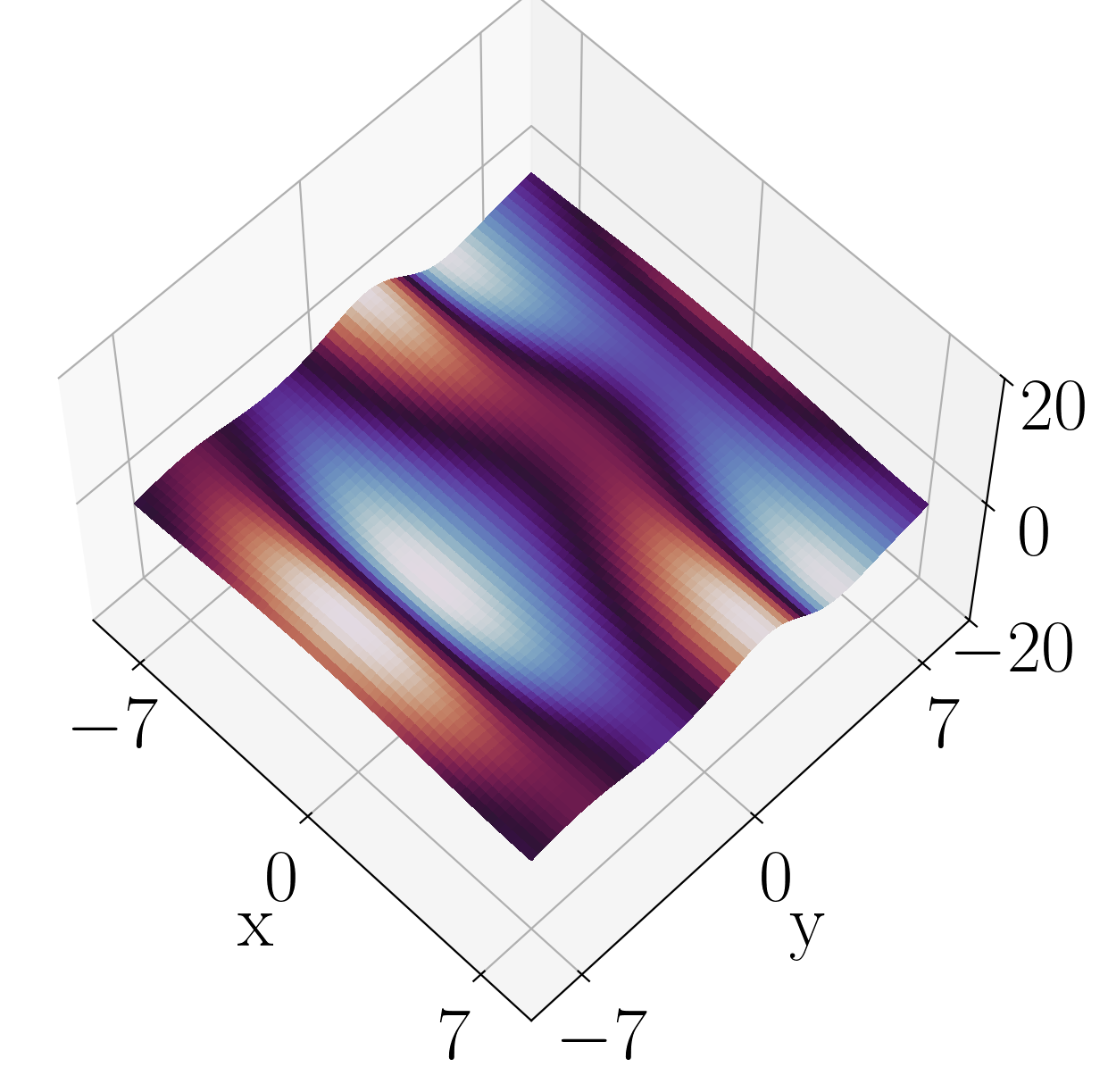} & 
\includegraphics[width=0.2\textwidth]{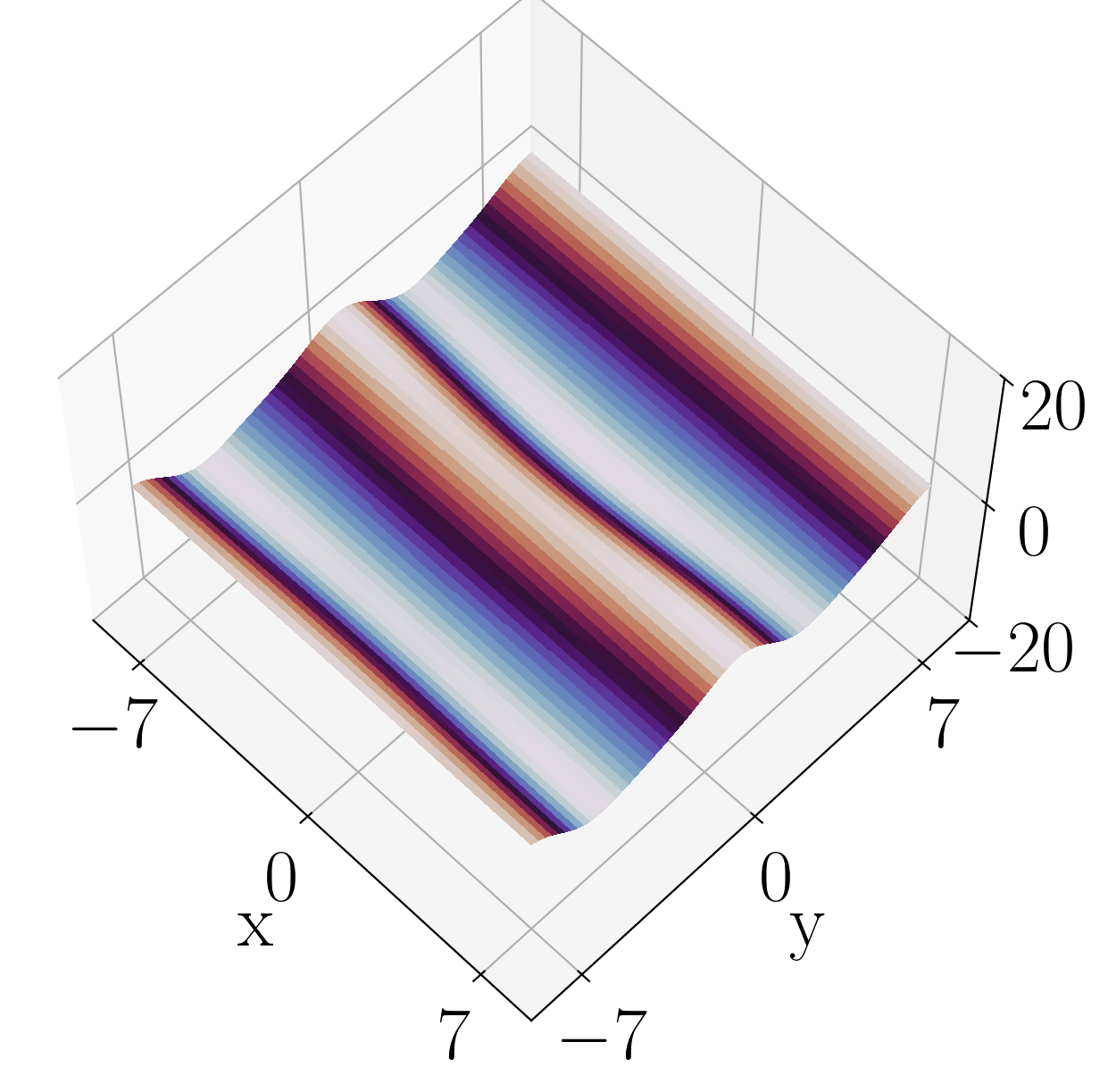} & 
\includegraphics[width=0.2\textwidth]{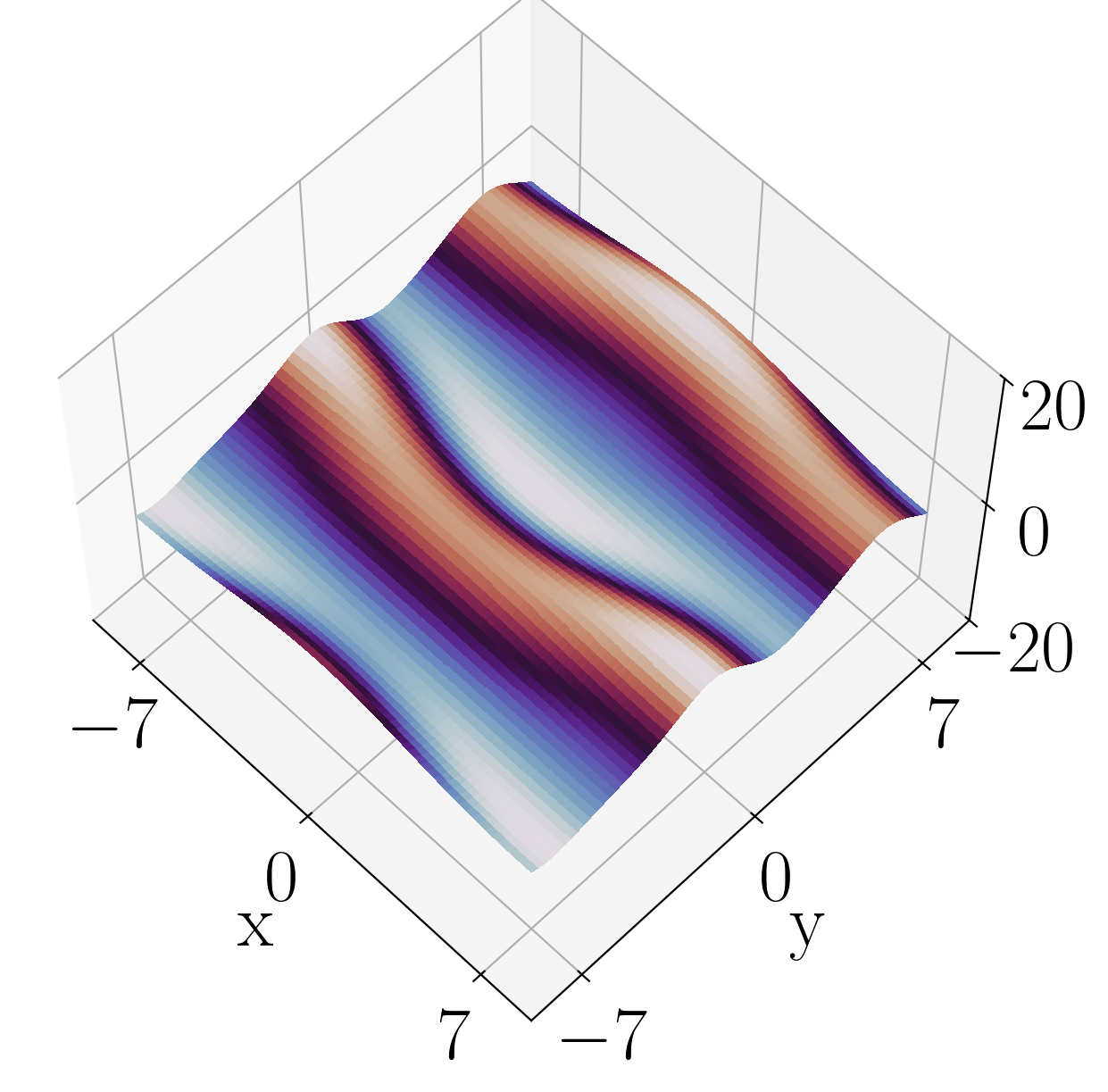} & 
\includegraphics[width=0.26\textwidth]{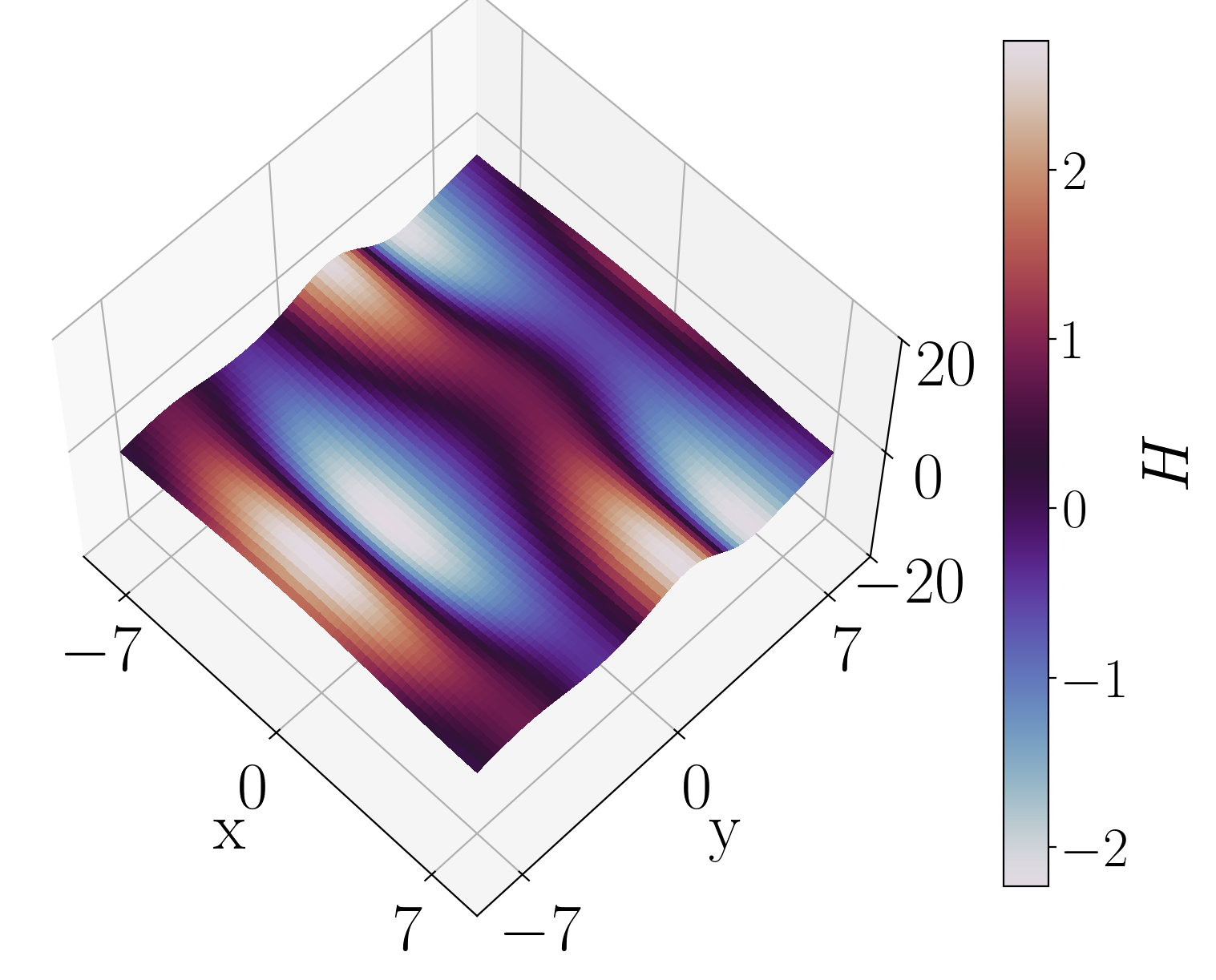} \\
{\footnotesize (e)} & {\footnotesize (f)} & {\footnotesize (g)} & {\footnotesize (h)} \\[8pt]

\multicolumn{4}{c}{\small Chaotic solution: $L=31.43$, $\delta=0.85771$} \\
\includegraphics[width=0.2\textwidth]{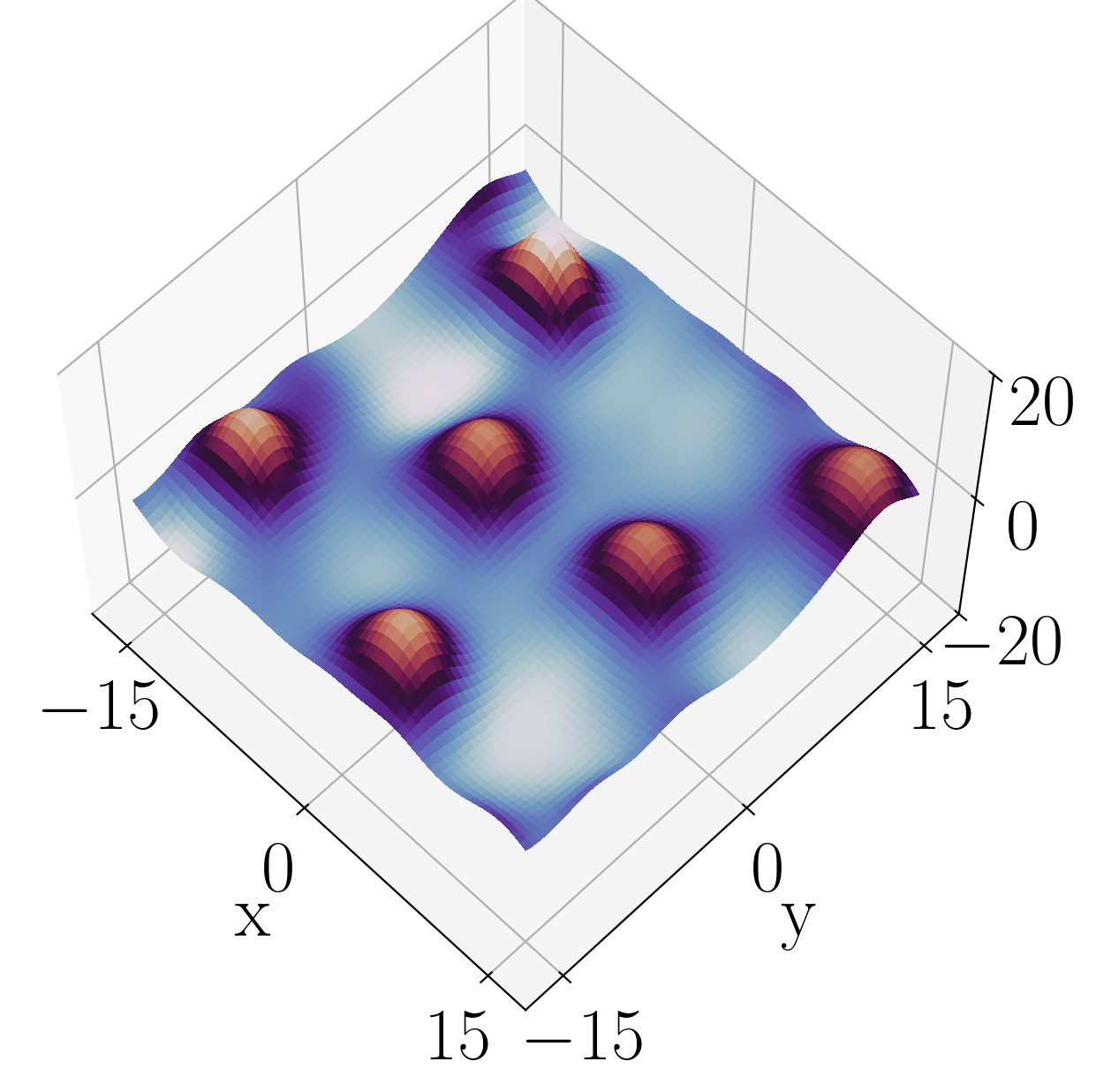} & 
\includegraphics[width=0.2\textwidth]{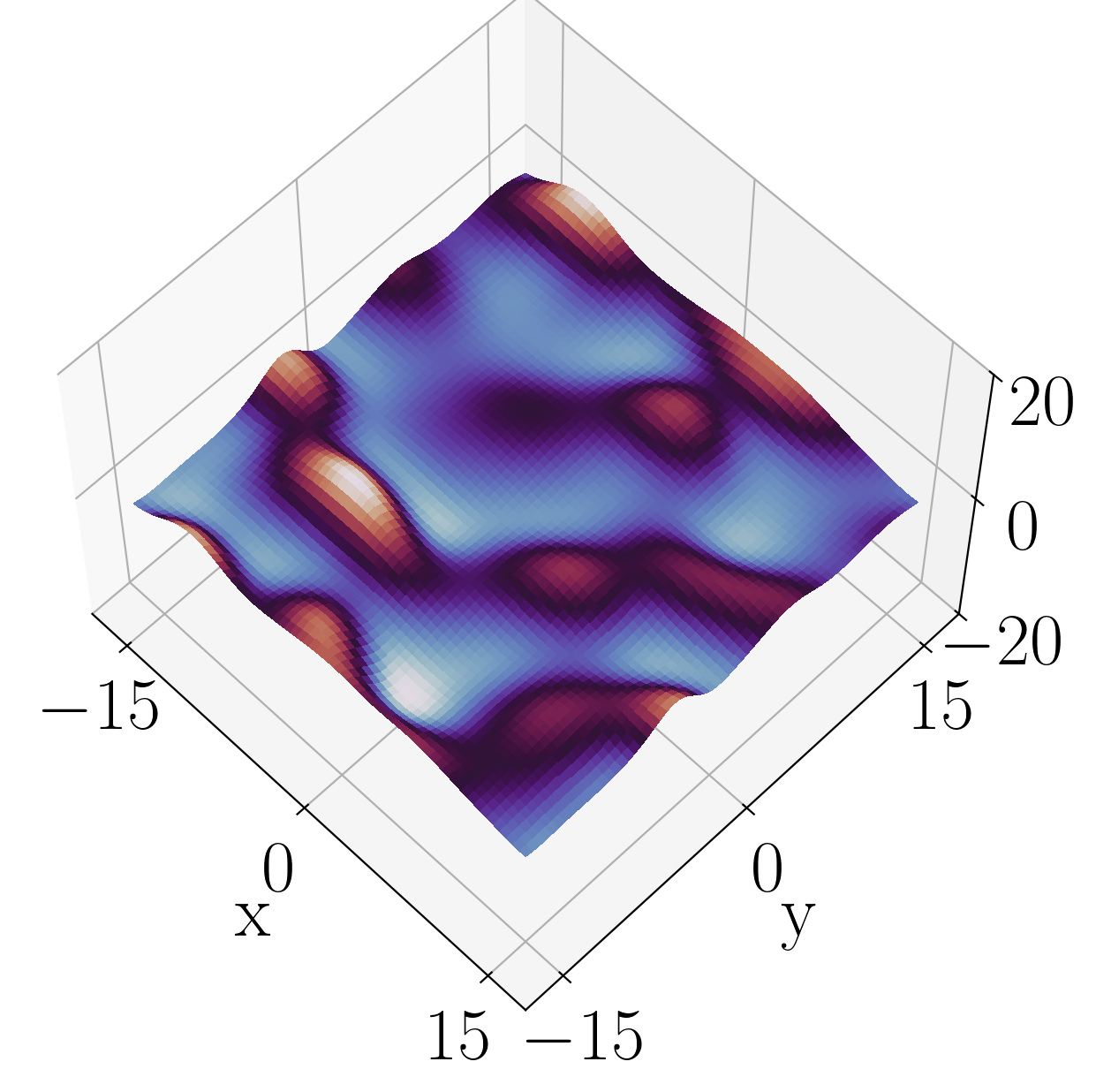} & 
\includegraphics[width=0.2\textwidth]{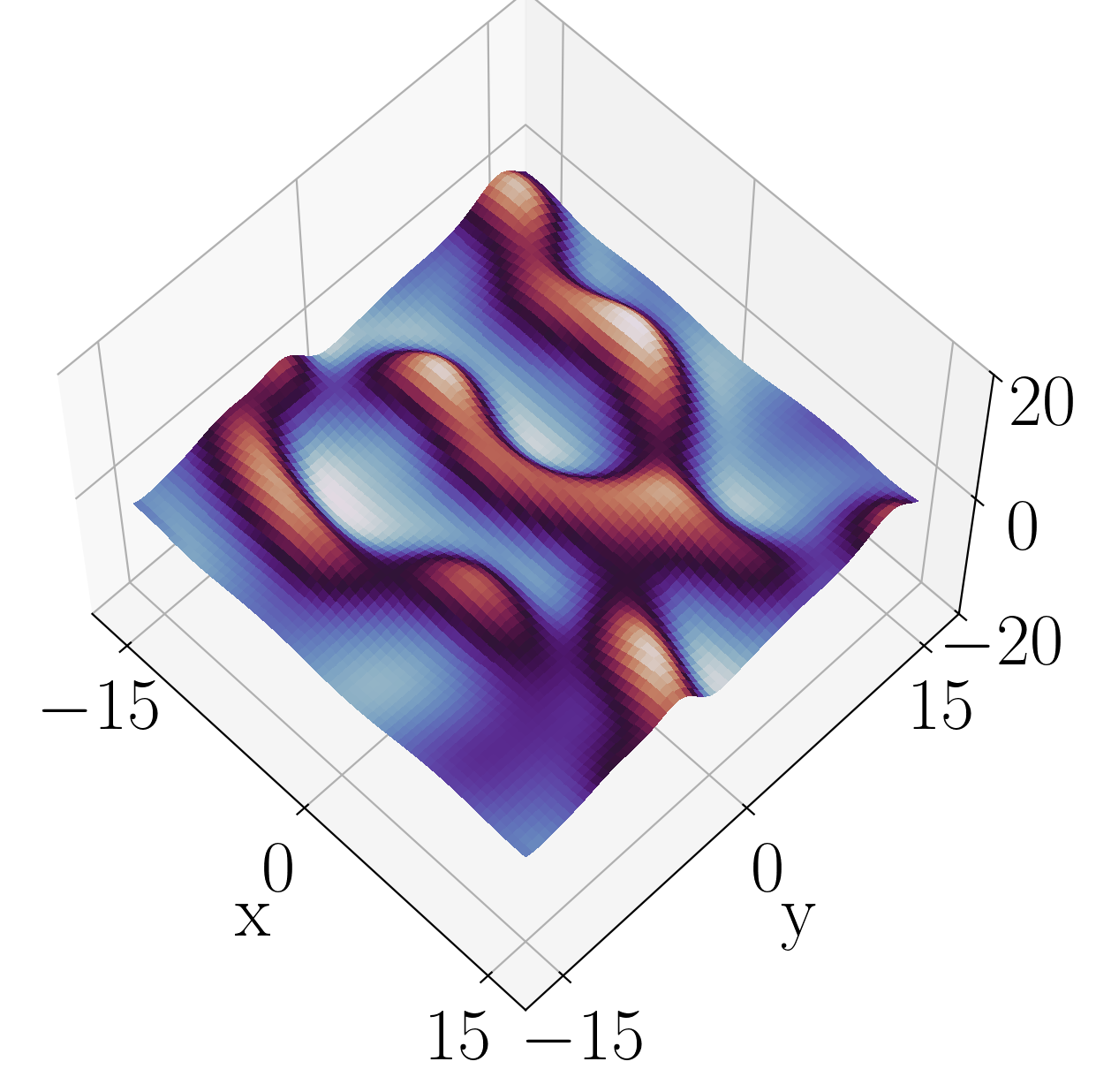} & 
\includegraphics[width=0.26\textwidth]{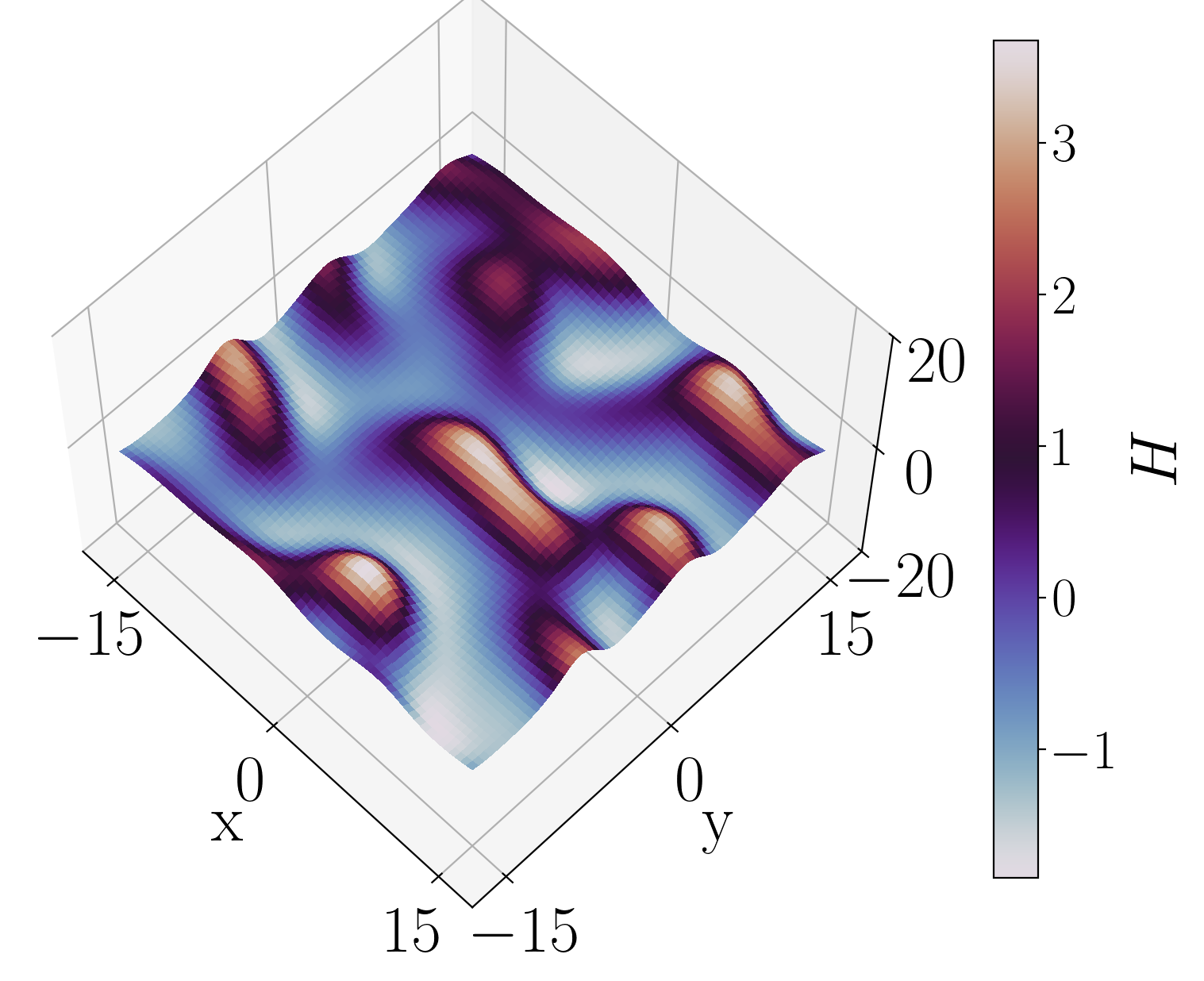} \\
{\footnotesize (i)} & {\footnotesize (j)} & {\footnotesize (k)} & {\footnotesize (l)} \\
\end{tabular}
\caption{  Spatio-temporal evolution of the interfacial dynamics for the cases shown in figure \ref{PowerSpectra}, representative of the different dynamical regimes identified in the regime map shown in figure \ref{fig:regime_map}.
(a-d) Travelling wave solution at $L=8.57$ and $\delta=1.14329$ at times $t=[0, 17.86, 35.73, 53.60]$.
(e-h)  Bursting travelling wave solution at $L=15.71,\delta=0.53135$ at times $t=[0, 11.53, 23.07,  34.61]$ respectively.
(i-l) Chaotic solution at $L=31.43 $ and $\delta=0.85771$ at times $t=[0, 33.33, 66.66, 100]$.
}\label{fig:snapshots}
\end{center}
\end{figure}

\begin{figure}
\begin{center}
 \includegraphics[width=\textwidth]{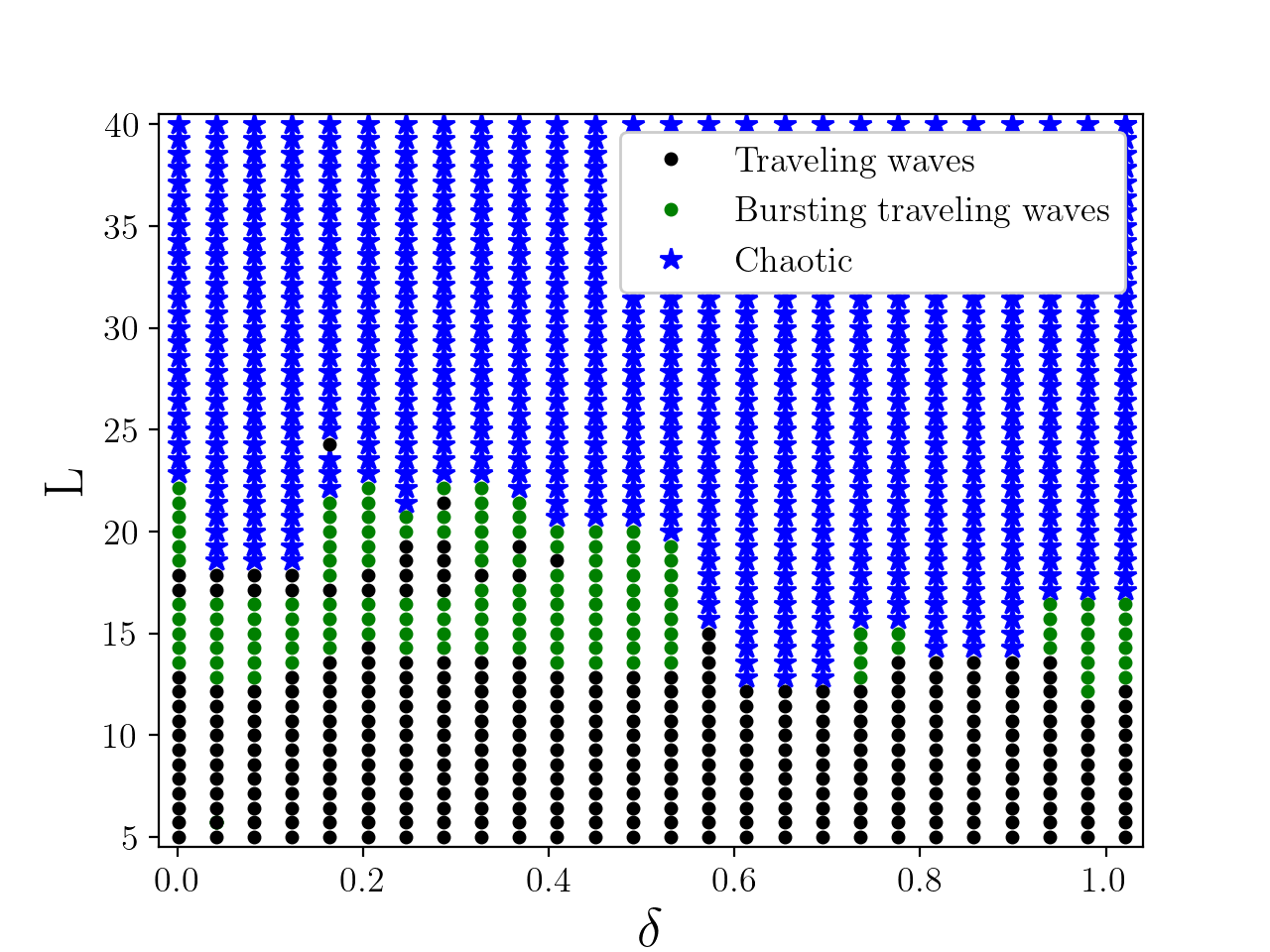}
\caption{Regime map of interfacial dynamics in falling films in the $(L-\delta)$ space. Here   $L\in[5,40]$ denotes the  domain length and and $\delta\in[0.001,1.2]$  controls the effective dissipation, corresponding to Reynolds numbers $Re \in[0.0541,35.38]$.
}
\label{fig:regime_map}
\end{center}
\end{figure}

 We begin the results section by presenting a comprehensive phenomenological regime map in the $(L,\delta)$ parameter space, constructed from nearly 2000 independent simulations of equation \ref{film_eq}, spanning $L\in [5,40]$ and  $\delta\in [0.002,1.1]$ (corresponding to effective Reynolds numbers $\Rey=35.38$ at $\delta=0.002$ and $\Rey=0.0541$ at $\delta=1.1$). Our exploration reveals a broad variety of dynamical behaviors, including travelling waves, bursting  travelling waves, and ultimately,  fully chaotic interfacial dynamics.

The classification of these different phenomena is based on
two different metrics. 
The first is the temporal evolution of the squared $L^2$ norm of the state,
$E(t) = \|\boldsymbol{x}(t)\|^2$,
which, in the present formulation, corresponds to the $L^2$ norm of the
discretised film height, and therefore provides a scalar measure
of the interfacial activity. 
The second is the temporal power spectrum of the  spatially averaged  film height, 
\begin{equation}
    P(\omega) = \frac{1}{N_T} \left| \int_{0}^{N_T} 
    \langle H(x,y,t) \rangle_{x,y}\, e^{-i\omega t}\, dt \right|^{2},
    \label{eq:power_spectrum}
\end{equation}
where 
$\langle \cdot \rangle_{x,y}$ represents a spatial average over the periodic domain. The integration is approximated using the FFT algorithm and performed over a sampling interval of duration $N_T=2000$ time units.

Figure \ref{PowerSpectra} shows $ P(\omega)$ for selected cases, illustrating the criteria used for $(L-\delta)$ regime classification. 
Figure \ref{PowerSpectra}a corresponds to a travelling wave solution at $L = 8.57$ and $\delta = 1.14329$, in which the $ P(\omega)$ shows a  single dominant peak at $\omega \approx \omega_0$, with a weak harmonic near $2\omega_0$.
The narrow and isolated peak indicates strictly periodic motion, 
consistent with a simple travelling wave whose structure propagates at constant phase speed. 
Figure \ref{PowerSpectra}b 
corresponds to a bursting travelling wave solution at $L=15.71 $ and $\delta=0.53135$.
Here, several sharp peaks appear in the low-frequency range, but unlike in figure \ref{PowerSpectra}a, they are not harmonically related. This indicates the presence of multiple active time scales of varying influence on the dynamics and intermittent amplitude bursts superimposed on a travelling-wave carrier. 
Figure \ref{PowerSpectra}c corresponds to a chaotic solution at $L = 31.43$ and $\delta = 0.85771$. The $ P(\omega)$ shows a continuous broadband spectrum with smoothly decaying energy toward higher frequencies, signifying a fully chaotic state with a continuous spectrum.
Finally, figure \ref{PowerSpectra}d-f show the  temporal evolution of the squared $L^2$ norm of the height
$E(t)$. These plots agree with what is observed  in the spectra such as a nearly constant value of energy, characteristic of a travelling wave (see figure \ref{PowerSpectra}d), periodical oscillations  with large amplitude for a bursting travelling wave (see figure \ref{PowerSpectra}e), and irregular  fluctuations with no clear pattern, indicating chaotic dynamics (see figure \ref{PowerSpectra}f).

Figure \ref{fig:snapshots} shows the spatiotemporal evolution of the film height for the cases presented in figure \ref{PowerSpectra}.
For the smallest domain ($L = 8.57$ and $\delta = 1.14329$), the dynamics converges to a  travelling wave in which the interface is spanwise uniform and periodic in the streamwise direction (see figure~\ref{fig:snapshots}a-d).  The wave propagates at constant speed with fixed amplitude and wavelength, and the surface consists of a single capillary ridge followed by a long, gently sloping trough.
When the domain is increased to $L=15.71$ with $\delta=0.53135$, the travelling wave pattern persists but becomes slowly modulated, exhibiting intermittent strengthening and weakening of the spanwise wavelength   (see figure~\ref{fig:snapshots}e-h). In particular, snapshots alternate between a nearly uniform travelling wave (figure~\ref{fig:snapshots}e) and states with pronounced amplitude modulation and localised deformation of the crests (figure~\ref{fig:snapshots}~e,g,h), consistent with a bursting travelling-wave regime. The interface shows clear departure from a strictly two-dimensional, single frequency travelling wave.
For the largest domain, $L=31.43$ at $\delta=0.85771$, the
film   displays strongly irregular, time-dependent patterns consisting of elongated ridges and troughs that continuously merge, split, and drift in both spatial directions (see figures \ref{fig:snapshots}i-l). 
This figure  shows that the transition from a coherent travelling wave to spatiotemporal chaos is enabled by the increasing domain size, which allows the coexistence and interaction of multiple nonlinear wavenumbers.

We now describe the phenomenology of the regime map shown in figure~\ref{fig:regime_map}. For small domains ($L \lesssim 13$) the system supports only travelling wave solutions over the entire range of $\delta$.  In this regime the dynamics settle into a single spatially periodic waveform of fixed shape translating at constant speed. The role of $\delta$ follows directly from the linear properties of (\ref{film_eq}).  Linearising about the flat state and considering Fourier modes $H\sim e^{\sigma t+i(k_x x+k_y y)}$ with $k^2=k_x^2+k_y^2$ gives $\sigma(k_x,k_y)=k_x^2-k^4+i\,\delta k^2 k_x$ . The real part of $\sigma$ determines the unstable band of long-wave modes, which is independent of $\delta$, while the imaginary part produces a drift of disturbances. The corresponding phase speed scales as $|c_p|=|\delta|k^2$, so increasing $\delta$ primarily modifies the propagation speed without affecting linear growth rates. For fixed $\delta$, increasing the domain size $L$ enriches the dynamics. Larger domains admit a denser set of Fourier modes within the unstable band, allowing nonlinear interactions between multiple modes through the advective term $HH_x$.  These interactions lead to modulated travelling waves and, for sufficiently large $L$, to fully chaotic dynamics.
The transitions between regimes are not monotonic in $\delta$.  Since $\delta$ alters only the imaginary part of the spectrum, it changes the relative phase speeds of interacting modes without modifying their growth rates, thereby affecting the coherence of the nonlinear wave interactions. For sufficiently large domains the system remains chaotic for all $\delta$.  In this regime we identify relative periodic orbits that capture recurrent patterns embedded in the chaotic dynamics.

\subsection{Characterization of the interfacial chaotic regime}

The remainder of this paper focuses on the fully chaotic regime identified in the $(L, \delta)$ parameter space. Our objective is to characterize this regime by
estimating the intrinsic manifold dimension of the attractor as a function of $L$ for a low value of $\delta$, as it gives a larger value of Reynolds, and identifying embedded ECS that organize the chaotic trajectories.

\subsubsection{Manifold dimension}\label{results:manifold_dimension}

\begin{figure}
\begin{center}
\begin{tabular}{cc}
\includegraphics[width=0.4\textwidth]{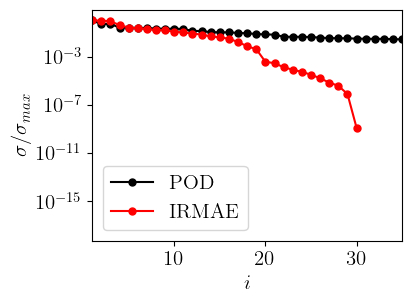} & 
\includegraphics[width=0.4\textwidth]{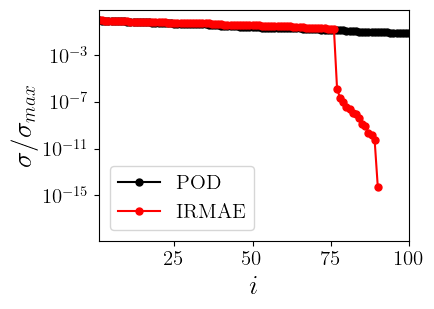} \\
 (a)    &  (b)\\
\includegraphics[width=0.5\textwidth]{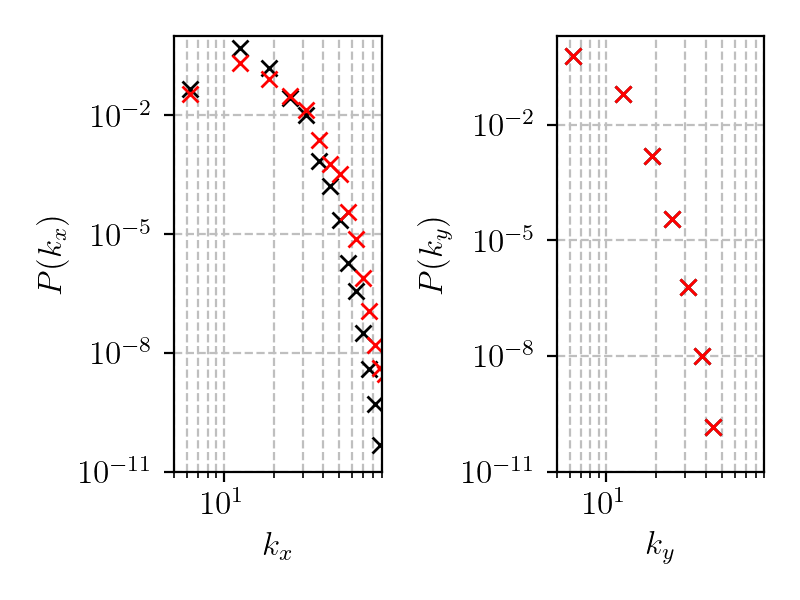} & 
\includegraphics[width=0.5\textwidth]{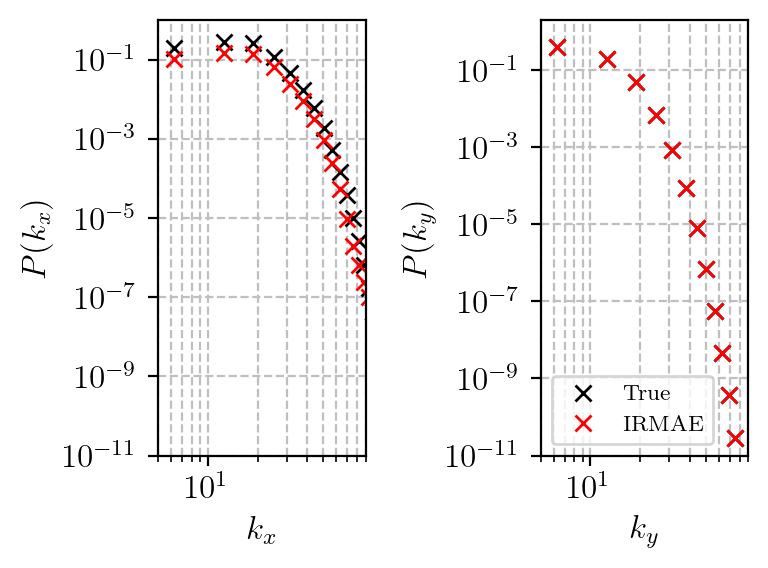} \\
 (c)    &  (d)\\
\end{tabular}
\caption{
Singular values, $\sigma$, normalized by the maximum singular value $\sigma_{max}$ of IRMAE learned latent spaces for falling films. Results are shown for representative cases: (a) $L=22$ and $\delta=0.002$, and (b) $L=30$ and $\delta=0.65373$.  The corresponding singular-value spectra obtained from POD are also shown in panels (a) and (b) for comparison. Panels (c) and (d) show the power spectra of the latent dynamics on the training dataset for cases (a) and (b), respectively. \label{sv_L22}}
\end{center}
\end{figure}

We estimate the manifold dimension for chaotic solutions following the procedure outlined in section \ref{AEs}, by examining the decay of the normalized singular values of the latent space data. A sharp drop in the singular values indicates the intrinsic dimension, $d_\mathcal{M}$, of the inertial manifold on which the dynamics evolve. 
Figure \ref{sv_L22}a shows the singular value spectrum for $L=22$ and $\delta=0.002$ ($Re=35.38$), and the expected manifold dimension is $d_\mathcal{M}=18$, as the
 singular values for $i > 18$ drop to $\approx 10^{-4}$.
In contrast, the singular value spectrum from a linear POD reduction decays only gradually, indicating that linear subspace approximations fail to capture the nonlinear geometry  of the attractor and thus substantially overestimate its intrinsic dimension.
For the case $\delta=0.65373$ ($Re=0.0541$) with $L=30,$ the intrinsic dimension increases to approximately $d_\mathcal{M}=76$. The IRMAE singular values decrease by more than six orders of magnitude  after $d_\mathcal{M}=76$, demonstrating clear spectral separation and confirming the existence of a compact yet higher-dimensional nonlinear manifold.


To assess the reconstruction fidelity of IRMAE, we compare the one-dimensional power spectra of the DNS fields with those of the reconstructed fields (figure~\ref{sv_L22}c,d for $L=22$ and $L=30$, respectively). In both cases the reconstructed spectra agree well with the DNS over the energetically dominant low- and intermediate-wavenumber ranges, indicating that the most energetic structures are accurately captured by the low-dimensional representation. Deviations appear only at the highest wavenumbers, where the reconstruction slightly underestimates the energy associated with short-wavelength oscillations. For the larger domain $L=30$, a modest under–representation is observed in the streamwise spectrum $P(k_x)$, while the spanwise spectrum $P(k_y)$ remains in closer agreement with the DNS. This difference is consistent with the structure of \eqref{film_eq}, in which nonlinear transfer and dispersion enter through streamwise derivatives (e.g., the terms $H H_x$ and $\delta \nabla^2 H_x$), making the streamwise gradients more challenging to reproduce in the low-dimensional representation.

\begin{figure}
\begin{center}
\includegraphics[width=0.5\textwidth]{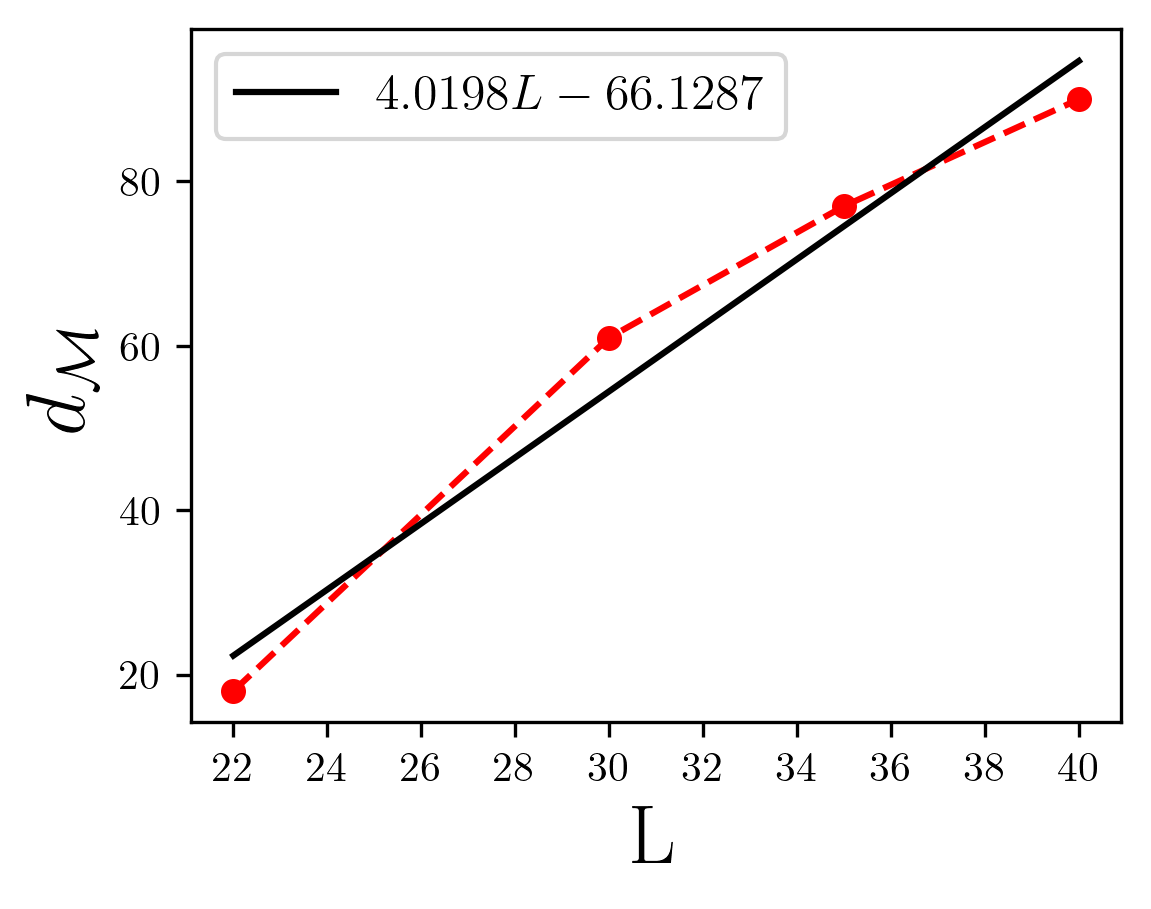}
\caption{
Estimated inertial manifold dimension 
$d_\mathcal{M}$  as a function of domain size $L$ with $\delta =0.002$.
}\label{L_dh}
\end{center}
\end{figure}

Figure~\ref{L_dh} shows the dependence of the estimated inertial manifold dimension $d_{\mathcal M}$ on the domain size $L$ for $\delta = 0.002$. 
The smaller domains fall below a linear trend because they are dominated by travelling-wave solutions with intrinsic dimension $d_{\mathcal M}=2$ (in a symmetry-reduced space). After $L=22$, a linear scaling emerges  once the dynamics become fully chaotic (excluding the travelling wave case at $L=25$), 
a linear fit yields
$d_{\mathcal M} \approx 4.01\,L - 66.1$.
For $L \gtrsim 22$, the approximately linear growth implies an effective dimension density $\rho_d \approx 4$ degrees of freedom per unit domain length, corresponding to a characteristic correlation length $\ell_c \sim 1/\rho_d \approx 0.25$. This behaviour is consistent with extensive spatiotemporal dynamics in which dynamically active degrees of freedom scale proportionally with system size, $d_{\mathcal M} \propto L$.  The approximately linear scaling with $L$ rather than $L^2$ suggests that the active degrees of freedom are organised primarily along the streamwise direction, indicating that the dynamics remain effectively quasi–one–dimensional despite the two–dimensional geometry. The observed scaling is  consistent with the 1D Kuramoto-Sivashinsky equation, where the attractor dimension scales linearly with the domain length \citep{Yang_prl,Takeuchi}.

\subsection{Identification of exact coherent states}

\begin{figure}
\centering
\begin{tabular}{c}
\includegraphics[width=0.5\textwidth]{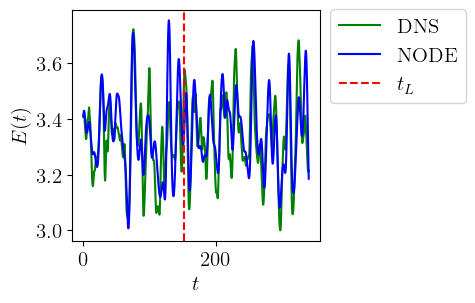}\\
(a)
\end{tabular}
\includegraphics[width=\textwidth]{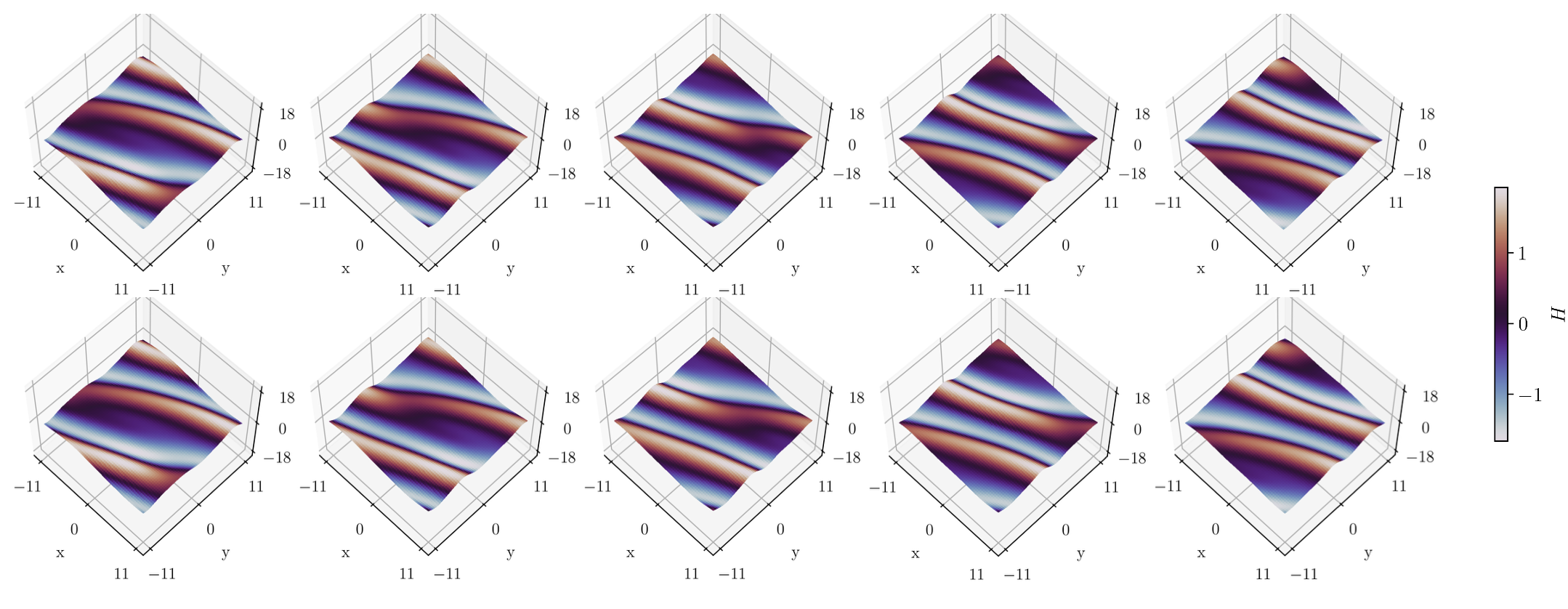}\\
(b)
\caption{ \textcolor{black}{(a) Temporal evolution of the squared $L^2$ norm of the film height $E(t)$  with $L=22$ and $\delta=0.002$}.
(b) Snapshots of the height field at times  $t=[0, 4, 8, 12, 16]$
illustrating the spatiotemporal complexity of the flow. The red dotted line indicates one Lyapunov time $t_L$ for this case.}
\label{fig:snapshots_22}
\end{figure}

The previous section shows that the chaotic attractor underlying the falling film dynamics is confined to a low-dimensional representation of finite dimension $d_\mathcal{M}$. 
To explore whether invariant solutions are embedded within this manifold, we train  neural ODE models in the manifold coordinates using different combinations of $L$ and $\delta$, which will be used to generate possible ICs to target ECS searches in DNS.

\begin{figure}
\centering
\begin{tabular}{ccc}
\includegraphics[width=0.33\textwidth]{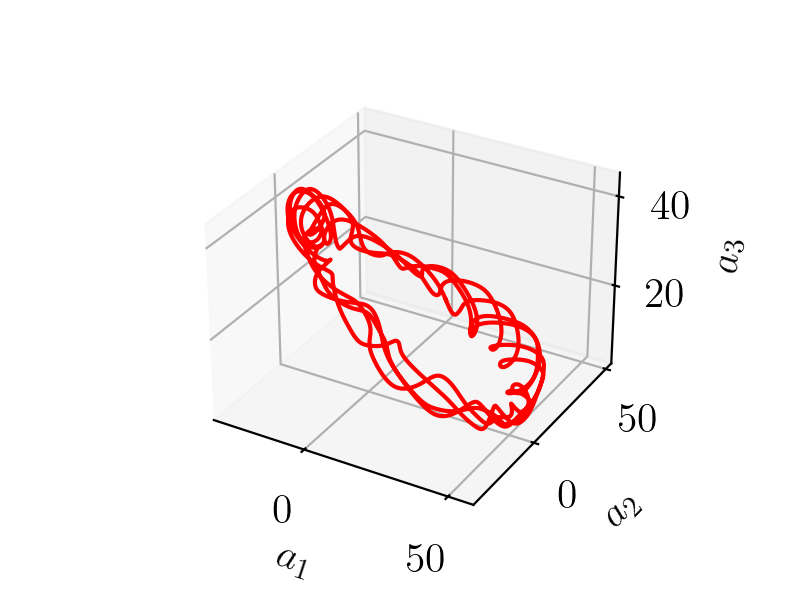}&  
\includegraphics[width=0.33\textwidth]{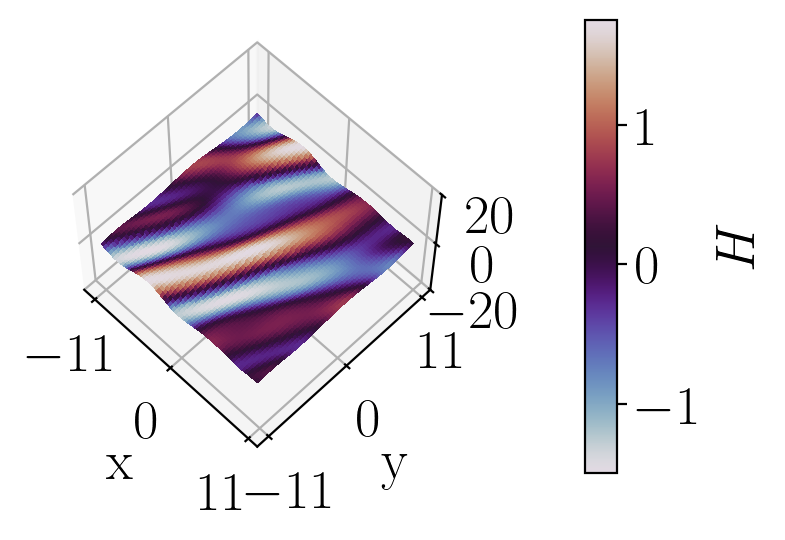}&  
\includegraphics[width=0.33\textwidth]{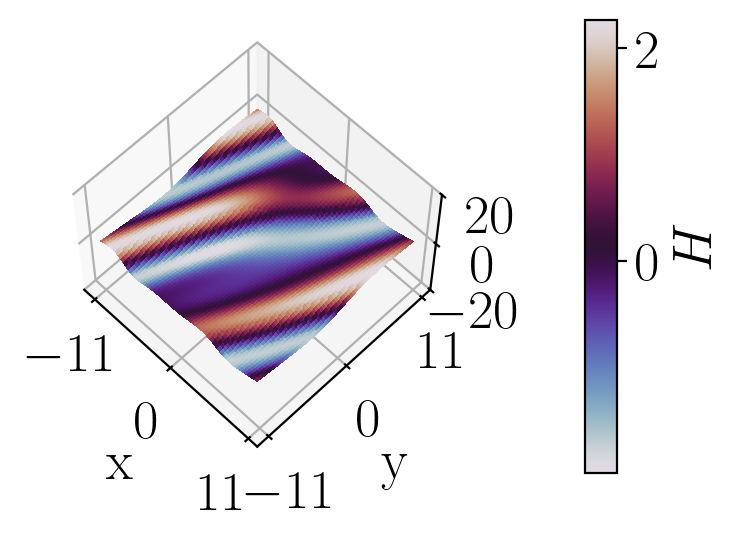}\\
(a) & (b) & (c)\\
\end{tabular}
\caption{
ECS search  for $RPO_{29.315}$.
(a) Invariant solution identified using the low-dimensional model. The visualization corresponds to a projection of  leading POD coefficients $(a_1,a_2,a_3)$. (b) Snapshots of the converged height field obtained from the low-dimensional model. (c) Snapshots of the corresponding invariant solution converged in the full state space using DNS.}
\label{fig:red_v_full_po}
\end{figure}

We first demonstrate that the low-dimensional models reproduce the intrinsic dynamics of the system in manifold coordinates for the case of  $L=22$ with $\delta=0.002$.
Figure \ref{fig:snapshots_22}a compares the temporal evolution of the model trajectories in manifold coordinates and their DNS counterpart from identical initial conditions. 
Figure~\ref{fig:snapshots_22}a shows the temporal evolution of the squared $L^2$ norm of the height fluctuations.
For the same initial condition, the low-dimensional model accurately reproduces the short time dynamics, with noticeable divergence only after $t \approx 150$.
The leading Lyapunov exponent (LLE) of the DNS for $L=22$ and $\delta=0.002$ is $\lambda_{\mathrm{LLE}}=0.006606$, corresponding to a Lyapunov time $t_L = 1/\lambda_{\mathrm{LLE}} \approx 151.4$ (see Appendix D for more details of the methodology for the LLE calculation). The NODE predictions can only forecast  over a time window that is only several times larger than one Lyapunov time, consistent with the expected behaviour of chaotic systems (see for example 
\cite{strogatz2001nonlinear,CRCA_pipe,linot2023dynamics}).
Figure~\ref{fig:snapshots_22}b displays
spatiotemporal evolution of the film height field  for the
model predictions and the corresponding DNS fields. The  interface shape exhibits oblique wavefronts that continuously drift, deform, and merge (not shown), reflecting the nonlinear behavior.
The agreement between the DNS and NODE predictions  confirms that the modelling in  manifold coordinates successfully captures the essential interfacial dynamics.
Details of the neural ODE architecture are given in Appendix B, while additional validation in terms of short-time tracking accuracy and long-time statistical properties is reported in Appendix C. We reiterate that the dynamical analysis presented here focuses on a 
representative case with $L=22$ and $\delta=0.002$. However, multiple NODE models were trained with different values of $L$ and  $\delta$ to facilitate a systematic search  for embedded ECS.  These models are not reported  as they are not the object of study but rather  computational tools enabling the identification of ECS.

Now that we have demonstrated that the low-dimensional model in manifold coordinates accurately captures the essential features of chaotic falling films, we now turn to exploring its use for the discovery of ECS. 
Generation of initial guesses in the DManD framework 
is done through the use of a recurrence function of the form

\begin{equation}
R(t;T) = 
\frac{
\|\boldsymbol{h}(t+T)-\boldsymbol{h}(t)\|_2}{\|\boldsymbol{h}(t)\|_2}.
\label{eq:recurrence}
\end{equation}

Figure \ref{fig:red_v_full_po}a shows the converged ECS in the manifold coordinates obtained by projecting onto the first three principal component modes, in which the trajectory collapses onto a RPO. When using this ECS in the DNS, the Newton-Krylov converges to $RPO_{29.315}$. Figures \ref{fig:red_v_full_po}b,c compare the film height of the ECS  converged in the  manifold coordinates with that obtained  from the full DNS. The two solutions exhibit close qualitative agreement,  indicating that both procedures converge to the same invariant solution  in the full state space. 
This demonstrates that the initial condition reconstructed from the  low dimensional representation lies within the Newton--Krylov  convergence neighbourhood of the ECS. It also confirms that the  reduced coordinates retain sufficient dynamical information to recover  a self-consistent invariant solution of the full system.

\begin{table}
\begin{center}
\begin{tabular}{cccccc}
\hline
Solution &$L$ & $\delta$ & $T$ & $r$ \\
$EQ_1$ & 22 & 0.002 & * & $8.020\times 10^{-9}$  \\
$EQ_2$ & 22 & 0.002 & * & $4.897\times 10^{-7}$  \\
$EQ_3$ & 30 & 0.65373 & * & $3.722\times10^{-5}$  \\

$TW_{85.291}$&10 & 1 & 85.291 & $1.003\times10^{-7}$  \\
$TW_{26.135}$&12 & 0.85 & 26.135 & $2.018\times 10^{-7}$  \\
$PO_{54.462} $& 30 & 0.65373 & 54.462 & $1.119\times 10^{-8}$  \\
$RPO_{37.995}$ & 14.29 & 0.08259 &  37.995  & $5.925\times10^{-7}$ \\
$RPO_{35.300}$&30 & 0.65373  & 35.443  & $6.602\times10^{-4}$   \\
$RPO_{187.385}$ & 22 & 0.002  & 187.385  & 
$1.381\times10^{-7}$ &  \\
$RPO_{28.840}$ & 22 & 0.002 & 28.840 & $8.409\times10^{-6}$ \\
$RPO_{45.01}$& 30 & 0.2  & 45.01  & $6.685\times10^{-9}$  \\
$RPO_{17.558}$& 30 & 0.65373  & 17.558 & $9.715\times10^{-4}$ \\
$RPO_{10.533}$& 22 & 0.002 & 10.5330 & $9.827\times10^{-5}$ \\
$RPO_{26.920}$& 22 & 0.002 & 26.920  & $1.143\times10^{-3}$ \\
$RPO_{20.503}$& 22 & 0.002 & 20.503 & $8.680\times10^{-6}$ \\
$RPO_{34.734}$& 19.29 & 0.65373 & 34.734  & $8.808\times10^{-7}$  \\
$RPO_{27.946}$& 19.29 & 0.65373  & 27.946  & $4.709\times10^{-8}$ \\
$RPO_{25.991}$& 22 & 0.002 & 25.991 & $5.018\times10^{-7}$ \\
$RPO_{25.788}$& 30 & 0.65373 & 25.788  & $6.470\times10^{-6}$  \\
$RPO_{29.315}$& 22 & 0.002 & 29.315 & $5.776\times10^{-5}$
\end{tabular}
\caption{
List of new invariant solutions embedded in the chaotic dynamics of falling films  using initial conditions generated by the low-dimensional model. The labels EQ, TW, and RPO correspond to equilibria, travelling waves, and relative periodic orbits, respectively. RPOs are labeled by their period ($T$).
The relative error $r$ is defined as $\|\text{shifted final state} - \text{initial state}\| / \|\text{initial state}\|$.
}\label{table:ecs_solutions}
\end{center}
\end{table}

Table~\ref{table:ecs_solutions} lists the first set of invariant solutions identified for chaotic falling films in the DNS (as far as the authors' knowledge). In total, twenty ECS are reported, including equilibria (EQ), travelling waves (TWs), and relative periodic orbits (RPOs). The solutions are organized by domain length $L$ and dispersion parameter $\delta$. RPOs are labeled by their period $T$, and the relative residual associated with each converged solution is provided. 
A clear dependence on the domain size is observed. For $L=22$ and $\delta=0.002$ the phase space contains multiple equilibria and a broad family of RPOs spanning more than an order of magnitude in period, indicating a strongly recurrent dynamics with several competing cycles. In contrast, for $L=30$ and $\delta=0.65373$ the set of invariant solutions is more sparse and is dominated by longer-wavelength structures, showing that the streamwise length controls the number and type of coherent states that can be sustained. The domain size therefore acts as a selection parameter for the dynamical skeleton of the flow.
The invariant solutions span a range of dispersion parameters.
At $L=22$  with short values of dispersion parameter  ($\delta=0.002$), long-period RPOs with $T \approx 180$ are observed, whereas at larger $\delta$ values the identified RPOs possess shorter recurrence times. This variation in $T$ reflects the increasing number of interacting wave structures and secondary modulations sustained in larger or more weakly stabilized domains.
From the full set, three representative RPOs are selected to illustrate the characteristic dynamics.

\begin{figure}
\centering
{\small  $RPO_{27.946}$ ($L=19.29,\quad \delta=0.65373$})\par\vspace{0.4em}
\begin{tabular}{c}
\includegraphics[width=0.19\textwidth]{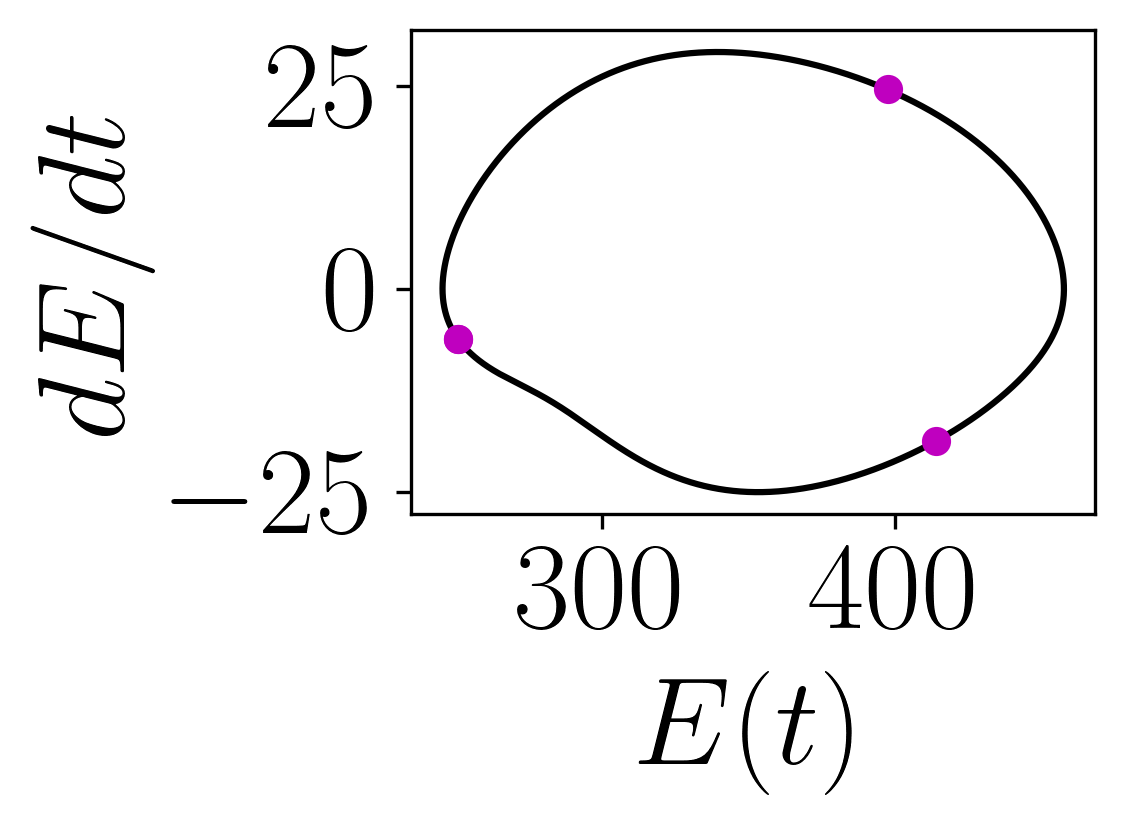}\\
\end{tabular}
\begin{tabular}{cccc}
\includegraphics[width=0.17\textwidth]{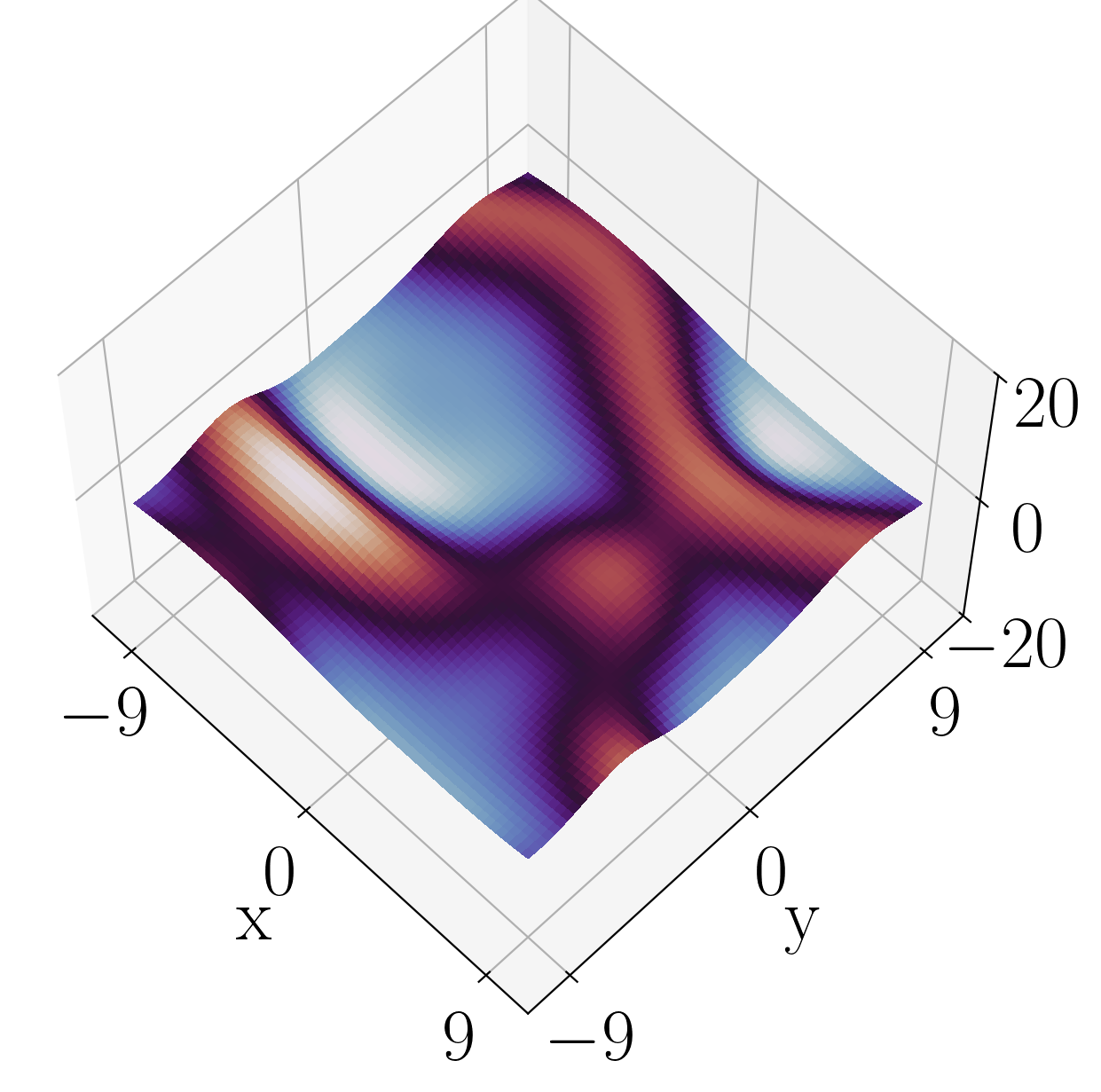}&  
\includegraphics[width=0.17\textwidth]{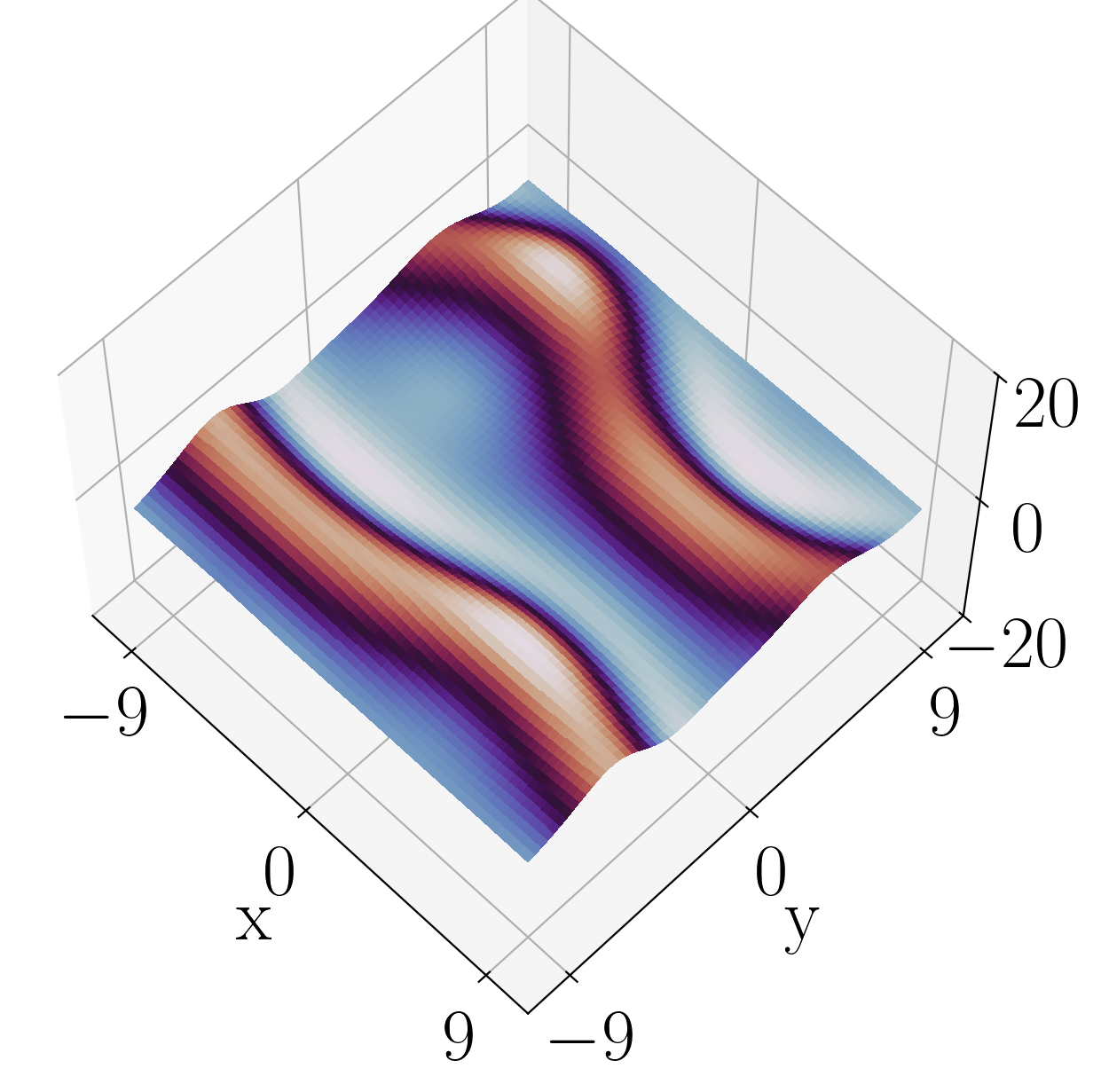}&  
\includegraphics[width=0.17\textwidth]{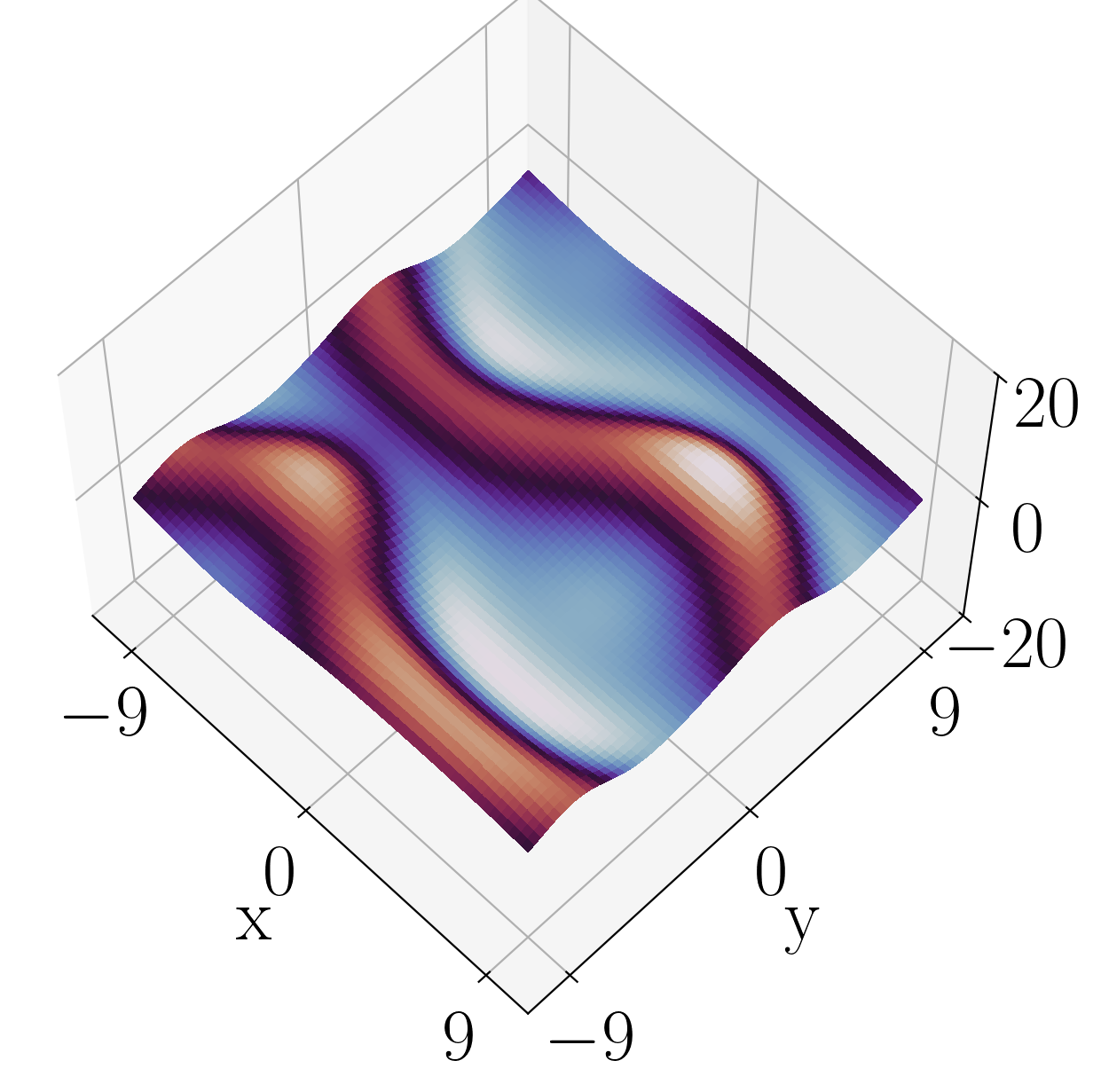}&
\includegraphics[width=0.215\textwidth]{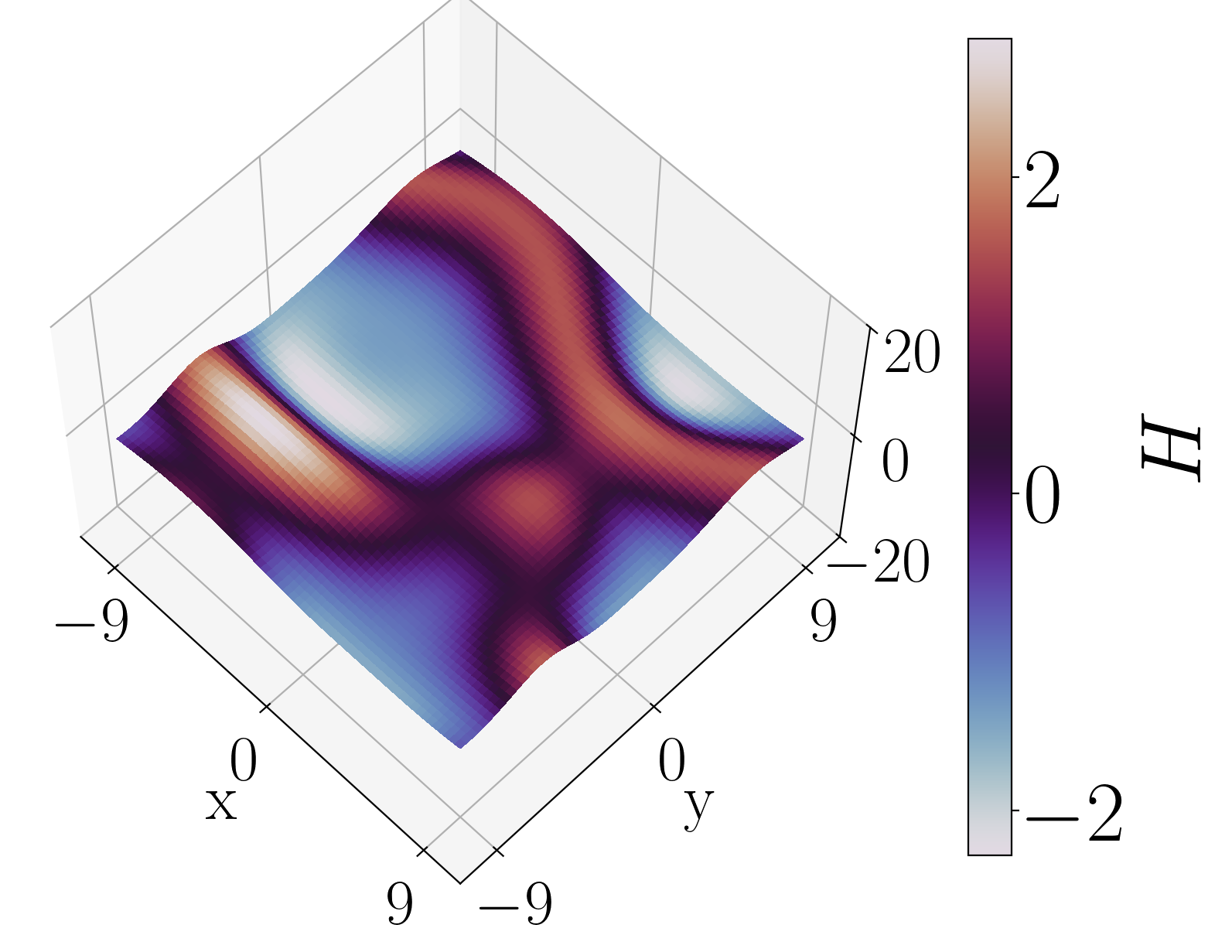}\\
\end{tabular}
{\small $RPO_{25.991}$  ($L=22,\quad \delta=0.002$)}\par\vspace{0.4em}
\begin{tabular}{c}
\includegraphics[width=0.195\textwidth]{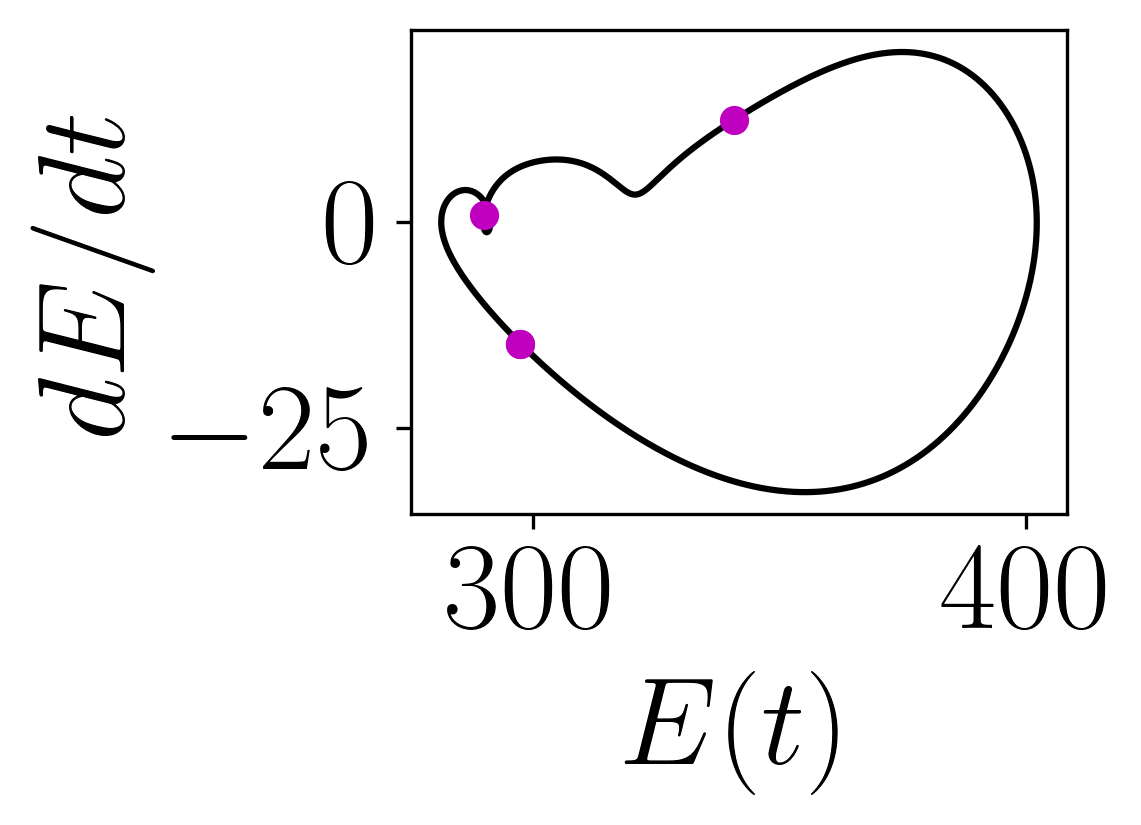}\\
\end{tabular}
\begin{tabular}{cccc}
\includegraphics[width=0.17\textwidth]{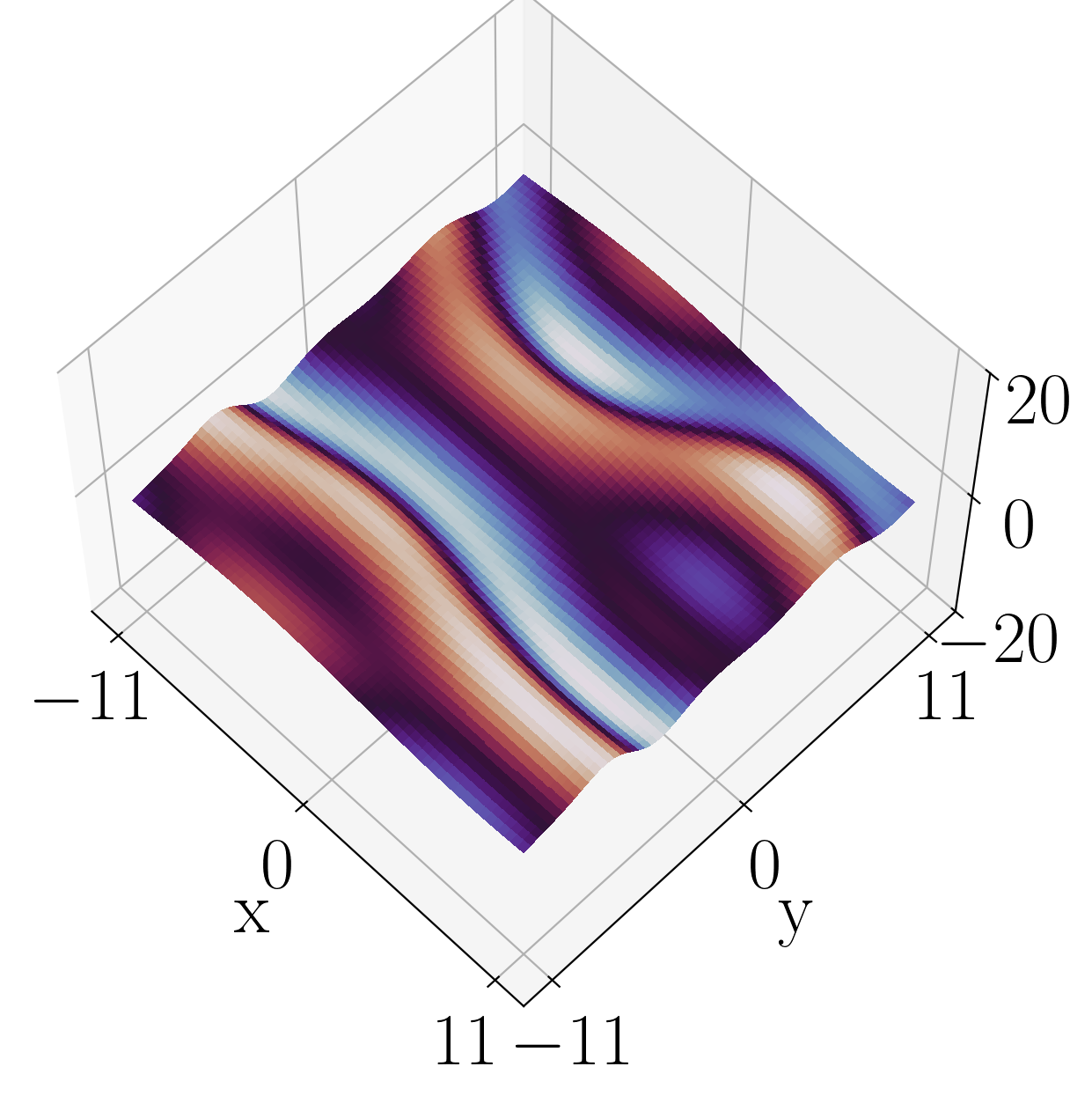}&  
\includegraphics[width=0.17\textwidth]{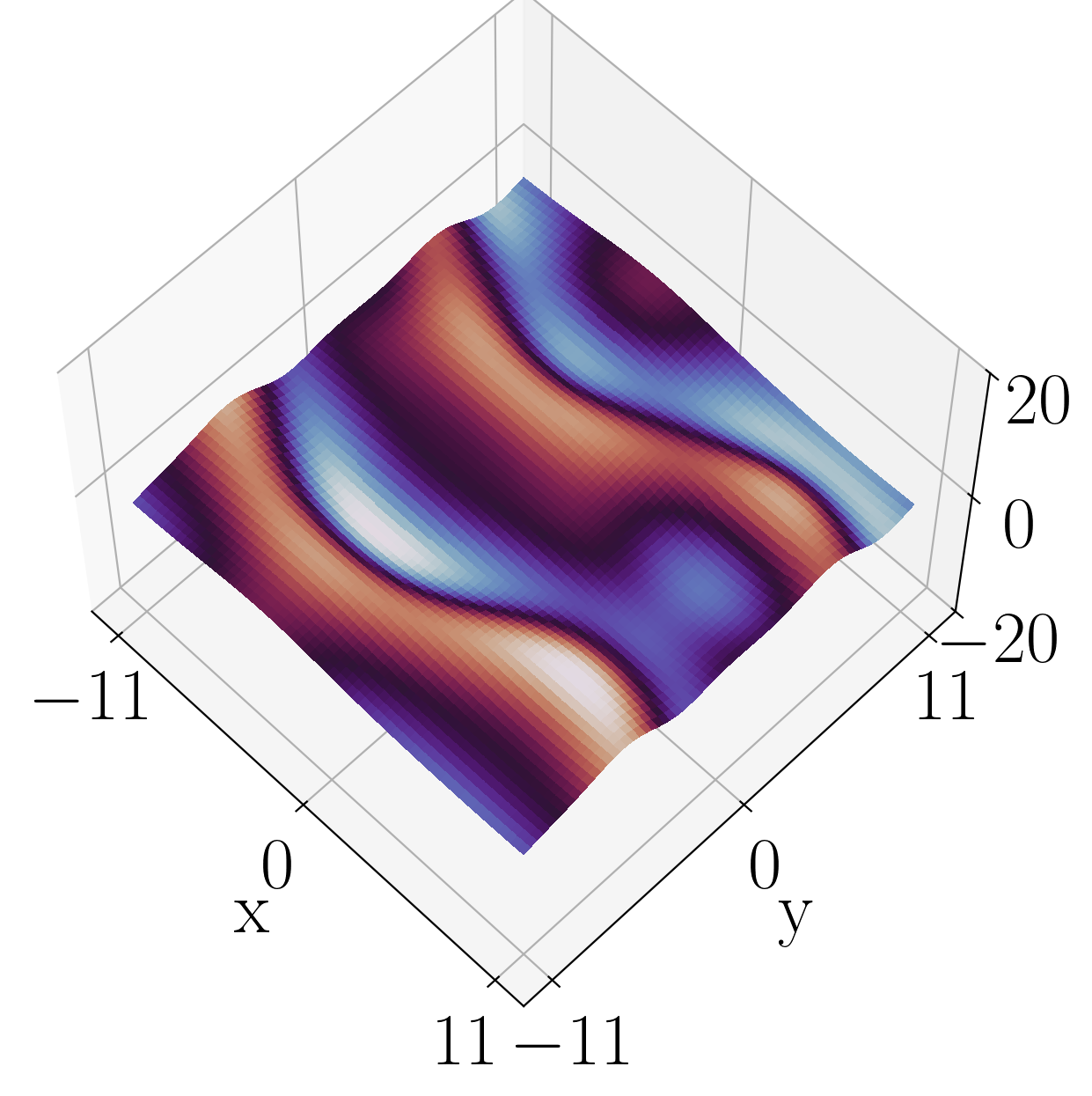}  
\includegraphics[width=0.17\textwidth]{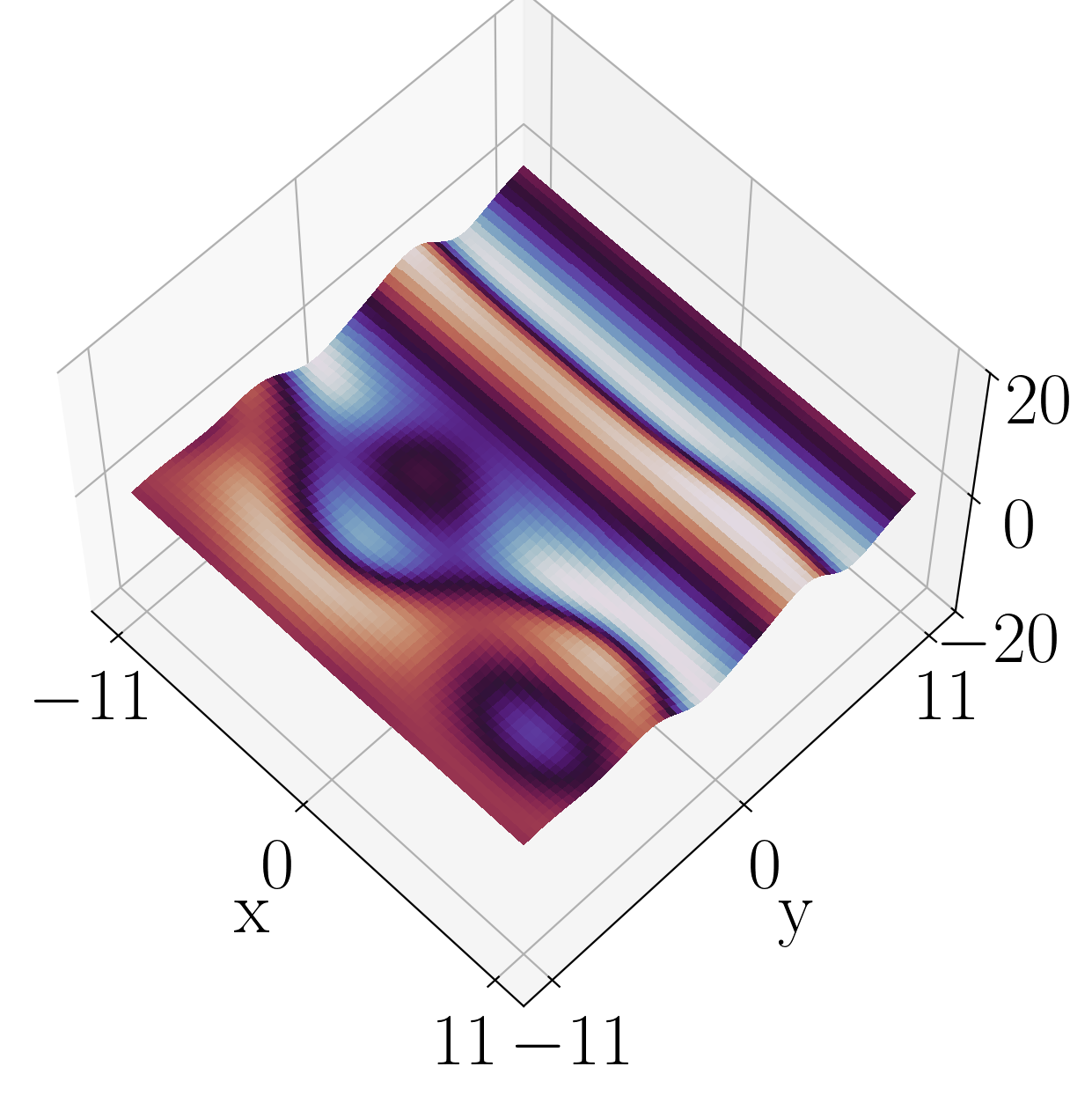}&
\includegraphics[width=0.215\textwidth]{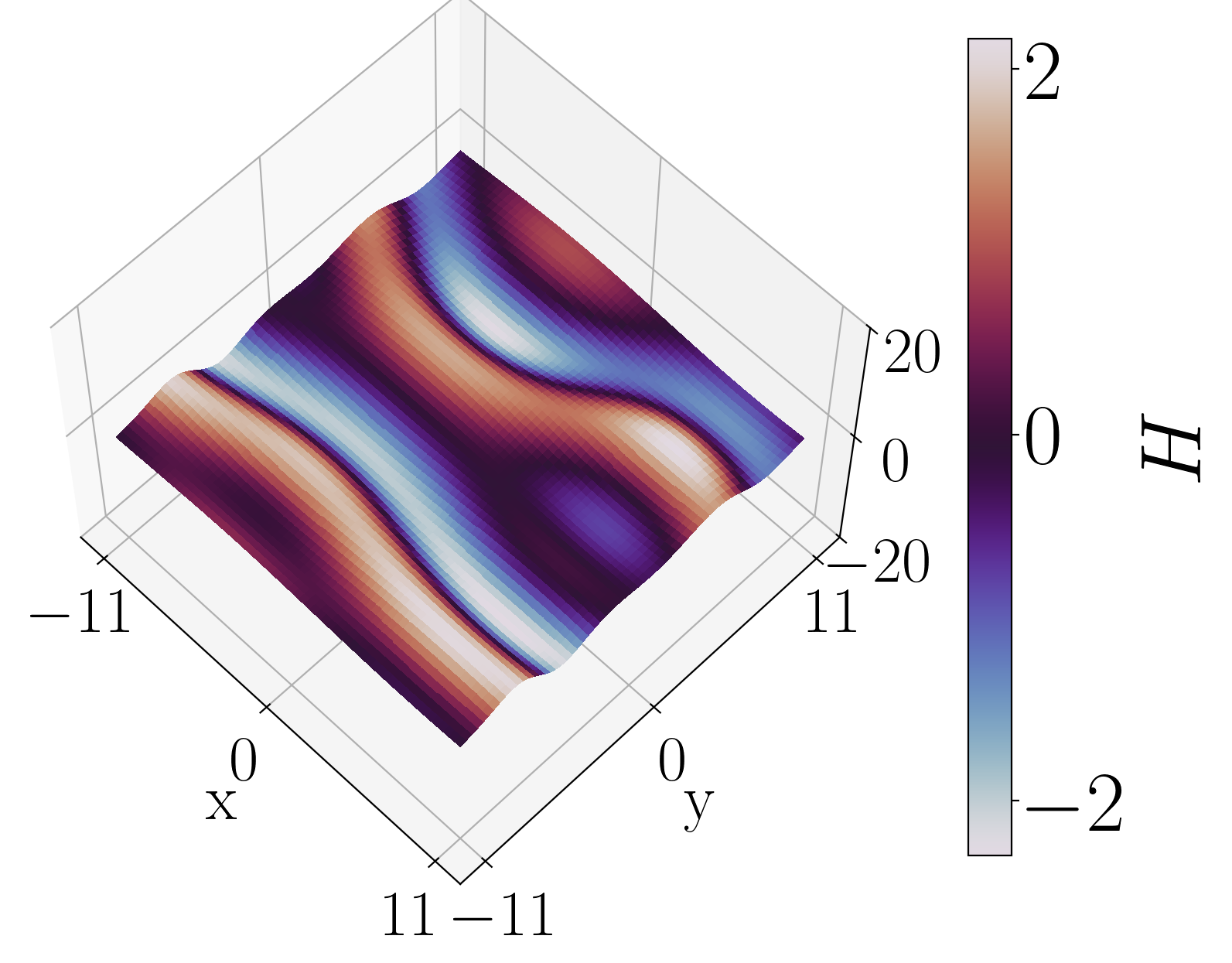}\\
\end{tabular}
{\small $RPO_{25.788}$  ($L=30,\quad \delta=0.65373$)}\par\vspace{0.4em}
\begin{tabular}{c}
\includegraphics[width=0.195\textwidth]{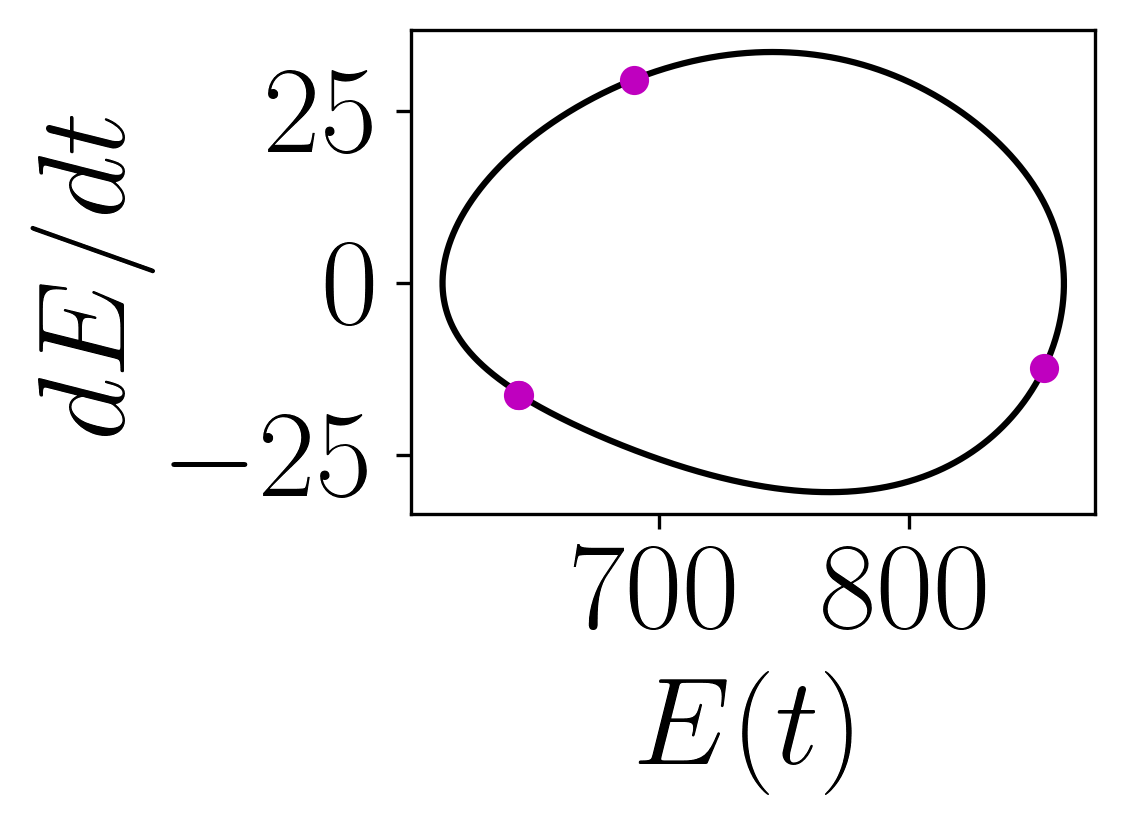}\\
\end{tabular}
\begin{tabular}{cccc}
\includegraphics[width=0.17\textwidth]{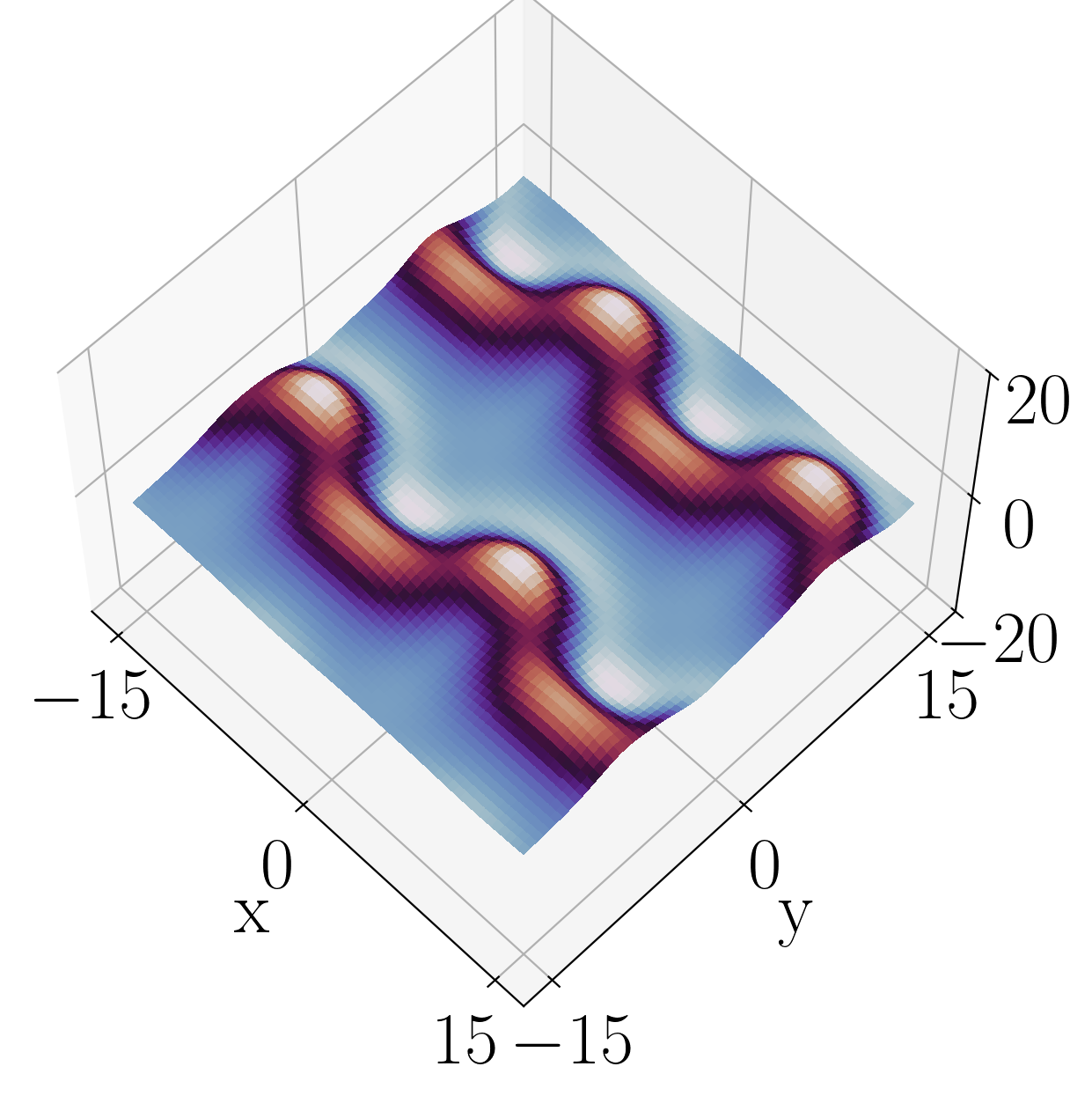}&  
\includegraphics[width=0.17\textwidth]{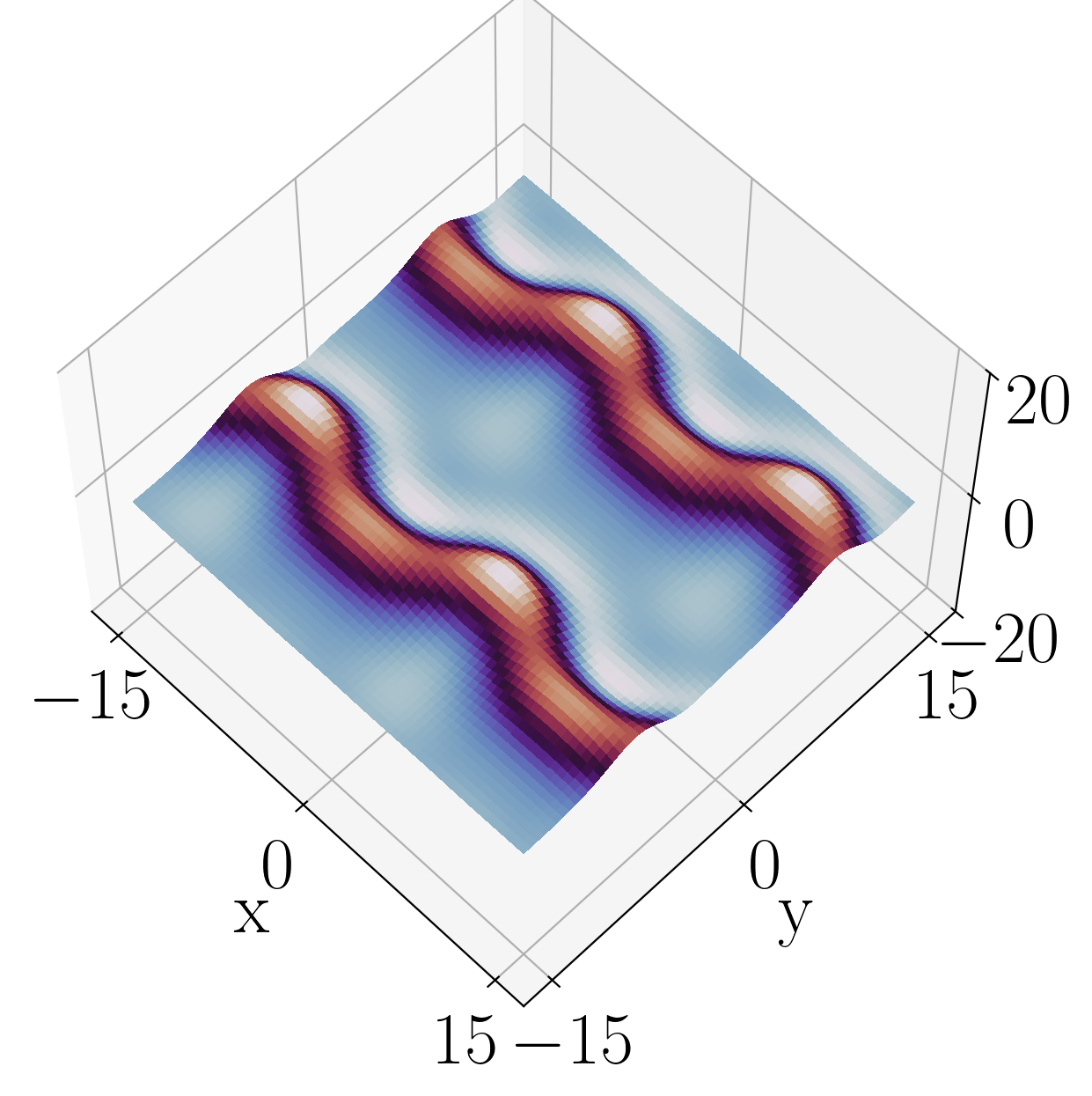}&  
\includegraphics[width=0.17\textwidth]{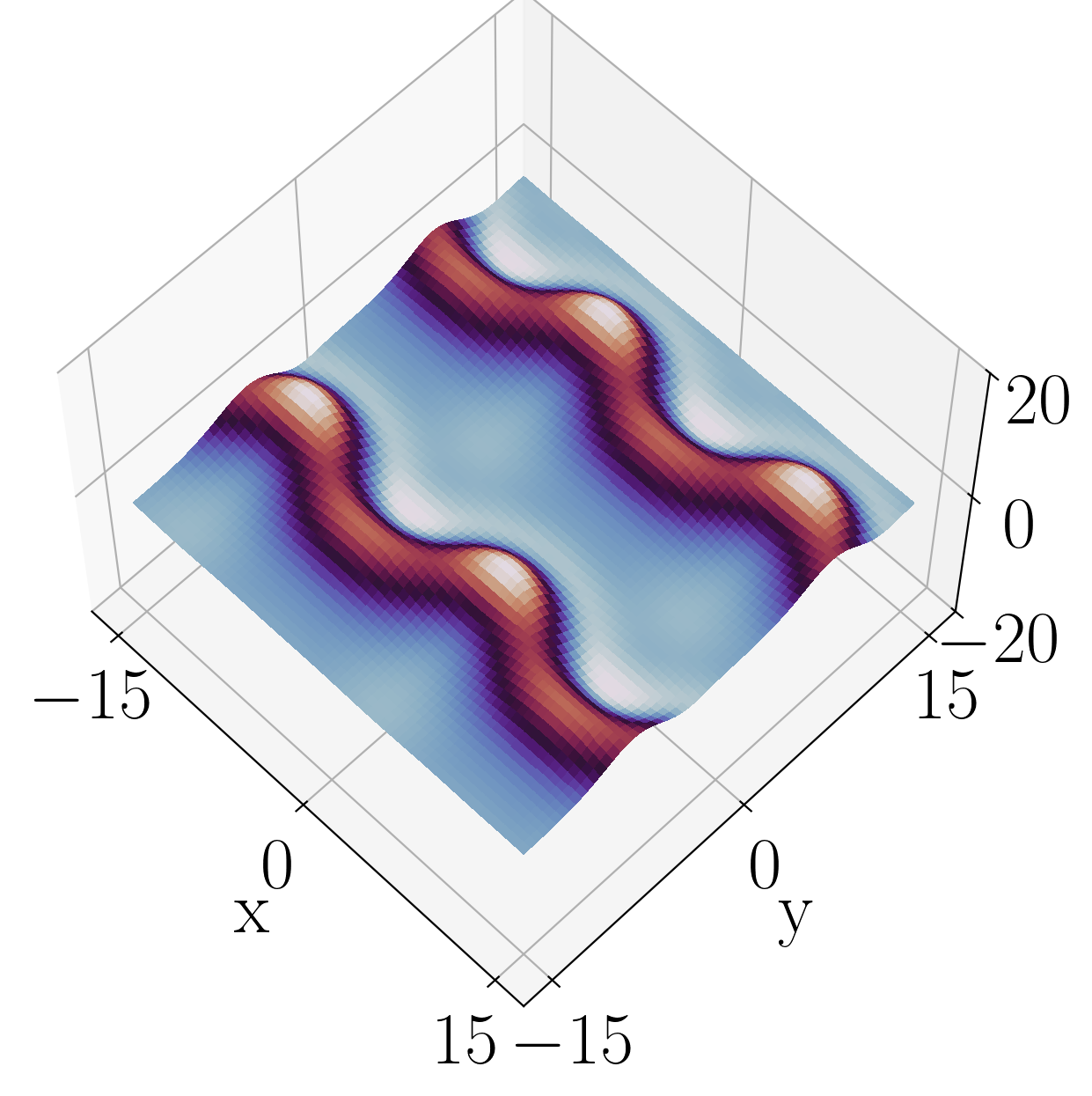}&
\includegraphics[width=0.205\textwidth]{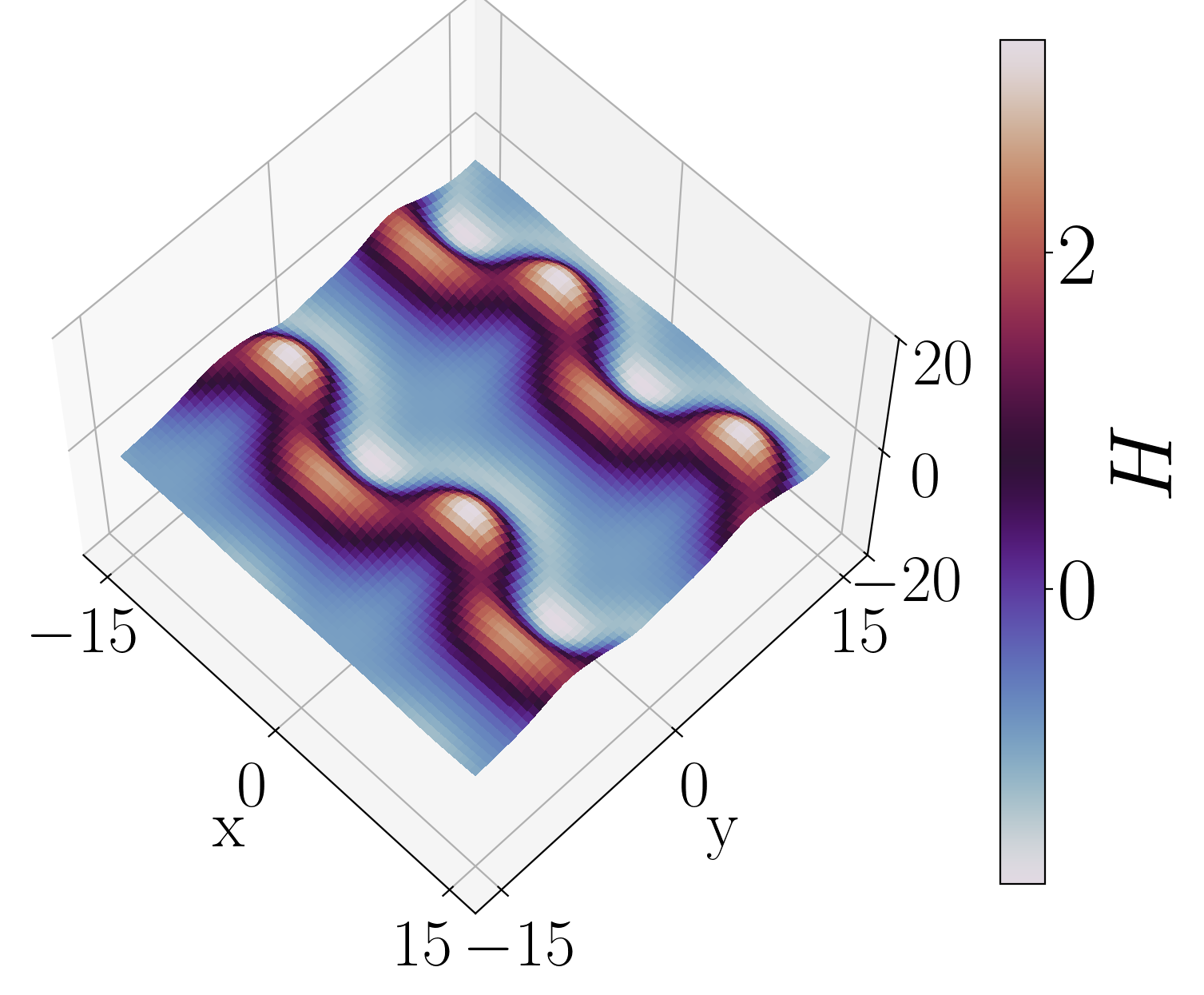}\\
\end{tabular}
\caption{Relative periodic orbits  for different domain sizes, presented in Table \ref{table:ecs_solutions}. In each row, the energy and phase diagrams, and interface snapshots at selected times within one period $T$. Red markers denote the corresponding snapshot times. 
First row: $RPO_{27.946}$ for 
$L=19.29$ and $ \delta=0.65373$ at  $t=[0,9.315,18.631,27.946]$
Middle row: $RPO_{25.991}$: for $L=22$ and $ \delta=0.002$ at  $t=[0,8.664,17.327,25.991].$
Bottom row: 
$RPO_{25.788}$ for $L=30$ and $ \delta=0.65373$ at $t=[0,8.596,17.192,25.788]$.
%
}\label{fig:three_pos}
\end{figure}

Figure~\ref{fig:three_pos} presents three converged relative periodic orbits for different domain sizes. In all cases the trajectory forms a closed loop in the $(E,\,\mathrm{d}E/\mathrm{d}t)$ phase plane, confirming that these solutions represent single-frequency dynamical cycles rather than quasiperiodic or chaotic motion.
The interfacial snapshots show that one period corresponds to a cyclic modulation in which the spanwise deformation strengthens, becomes spatially concentrated and subsequently redistributes. The RPOs therefore capture the fundamental nonlinear cycle associated with these interfacial dynamics and are consistent with the recurrent large-amplitude events observed in the chaotic regime. A clear dependence on the domain size is observed. In smaller domains a dominant wavelength covers the entire spanwise length, whereas in larger domains several spatially separated structures are accommodated within a single period. The streamwise extent thus selects both the spatial organisation and the period of the cycle, indicating that the broadband chaotic dynamics is composed of recurrent nonlinear processes associated with the embedded RPOs.

\begin{figure}
\centering
\includegraphics[scale=0.7]{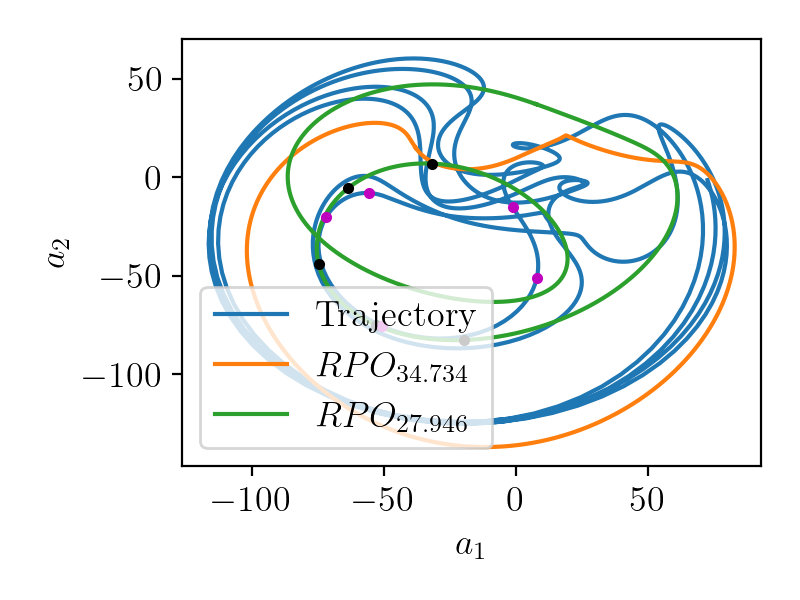}\\
(a)
\includegraphics[width=\textwidth]{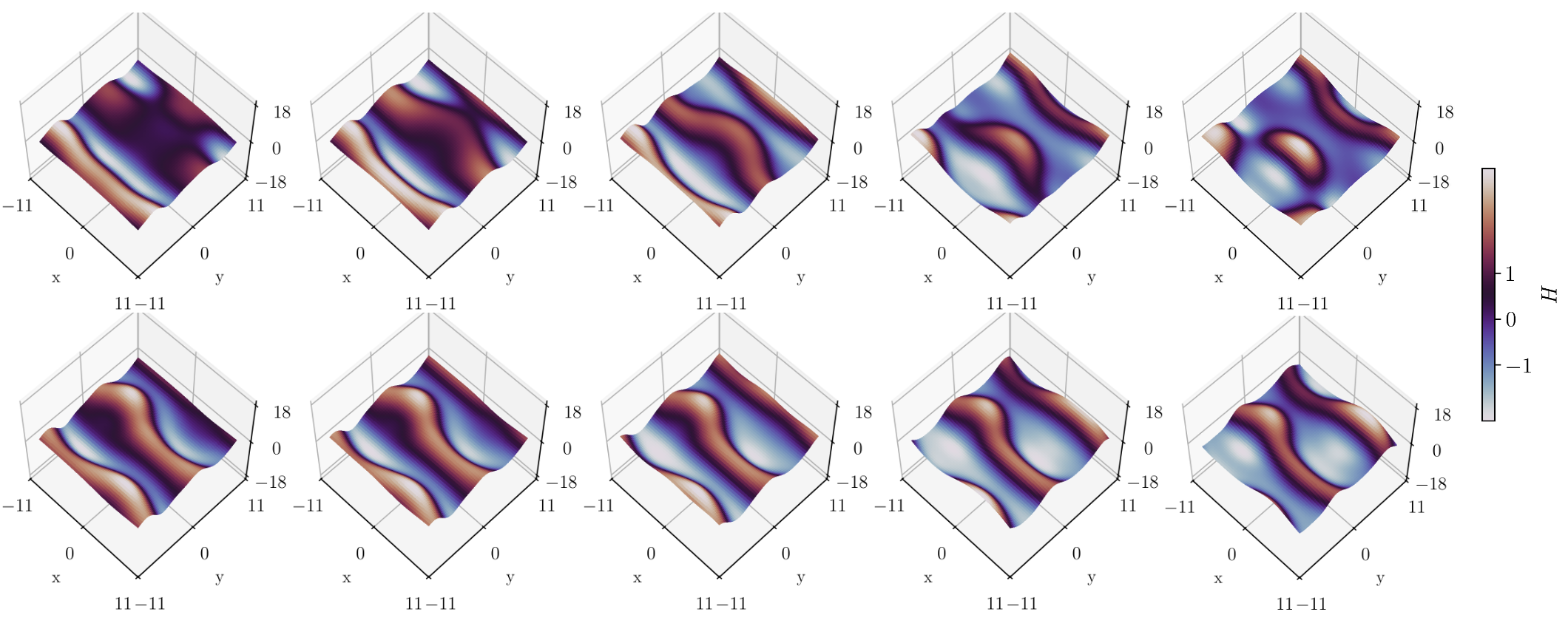}\\
(b)
\begin{tabular}{cc}
\includegraphics[width=0.5\textwidth]{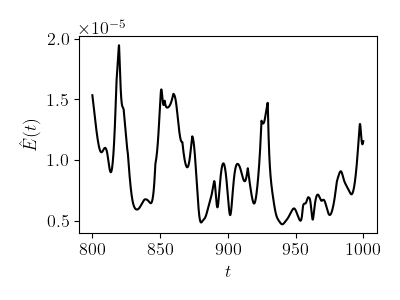} & 
\includegraphics[width=0.5\textwidth]{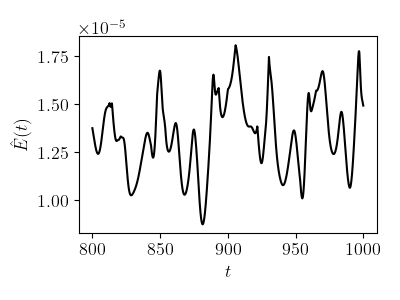}\\
(c) & (d)
\end{tabular}
\caption{ (a) DNS generic trajectory in the regime $L=19.29$, $\delta=0.65373$ projected onto the first two POD components $(a_1,a_2)$, along with two representative RPO solutions. The top set of snapshots show the generic trajectory at times $t=[815,817,819,821,823]$, from left to right. The bottom set show the $RPO_{27.946}$ solution at times $t=[15.72, 20.72, 25.72, 3.72,8.72]$, modulo the period of the RPO. (c-d) Computed distances $\hat{E}(t)$ for $RPO_{34.734}$ and $RPO_{27.946}$ as a function of time for times $800<T<1000.$ }
\label{fig:shadowing1}
\end{figure}

\begin{figure}
\centering
\includegraphics[scale=0.7]{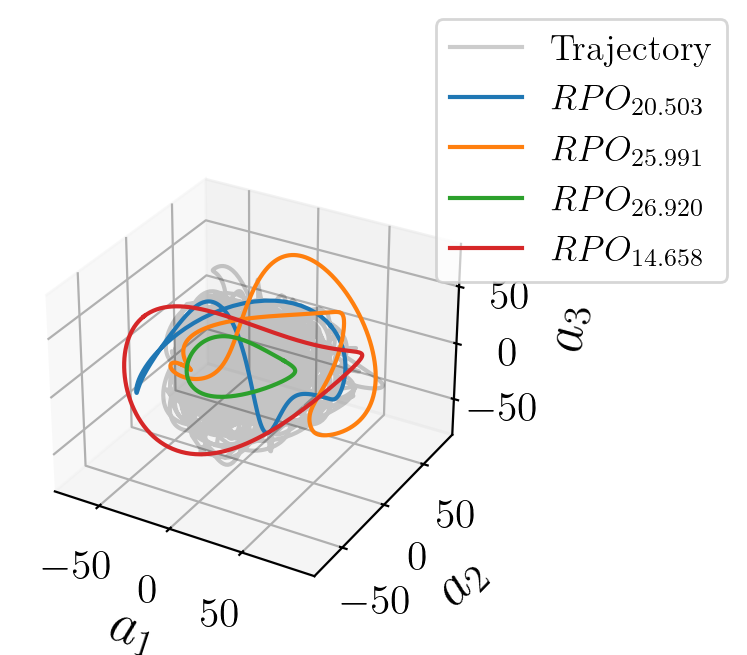}\\
(a)\\
\begin{tabular}{cc}
\includegraphics[width=0.5\textwidth, trim={0 10 0 60}, 
clip]{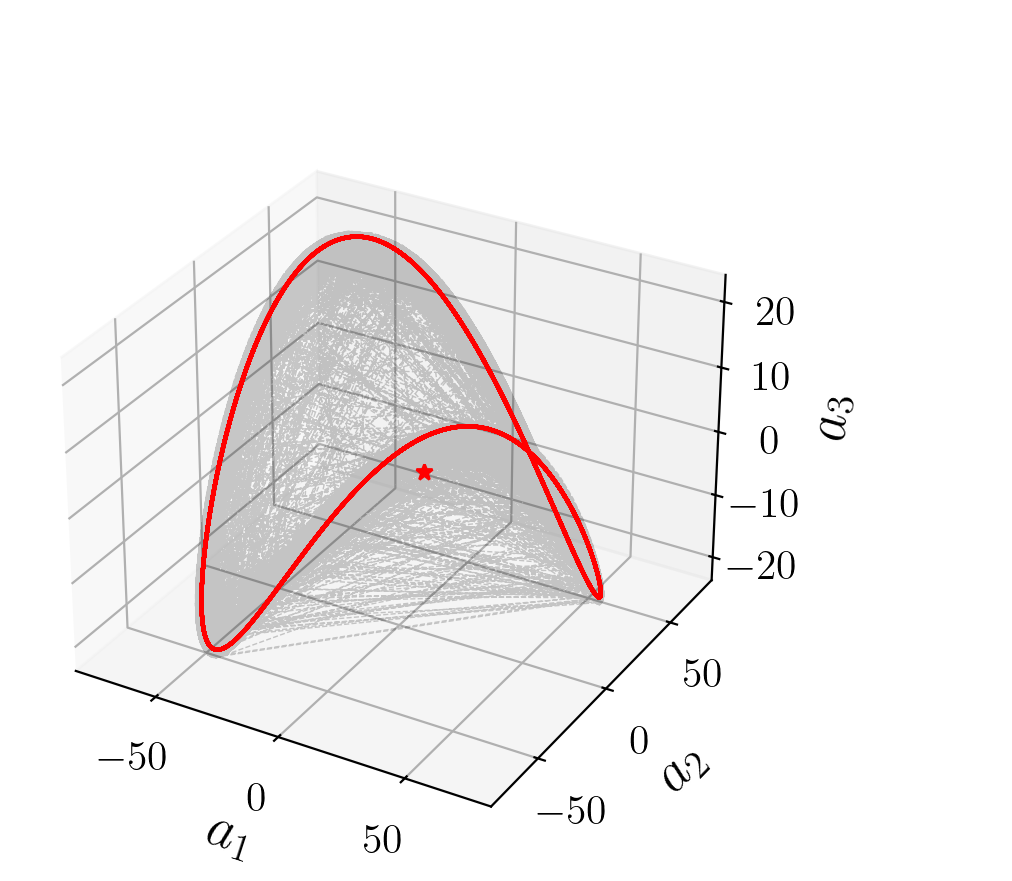}
&\includegraphics[width=0.45\textwidth]{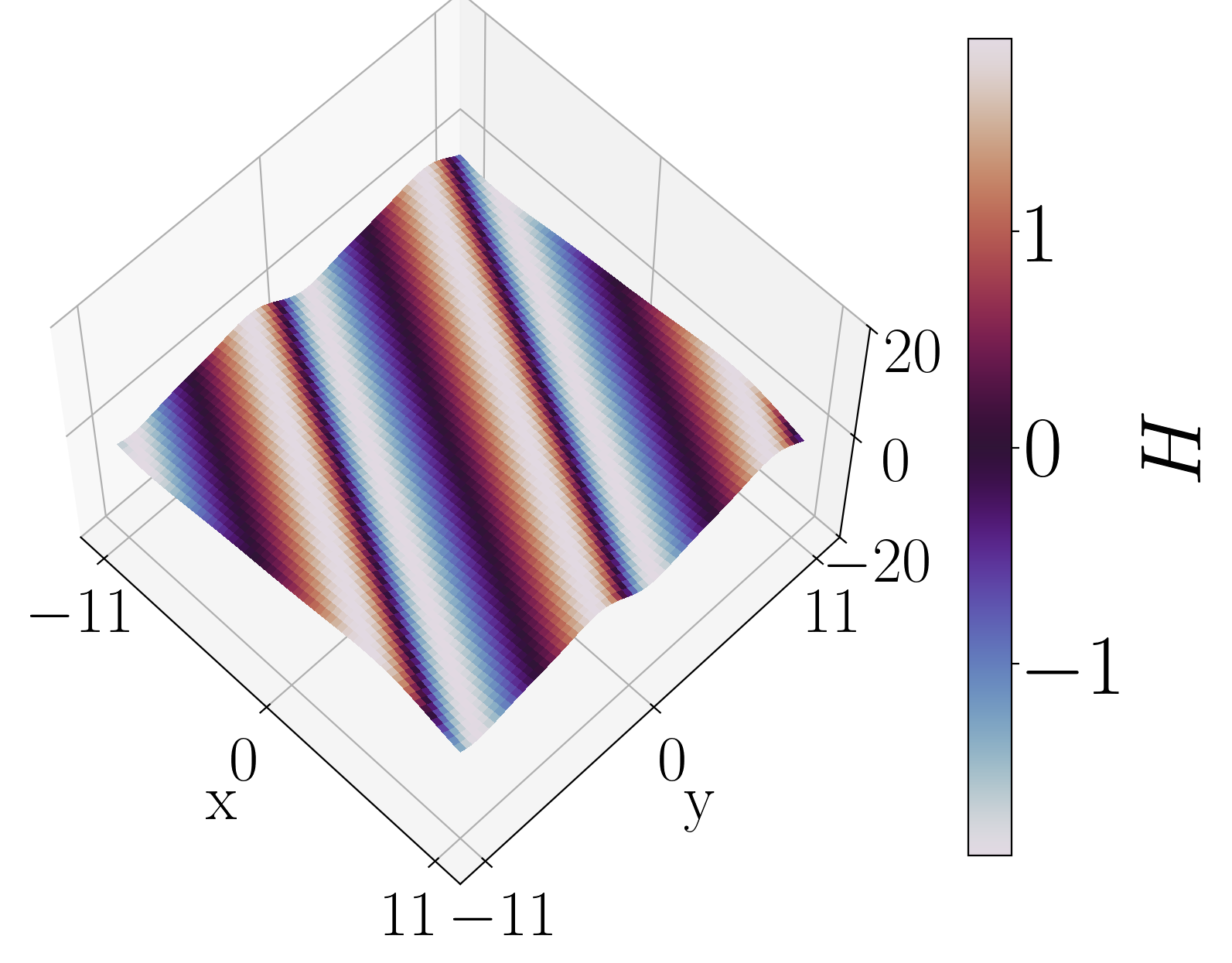}  \\
  (b)   &  (c)\\
\end{tabular}
\caption{
(a) Selected relative periodic orbits  for the case $L=22$ and $\delta=0.002$, superimposed on a representative chaotic falling-film trajectory, and projected onto the first three principal components $(a_1,a_2,a_3)$ obtained from POD.
(b) Three-dimensional projection of $RPO_{187.315}$ onto the leading principal components $(a_1,a_2,a_3)$ (red), together with the equilibrium $EQ_1$ (red star) and a trajectory initiated from a $10^{-6}$ perturbation of the RPO (grey). The perturbed trajectory departs from the orbit and subsequently approaches $EQ_1$, indicating that the RPO is dynamically unstable. 
(c) Snapshots of the equilibrium height solution $EQ_1$.
}
\label{fig:limcycle}
\end{figure}

We next examine whether these RPOs contribute to organizing the dynamics of an interfacial chaotic trajectory. 
Figure~\ref{fig:shadowing1}a,b  shows
a generic DNS trajectory for $L=19.29$ and $\delta=0.65373$ projected onto the plane spanned by the first two POD modes, together with two representative RPOs (e.g., $RPO_{34.734}$ and $RPO_{27.946}$). Visual inspection already suggests repeated close approaches of the DNS trajectory to the neighbourhood of the RPOs in the symmetry-reduced state space. To quantify these events we compute the instantaneous projected distance
$\hat E(t)=\min_{t'}\|\boldsymbol{a}(t)-\boldsymbol{a}'(t')\|_2,$
where $\boldsymbol{a}(t)$ and $\boldsymbol{a}'(t')$ denote the POD coordinates of the generic trajectory and the RPO, respectively.
The time series of $\hat E(t)$, shown in figure~\ref{fig:shadowing1}(c,d), is computed over a time interval of approximately $200$ time units during which the DNS trajectory evolves in the neighbourhood of the RPO in state space. Over this interval $\hat E(t)$ attains values of $\mathcal{O}(10^{-5})$, indicating very close visits of the DNS trajectory to the relative periodic orbits.

While $\hat E(t)$ characterizes the temporal proximity of these visits, we will quantify the degree to which a generic trajectory shadows a given RPO. Thus, we compare their representations in the plane spanned by the first two POD modes. Let $\boldsymbol{x}(t)\in\mathbb{R}^{N}$ denote a generic trajectory sampled at times $\{t_j\}_{j=1}^{N_T}$ over a time window of duration $\tau$, and let $\boldsymbol{x}'(t)$ denote an RPO of period $T$ sampled at times $\{t'_k\}_{k=1}^{N_T}$ over one period, where $N_T$ is the number of samples over one period of the RPO. Let $\mathbf{U}\in\mathbb{R}^{N\times 2}$ contain the first two POD modes, and define the projected coordinates
\begin{equation}
\boldsymbol{a}(t_j)=\mathbf{U}^\top\boldsymbol{x}(t_j),\qquad
\boldsymbol{a}'(t'_k)=\mathbf{U}^\top\boldsymbol{x}'(t'_k)\in\mathbb{R}^{2}.
\end{equation}
We define the normalized shadowing distance as
\begin{equation}
d_s(\boldsymbol{a},\boldsymbol{a}')
\equiv \frac{1}{R_a}\frac{1}{N_T} \sum_{j=1}^{N_T} \min_{1\le k\le N_\tau} \left\|
\boldsymbol{a}'(t'_k)-\boldsymbol{a}(t_j)
\right\|_2,
\label{eq:shadow_norm}
\end{equation}
where $\|\cdot\|_2$ denotes the Euclidean norm in $\mathbb{R}^2$ and $R_a$ is a characteristic attractor length scale in the same plane,
defined by
\begin{equation}
R_a = \left\|
\max_{1\le j\le N_T}\boldsymbol{a}(t_j) - \min_{1\le j\le N_T}\boldsymbol{a}(t_j) \right\|_2,
\label{eq:attractor_radius}
\end{equation}
with $\min$ and $\max$ taken componentwise. We additionally verified that close projected distances correspond to small full-state distances 
$\min_{\boldsymbol{s},\,t'}
\|\boldsymbol{x}(t)-\mathbf{G}(\boldsymbol{s})\boldsymbol{x}'(t')\|_{2}$ 
over the same intervals.
We note that the distance metric employed here follows the same rationale as  introduced by \citet{Crowley, crowley2022turbulence}. Low values identify time intervals during which both the instantaneous state and its temporal evolution are accurately represented by the selected recurrent solution.

For the trajectory shown in figure~\ref{fig:shadowing1}c and \ref{fig:shadowing1}d, evaluated over the same time interval,  the normalized shadowing distance is  ${d}_s(\boldsymbol{a},\boldsymbol{a}') = 0.0177$ for $\mathrm{RPO}_{34.734}$  and ${d}_s(\boldsymbol{a},\boldsymbol{a}') = 0.0269$ for $\mathrm{RPO}_{27.946}$.  Both values are $\mathcal{O}(10^{-2})$ relative to the attractor extent,  showing that the DNS trajectory remains within a small phase-space neighbourhood of these recurrent solutions.  These small values  provide quantitative evidence that the extracted RPOs form part of the dynamical skeleton underlying the interfacial evolution.

Figure~\ref{fig:limcycle}a shows a trajectory of the chaotic falling–film dynamics for $L=22$ and $\delta=0.002$ together with several relative periodic orbits embedded in the same region of state space.  When projected onto the leading principal components, the chaotic trajectory  passes near the projections of these invariant solutions.  This indicates that the RPOs lie within the region explored by the attractor and correspond to dynamically relevant states visited by the chaotic trajectory. The dynamical connection between the equilibrium state $\mathrm{EQ}_1$ and the $RPO_{187.385}$ is shown in figure~\ref{fig:limcycle}b.
Starting from $RPO_{187.385}$, we apply a perturbation of amplitude $10^{-6}$ along its most unstable Floquet direction.  The perturbed trajectory departs from the periodic orbit and subsequently approaches the neighbourhood of $\mathrm{EQ}_1$ in the $(a_1,a_2,a_3)$ projection.
This observation suggests a dynamical pathway
in which the unstable manifold of the RPO intersects the basin (or stable manifold) of $\mathrm{EQ}_1$,  placing both solutions within the same region of state space explored by the attractor.
The interfacial structure associated with $\mathrm{EQ}_1$ is shown in figure~\ref{fig:limcycle}c.
$\mathrm{EQ}_1$ corresponds to a finite-amplitude, streamwise-modulated wave state rather than the flat Nusselt film, indicating that the chaotic dynamics explore strongly deformed coherent interfacial configurations.

These results support a state-space interpretation of falling-film chaos in which equilibria and relative periodic orbits form a network of invariant solutions that organizes the recurrent large-amplitude deformations of the interface.


\section{Conclusion \label{conclusion_Section}}

Dynamical systems approaches to spatiotemporal chaos have been developed primarily in the context of single-phase flows. In this work, we  extend these ideas to two-phase flows in falling films as the dynamics are intrinsically coupled to the evolution of a deforming interface.
Starting from the Navier–Stokes equations, we revisit the classical long-wave reduction for falling films and recover an interface evolution equation with the structure of a two-dimensional Kuramoto-Sivashinsky-type model. This reduced model retains the dominant physical mechanisms governing the flow through parameters depending on the Reynolds, Weber and Froude numbers.
Using this formulation, we perform an extensive parametric study comprising over $2000$ simulations and construct regime maps in the $(L,\delta)$ parameter space. The resulting maps reveal a rich hierarchy of interfacial behaviours, including travelling waves, bursting travelling waves and fully chaotic regimes. The transition to chaos occurs primarily through the increase of the domain size, which allows nonlinear interactions among multiple unstable modes. This systematic characterization provides a unified description linking the governing physical parameters to the observed interfacial dynamics.

We then constructed reduced dynamical models in the learned manifold coordinates using neural ODEs. Although these models do not reproduce the full dynamics indefinitely, they accurately capture short-time trajectory evolution and long-time statistical behaviour. Their primary role is to generate dynamically consistent initial conditions for Newton--Krylov searches in the full system. Using this approach, we identified multiple invariant solutions embedded within the chaotic attractor, including equilibria, travelling waves and relative periodic orbits. Shadowing analysis shows that chaotic trajectories repeatedly approach the neighbourhoods of several of these relative periodic orbits in state space, indicating that the recurrent interfacial patterns observed in the dynamics correspond to visits to these invariant solutions. The chaotic evolution can therefore be interpreted as motion on the attractor that is organized by a network of unstable coherent states. To the best of our knowledge, these constitute the first exact coherent structures reported for chaotic falling-film flows, demonstrating that the state-space framework developed for single-phase turbulence can be extended to nonlinear interfacial systems.

The present study also has several limitations that point naturally to directions for future work. First, the interface evolution equation employed here relies on the long-wave approximation, which is valid when the characteristic interfacial wavelengths are large compared to the film thickness. While the resulting model retains the dominant mechanisms responsible for instability, nonlinear saturation and dissipation, extending the present state-space framework to models beyond the long-wave limit remains an important objective. Second, the extensivity analysis of the inertial-manifold dimension was carried out primarily for a representative low value of the dispersion parameter, $\delta=0.002$, corresponding to strongly chaotic dynamics. A systematic exploration of the scaling $d_{\mathcal M}(L,\delta)$ across the dispersion parameter space remains an open question. Finally, the present simulations were restricted to square domains with $L_x=L_y=L$ in order to isolate the effect of system size. Allowing independent variation of the streamwise and spanwise domain lengths would enable investigation of anisotropic scaling and the possible emergence of distinct correlation lengths in the two directions. Extending the framework to fully three-dimensional flows is also a natural next step, where additional interfacial modes and spanwise instabilities are expected to increase the dimensionality of the attracting set while preserving the dissipative structure of the dynamics. \\


\noindent \textbf{Declaration of Interests}. The authors report no conflict of interest. \\

\noindent \textbf{ Data availability}. The code and data (in compressed format) used in the paper is available in the group GitHub repository.
\\

This research used the Delta advanced computing and data resource which is supported by the National Science Foundation (award OAC 2005572) and the State of Illinois.\\

\begin{appen}

\section{Derivation of the chaotic falling film equation}\label{appA}

For completeness, and to aid readers who may not be familiar with long-wave reductions for falling films, we summarise below the main steps leading from the three-dimensional Navier-Stokes equations on an inclined plane to the reduced Kuramoto-Sivashinsky-type equation employed in this study. While the original work of 
\citet{topper_kawahara} introduced this model, the underlying derivation was presented only in outline form. Given the continued use of this equation in the study of spatiotemporal chaos in falling films, a transparent and fully written derivation is useful both for clarity and as a future reference point for researchers entering the field.

Starting with the momentum equations in three dimensions on an inclined plane with angle $\theta$

\begin{equation}\frac{\partial u}{\partial t}+u\frac{\partial u}{\partial x}+v\frac{\partial u}{\partial y}+w\frac{\partial u}{\partial z}=-\frac{1}{\rho}\frac{\partial P}{\partial x}+g\sin\theta+\nu\left(\frac{\partial^2}{\partial x^2}+\frac{\partial^2}{\partial y^2}+\frac{\partial^2}{\partial z^2}\right)u,\end{equation}
\begin{equation}\frac{\partial v}{\partial t}+u\frac{\partial v}{\partial x}+v\frac{\partial v}{\partial y}+w\frac{\partial v}{\partial z}=-\frac{1}{\rho}\frac{\partial P}{\partial y}-g\cos\theta+\nu\left(\frac{\partial^2}{\partial x^2}+\frac{\partial^2}{\partial y^2}+\frac{\partial^2}{\partial z^2}\right)v,\end{equation}
\begin{equation}\frac{\partial w}{\partial t}+u\frac{\partial w}{\partial x}+v\frac{\partial w}{\partial y}+w\frac{\partial w}{\partial z}=-\frac{1}{\rho}\frac{\partial P}{\partial z}+\nu\left(\frac{\partial^2}{\partial x^2}+\frac{\partial^2}{\partial y^2}+\frac{\partial^2}{\partial z^2}\right)w,\end{equation}
\begin{equation}\frac{\partial u}{\partial x}+\frac{\partial v}{\partial y}+\frac{\partial w}{\partial z}=0\end{equation}
\\
we non-dimensionalize using the following transformations:
\begin{align*}  
&x'=l_0x,\hspace{1mm}y'=h_0y,\hspace{1mm}z'=l_0z,\hspace{1mm}U'=U_0U,\hspace{1mm}u'=U_0u,\hspace{1mm}v'=U_0\frac{h_0}{l_0}v,\hspace{1mm}\\&w'=U_0w,\hspace{1mm}h'=h_0(1+\eta),\hspace{1mm}t'=\frac{l_0}{U_0}t.
\end{align*}

Here, $U_0=\sqrt{gh_0\cos\theta}$ is the gravitational fluid velocity. We define $h=h(x,z,t)$ to be the deviation of the wave from its base state $h_0.$ The non-dimensionalizations result in \\

\textbf{x-momentum:}
\begin{align*} &\frac{U_0^2}{l_0}\left(\frac{\partial u}{\partial t}+u\frac{\partial u}{\partial x}+v\frac{\partial u}{\partial y}+w\frac{\partial u}{\partial z}\right)\\&=-\frac{1}{\rho l_0}\frac{\partial P}{\partial x}+g\sin\theta+\nu\left(\frac{U_0}{l_0^2}\frac{\partial^2u}{\partial x^2}+\frac{U_0}{h_0^2}\frac{\partial^2u}{\partial y^2}+\frac{U_0}{l_0^2}\frac{\partial^2u}{\partial z^2}\right)
\end{align*}
\\
Substituting the expression for $U_0$ gives us
\begin{align*}&\frac{gh_0\cos\theta}{l_0}\left(\frac{\partial u}{\partial t}+u\frac{\partial u}{\partial x}+v\frac{\partial u}{\partial y}+w\frac{\partial u}{\partial z}\right)\\&=-\frac{1}{\rho l_0}\frac{\partial P}{\partial x}+g\cos\theta+\nu\sqrt{gh_0\cos\theta}\left(\frac{1}{l_0^2}\frac{\partial^2u}{\partial x^2}+\frac{1}{h_0^2}\frac{\partial^2u}{\partial y^2}+\frac{1}{l_0^2}\frac{\partial^2u}{\partial z^2}\right).
\end{align*}
\\
Dividing through by $g\cos\theta$ gives
\begin{align*}&\frac{h_0}{l_0}\left(\frac{\partial u}{\partial t}+u\frac{\partial u}{\partial x}+v\frac{\partial u}{\partial y}+w\frac{\partial u}{\partial z}\right)\\&=-\frac{1}{\rho g l_0\cos\theta}\frac{\partial P}{\partial x}+\tan\theta+\nu\sqrt{\frac{h_0}{g\cos\theta}}\left(\frac{1}{l_0^2}\frac{\partial^2u}{\partial x^2}+\frac{1}{h_0^2}\frac{\partial^2u}{\partial y^2}+\frac{1}{l_0^2}\frac{\partial^2u}{\partial z^2}\right).
\end{align*}
\\
Factoring $h_0^2$ from the viscous terms gives
\begin{align*}&\frac{h_0}{l_0}\left(\frac{\partial u}{\partial t}+u\frac{\partial u}{\partial x}+v\frac{\partial u}{\partial y}+w\frac{\partial u}{\partial z}\right)\\&=-\frac{1}{\rho g l_0\cos\theta}\frac{\partial P}{\partial x}+\tan\theta+\frac{\nu}{h_0\sqrt{gh_0\cos\theta}}\left(\frac{h_0^2}{l_0^2}\frac{\partial^2u}{\partial x^2}+\frac{\partial^2u}{\partial y^2}+\frac{h_0^2}{l_0^2}\frac{\partial^2u}{\partial z^2}\right).\end{align*}
\\
Defining the Reynolds number as $\text{Re\hspace{1mm}}=U_0h_0/\nu$, writing $\tan\theta=h_x$, and rewriting the constants in front of the pressure term gives
\begin{align*}&\frac{h_0}{l_0}\left(\frac{\partial u}{\partial t}+u\frac{\partial u}{\partial x}+v\frac{\partial u}{\partial y}+w\frac{\partial u}{\partial z}\right)\\&=-\frac{h_0}{\rho g h_0l_0\cos\theta}\frac{\partial P}{\partial x}+h_x+\frac{1}{\text{Re}}\left(\frac{h_0^2}{l_0^2}\frac{\partial^2u}{\partial x^2}+\frac{\partial^2u}{\partial y^2}+\frac{h_0^2}{l_0^2}\frac{\partial^2u}{\partial z^2}\right).\end{align*}

\textbf{y-momentum}: 
\begin{align*}&\frac{gh_0^2\cos\theta}{l_0^2}\left(\frac{\partial v}{\partial t}+u\frac{\partial v}{\partial x}+v\frac{\partial v}{\partial y}+w\frac{\partial v}{\partial z}\right)\\&=-\frac{1}{\rho h_0}\frac{\partial P}{\partial y}-g\cos\theta+\frac{h_0}{l_0}\sqrt{gh_0\cos\theta}\left(\frac{1}{l_0^2}\frac{\partial^2v}{\partial x^2}+\frac{1}{h_0^2}\frac{\partial^2v}{\partial y^2}+\frac{1}{l_0^2}\frac{\partial^2v}{\partial z^2}\right)\end{align*}
\begin{align*}&\Rightarrow\frac{h_0^2}{l_0^2}\left(\frac{\partial v}{\partial t}+u\frac{\partial v}{\partial x}+v\frac{\partial v}{\partial y}+w\frac{\partial v}{\partial z}\right)\\&=-\frac{1}{\rho g h_0\cos\theta}\frac{\partial P}{\partial y}-1+\nu\frac{h_0}{l_0}\sqrt{\frac{h_0}{g\cos\theta}}\left(\frac{1}{l_0^2}\frac{\partial^2v}{\partial x^2}+\frac{1}{h_0^2}\frac{\partial^2v}{\partial y^2}+\frac{1}{l_0^2}\frac{\partial^2v}{\partial z^2}\right)\end{align*}
\begin{align*}&\Rightarrow\frac{h_0^2}{l_0^2}\left(\frac{\partial v}{\partial t}+u\frac{\partial v}{\partial x}+v\frac{\partial v}{\partial y}+w\frac{\partial v}{\partial z}\right)\\&=-\frac{1}{\rho g h_0\cos\theta}\frac{\partial P}{\partial y}-1+\nu\frac{h_0}{l_0}\frac{1}{\text{Re}}\left(\frac{h_0^2}{l_0^2}\frac{\partial^2v}{\partial x^2}+\frac{\partial^2v}{\partial y^2}+\frac{h_0^2}{l_0^2}\frac{\partial^2v}{\partial z^2}\right)\end{align*}

The z-momentum equation has the same form as the x-momentum equation, lacking the force balance term:
\begin{align*}&\frac{h_0}{l_0}\left(\frac{\partial w}{\partial t}+u\frac{\partial w}{\partial x}+v\frac{\partial w}{\partial y}+w\frac{\partial w}{\partial z}\right)\\&=-\frac{h_0}{\rho g h_0l_0\cos\theta}\frac{\partial P}{\partial z}+\frac{1}{\text{Re}}\left(\frac{h_0^2}{l_0^2}\frac{\partial^2w}{\partial x^2}+\frac{\partial^2w}{\partial y^2}+\frac{h_0^2}{l_0^2}\frac{\partial^2w}{\partial z^2}\right)\end{align*}
\\
The continuity equation remains unchanged, as

$$\frac{\partial u}{\partial x}+\frac{\partial v}{\partial y}+\frac{\partial w}{\partial z}=0\rightarrow \frac{U_0}{l_0}\frac{\partial u}{\partial x}+\frac{U_0h_0}{l_0h_0}\frac{\partial v}{\partial y}+\frac{U_0}{l_0}\frac{\partial w}{\partial z}=0\rightarrow \frac{\partial u}{\partial x}+\frac{\partial v}{\partial y}+\frac{\partial w}{\partial z}=0$$
\\
We introduce the long-wave parameter $\delta=h_0/l_0,$ where $l_0$ is the characteristic streamwise wavelength of the interfacial disturbances. We also define the material derivative $D/Dt=\partial/\partial t+(u\cdot\nabla)$, and scale the pressure as $P'=P_0+\rho gh_0\cos\theta P$, where $P$ is now the non-dimensional pressure fluctuation about hydrostatic balance. Thus,the  Navier-stokes equations reduce to the following non-dimensional form

\begin{align}\text{Re}\hspace{1mm}\delta\frac{Du}{Dt}=-\text{Re}\delta\frac{\partial P}{\partial x}+\text{Re}h_x+\delta^2\frac{\partial^2u}{\partial x^2}+\frac{\partial^2u}{\partial y^2}+\delta^2\frac{\partial^2u}{\partial z^2}\label{nsx}\end{align}
\begin{align}\text{Re}\hspace{1mm}\delta^2\frac{Dv}{Dt}=-\text{Re}\frac{\partial P}{\partial y}-\text{Re}+\delta^3\frac{\partial^2v}{\partial x^2}+\delta\frac{\partial^2v}{\partial y^2}+\delta^3\frac{\partial^2v}{\partial z^2}\label{nsy}\end{align}
\begin{align}\text{Re}\hspace{1mm}\delta\frac{Dw}{Dt}=-\text{Re}\delta\frac{\partial P}{\partial z}+\delta^2\frac{\partial^2w}{\partial x^2}+\frac{\partial^2w}{\partial y^2}+\delta^2\frac{\partial^2w}{\partial z^2}\label{nsz}\end{align}
\\
We next introduce the kinematic boundary condition, which states that there must be no advection between the boundary and the surrounding air:

$$\frac{D}{Dt}(y-h) = 0.$$
\\
Expansion of this term gives
$$\frac{\partial y}{\partial t}-\left(\frac{\partial h}{\partial t}+u\frac{\partial h}{\partial x}+w\frac{\partial h}{\partial z}\right)=0.$$
\\
From this we get
$$v-h_t-uh_x-wh_y=0\hspace{1.5mm} \text{at}\hspace{1.5mm} y=h.$$
\\
Nondimensionalization of this equation gives
$$\delta U_0(v-\eta_t-u\eta_x-w\eta_z)=0$$
\begin{equation}\Rightarrow v=\eta_t+u\eta_x+w\eta_z\hspace{1.5mm} \text{at}\hspace{1.5mm} y=1+\delta\eta.\label{kinbc}\end{equation}
\\
The no-slip boundary condition at $y=0$ remains unchanged. We now consider the pressure boundary conditions, of which there are three:
\begin{equation}P_0-P+(2\mu\mathbf{D}\cdot\mathbf{n})\cdot\mathbf{n}=2\sigma K(h) \hspace{1.5mm}\text{at}\hspace{1.5mm} y=h,\label{npbc}\end{equation}
\begin{equation}(2\mu\mathbf{D}\cdot\mathbf{n})\cdot\bm{\tau}_i=\nabla_s\sigma\cdot\bm{\tau}_i\hspace{1.5mm}\text{at}\hspace{1.5mm} y=h.\label{tpbc}\end{equation}
\\
Here, $D_{ij}=1/2(\partial u_i/\partial x_j+\partial u_j/\partial x_i)$ is the rate-of-deformation tensor, $\mathbf{n}$ is the free surface unit normal vector $(1+h_x^2+h_y^2)^{-1/2}[-h_x,1,-h_z]^T,$ $K(h)$ is the mean free surface curvature, $\bm{\tau}_1=(h_x^2+h_{x}^2)^{-1/2}[h_{x},0,-h_z]^T$ and $\bm{\tau}_2=(h_x^2+h_{z}^2)^{-1/2}[h_x,0,h_z]^T$  are the free surface tangential unit vectors, and $\nabla_s$ is the surface gradient operator. We note that in the case of an isothermal fluid, the right-hand side of (\ref{tpbc}) vanishes.
\\
Removing pressure from the $x$ and $z$ momentum equations is done by taking the $z$-derivative of (\ref{nsx}) and the $x$-derivative of (\ref{nsz}) and taking their difference:
$$\text{Re}\delta[u_t+uu_x+vu_y+wu_z]_z=-\text{Re}P_{xz}+\delta^2(u_{xxz}+u_{zzz})+u_{yyz}$$
$$\text{Re}\delta[w_t+uw_x+vw_y+ww_z]_x=-\text{Re}P_{xz}+\delta^2(w_{xxx}+w_{xzz})+w_{xyy}$$
\begin{align*}
\Rightarrow&(u_{z})_{yy}-(w_x)_{yy}+\delta^2(u_{xxz}+u_{zzz}-w_{xxx}-w_{xzz})\\&-\text{Re}\delta[u_t+uu_x+vu_y+wu_z]_z+\text{Re}\delta[w_t+uw_x+vw_y+ww_z]_x=0.
\end{align*}
Rewriting, and introducing $u'=U(y)+u$ we arrive at
\begin{align*}
&(u_z-w_x)_{yy}+\delta^2\Delta(u_z-w_x)-\text{Re}\delta[(u_z-w_x)_t+U(u_{z}-w_{x})_z+U_yv_z\\&+(uu_x+vu_y+wu_z)_z+(wu_x+vw_y+ww_z)_x]=0
\end{align*}
To remove the pressure from the $y$ momentum equation, we take the $xy$ derivative of (\ref{nsx}), the $yz$ derivative of (\ref{nsz}), apply the operator $\Delta = \partial^2/\partial x^2+\partial^2/\partial z^2$ to (\ref{nsy}) and subtract the resulting $y$ momentum equation from the resulting $x$ and $z$ momentum equations:
\begin{align*}
&\frac{1}{h_0l_0}\text{Re}\delta[u_t+uu_x+vu_y+wu_z]_{xy}=\frac{1}{h_0l_0}\left[-\text{Re}\delta P_{xxy}+\delta^2(u_{xxxy}+u_{xyzz})+u_{xyyy}\right]
\\&\frac{1}{h_0l_0}\text{Re}\delta[w_t+uw_x+vw_y+ww_z]_{yz}=\frac{1}{h_0l_0}\left[-\text{Re}\delta P_{yzz}+\delta^2(w_{xxyz}+w_{yzzz})+w_{yyyz}\right]
\\&\frac{1}{l_0^2}\text{Re}\delta^2\Delta[v_t+uv_x+vv_y+wv_z]=\frac{1}{l_0^2}\bigl[-\text{Re}(P_{xxy}+P_{yzz})
\\&+\frac{1}{l_0^2}\delta^3(v_{xxxx}+v_{xxzz}+v_{xxzz}+v_{zzzz})+\delta(v_{xxyy}+v_{yyzz})\bigr]
\end{align*}
\\
We first condense the $y$ momentum equation:
\begin{align*}
&\frac{1}{l_0^2}\text{Re}\delta^2\Delta[v_t+uv_x+vv_y+wv_z]=\frac{1}{l_0^2}\bigl[-\text{Re}\delta(P_{xxy}+P_{yzz})
\delta^3\Delta^2v+\delta\Delta v_{yy}\bigr]
\end{align*}
Noticing that multiplying the above equations by a factor of $h_0l_0$ will make the order of the scaling variables in front of the pressures commute:
\begin{align*}
&\text{Re}\delta[u_t+uu_x+vu_y+wu_z]_{xy}=\frac{1}{h_0l_0}\left[-\text{Re}\delta P_{xxy}+\delta^2(u_{xxxy}+u_{xyzz})+u_{xyyy}\right]
\\&\frac{1}{h_0l_0}\text{Re}\delta[w_t+uw_x+vw_y+ww_z]_{yz}=\frac{1}{h_0l_0}\left[-\text{Re}\delta P_{yzz}+\delta^2(w_{xxyz}+w_{yzzz})+w_{yyyz}\right]
\\&\text{Re}\delta^3\Delta[v_t+uv_x+vv_y+wv_z]=-\text{Re}\delta(P_{xxy}+P_{yzz})+
\delta^4\Delta^2v+\delta^2\Delta v_{yy}
\end{align*}
Subtracting the resulting $y$ momentum equation will now remove the pressures. We also notice, using the continuity equation, that $u_x+w_z=-v_y$ allows us to recast viscous terms into terms only including $v$:
\begin{align*}
&\text{Re}\delta[u_t+uu_x+vu_y+wu_z]_{xy}+\text{Re}\delta[w_t+uw_x+vw_y+ww_z]_{yz}\\&
-\text{Re}\delta^3\Delta[v_t+uv_x+vv_y+wv_z]=-\delta^2[v_{xxyy}+v_{yyzz}]-v_{yyyy}-\delta^4\Delta ^2v-\delta^2\Delta v_{yy}
\end{align*}
Rearranging terms and introducing $u'=U+u$ gives us the final result:
\begin{align}
&v_{yyyy}+\text{Re}\delta[u_t+(U+u)u_x+v(U_y+u_y)+wu_z]_{xy}\nonumber\\&+\text{Re}\delta[w_t+(U+u)w_x+vw_y+ww_z]_{yz}\nonumber
\\&-\text{Re}\delta^3\Delta[v_t+(U+u)v_x+vv_y+wv_z]+2\delta^2\Delta v_{yy}+\delta^4\Delta^2v=0
\end{align}
To non-dimensionalize the stress boundary conditions, we proceed first by expanding (\ref{npbc}):
$$(2\mu\mathbf{D}\cdot\mathbf{n})\cdot\mathbf{n}=\frac{\mu}{|n|^2}\begin{bmatrix}2u_x &u_y+v_x&u_z+w_x \\ u_y+v_x&2v_y& v_z+w_y\\u_z+w_x&v_z+w_y&2w_z\end{bmatrix}\begin{bmatrix}-h_x\\1\\-h_z\end{bmatrix}\cdot\begin{bmatrix}-h_x\\1\\-h_x\end{bmatrix}$$
$$=\frac{\mu}{|n|^2}\begin{bmatrix}-2u_xh_x+u_y+v_x-h_z(u_z+w_x)\\-h_x(v_x+u_y)+2v_y-h_z(v_z+w_y)\\-h_x(u_z+w_x)+v_z+w_y-2h_zw_z\end{bmatrix}\cdot\begin{bmatrix}-h_x\\1\\-h_z\end{bmatrix}$$
$$=\frac{2\mu}{|n|^2}\left[h_x^2u_x+h_z^2w_z-h_x(u_y+v_x)-h_z(v_z+w_y)+h_xh_z(u_z+w_x)+v_y\right]$$
\\
The mean curvature $K(h)$ is taken to be $-1/2\nabla_s\cdot\mathbf{n}=-1/2(\mathbf{I}-\mathbf{n}\otimes\mathbf{n})\cdot\nabla\cdot\mathbf{n}.$ The calculation proceeds as follows:
$$K(h)=-\frac{1}{2}\left(\begin{bmatrix}1&0&0\\0&1&0\\0&0&1\end{bmatrix}-\frac{1}{|n|^2}\begin{bmatrix}-h_x\\1\\-h_z\end{bmatrix}\begin{bmatrix}-h_x&1&-h_z\end{bmatrix}\right)\cdot\begin{bmatrix}\partial_x\\\partial_y\\\partial_z\end{bmatrix}\cdot\begin{bmatrix}-h_x\\1\\-h_z\end{bmatrix}$$
$$=-\frac{1}{2|n|^3}\begin{bmatrix}|n|^2-h_x^2&h_x&-h_xh_z\\h_x&|n|^2-1&h_z\\-h_xh_z&h_z&|n|^2-h_z^2\end{bmatrix}\begin{bmatrix}\partial_x\\\partial_y\\\partial_z\end{bmatrix}\begin{bmatrix}-h_x\\1\\-h_z\end{bmatrix}$$
$$=-\frac{1}{2|n|^3}\begin{bmatrix}(|n|^2-h_x^2)\partial_x+h_x\partial_y-h_xh_z\partial_z\\h_x\partial_y+(|n|^2-1)+h_x\partial_z\\h_xh_z\partial_x+h_x\partial_y+(1-h_z^2)\partial_z\end{bmatrix}\cdot\begin{bmatrix}-h_x\\1\\-h_z\end{bmatrix}$$
$$=\frac{1}{2|n|^3}\left[(1+h_z^2)h_{xx}-2h_xh_zh_{xz}+(1+h_x^2)h_{zz}\right]$$
\\
In its full form, we have
$$K(h)=\frac{1}{2}\frac{(1+h_z^2)h_{xx}-2h_xh_zh_{xz}+(1+h_x^2)h_{zz}}{(1+h_x^2+h_z^2)^{3/2}}.$$
\\
Our normal pressure boundary condition then becomes
\begin{align}&-P+\frac{2\mu}{1+h_x^2+h_z^2}\left[h_x^2u_x+h_z^2w_z-h_x(u_y+v_x)\right.\nonumber\\&-\left.h_z(v_z+w_y)+h_xh_z(u_z+w_x)+v_y\right]\nonumber\\&=\sigma\frac{(1+h_z^2)h_{xx}-2h_xh_zh_{xz}+(1+h_x^2)h_{zz}}{(1+h_x^2+h_z^2)^{3/2}}\hspace{1.5mm} \text{at}\hspace{1.5mm}y=h.\end{align}
\\
Non-dimensionalizing this equation gives us
\begin{align*}
&\rho gh_0\cos\theta(P_0-P)+\frac{2\mu U_0}{1+\delta^2\eta_x^2+\delta^2\eta_z^2}\left[\delta^3\eta_x^2u_x+\delta^3\eta_z^2w_z-\delta\eta_x\left(u_y+\delta^2v_x\right)\nonumber\right.\\&-\left.\delta\eta_z\left(\delta^2v_z+w_y\right)+\delta^3\eta_x\eta_z\left(u_z+w_x\right)+\delta v_y\right]\\&=\sigma\frac{\left(1+\delta^2\eta_z^2\right)\delta^2\eta_{xx}-2\delta^4\eta_x\eta_z\eta_{xz}+(1+\delta^2\eta_x^2)\delta^2\eta_{zz}}{(1+\delta^2\eta_x^2+\delta^2\eta_z^2)^{3/2}}
\end{align*}
\begin{align}
&\Rightarrow P_0-P+\frac{2}{\text{Re}(1+\delta^2\eta_x^2+\delta^2\eta_z^2)^{3/2}}\left[\delta^3\eta_x^2u_x+\delta^3\eta_z^2w_z-\delta\eta_x\left(u_y+\delta^2v_x\right)\right.\nonumber\\&-\left.\delta\eta_z\left(\delta^2v_z+w_y\right)+\delta^3\eta_x\eta_z\left(u_z+w_x\right)+\delta v_y\right]\nonumber\\&=\text{We}*h_0\frac{\left(1+\delta^2\eta_z^2\right)\delta^2\eta_{xx}-2\delta^4\eta_x\eta_z\eta_{xz}+(1+\delta^2\eta_x^2)\delta^2\eta_{zz}}{(1+\delta^2\eta_x^2+\delta^2\eta_z^2)^{3/2}}\label{pbc2}
\end{align}
We then take the $x$ derivative of (\ref{pbc2}) and substitute into the $x$-momentum equation:
\begin{align}
P_x&=\Biggl\{\frac{2\delta^2}{\text{Re}(1+\delta^2\eta_x^2+\delta^2\eta_z^2)^{3/2}}\left[\delta^2\eta_x^2u_x+\delta^2\eta_z^2w_z-\eta_x\left(u_y+\delta^2v_x\right)\right.\nonumber\\&-\left.\eta_z\left(\delta^2v_z+w_y\right)+\delta^2\eta_x\eta_z\left(u_z+w_x\right)+v_y\right]\biggr\}_x\nonumber\\&-\text{We}*\delta\Biggl\{\frac{\left(1+\delta^2\eta_z^2\right)\delta^2\eta_{xx}-2\delta^4\eta_x\eta_z\eta_{xz}+(1+\delta^2\eta_x^2)\delta^2\eta_{zz}}{(1+\delta^2\eta_x^2+\delta^2\eta_z^2)^{3/2}}\Biggr\}_x
\end{align}
\begin{align}
\Rightarrow &\frac{1}{\text{Re}}\left(u_{yy}+\delta^2(u_{xx}+u_{zz})\right)-\delta[u_t+uu_x+vu_y+wu_z]\nonumber\\&-\Biggl\{\frac{4}{\text{Re}(1+\delta^2\eta_x^2+\delta^2\eta_z^2)^{3/2}}\left[\delta^3\eta_x^2u_x+\delta^3\eta_z^2w_z-\delta\left(u_y+\delta^2v_x\right)\right.\nonumber\\&-\left.\delta^2\left(u_z+w_x\right)+\delta^3\eta_x\eta_z\left(\delta v_z+w_y\right)\right]\biggr\}_x\nonumber\\&+\text{We}\Biggl\{\frac{\left(1+\delta^2\eta_z^2\right)\delta^2\eta_{xx}-2\delta^4\eta_x\eta_z\eta_{xz}+(1+\delta^2\eta_x^2)\delta^2\eta_{zz}}{(1+\delta^2\eta_x^2+\delta^2\eta_z^2)^{3/2}}\Biggr\}_x=0\label{pbc3}
\end{align}
\\
Applying the same procedure with the operator $h_x\partial/\partial y$ applied to the $y$-momentum equation and non-dimensionalizing gives

\begin{align}
&\frac{1}{\text{Re}}\left(\delta^2\eta_xv_{yy}+\delta^4\eta_x(v_{xx}+v_{zz})\right)-\delta^3\eta_x[v_t+uv_x+vv_y+wv_z]\nonumber\\&-\eta_x\Biggl\{\frac{2\delta^2}{\text{Re}(1+\delta^2\eta_x^2+\delta^2\eta_z^2)^{3/2}}\left[\delta^2\eta_x^2u_x+\delta^2\eta_z^2w_z-\eta_x\left(u_y+\delta^2v_x\right)\right.\nonumber\\&-\left.\eta_z\left(\delta^2v_z+w_y\right)+\delta^2\eta_x\eta_z\left(\delta u_z+w_x\right)\right]\biggr\}_y=0\label{pbc4}.
\end{align}
\\
Adding (\ref{pbc3}) and (\ref{pbc4}) and defining the operator $\tilde{\nabla}=\partial/\partial x+\eta_x\partial/\partial y$ and $\Delta=\partial^2/\partial x^2+\partial^2/\partial y^2$ gives
\begin{align}
&u_{yy}+\delta^2\Delta u-\text{Re}\delta[u_t+u_x+vu_y+wu_z]\\&-\tilde{\nabla}\Biggl\{\frac{2\delta^2}{1+\delta^2\eta_x^2+\delta^2\eta_z^2}[\delta^2\eta_x^2u_x+\delta^2\eta_z^2w_z-\eta_x(u_y+\delta^2v_x)-\delta\eta_z(u_z+w_x)]\Biggr\}\\&+\text{ReWe}\delta^3\Biggl\{\frac{\left(1+\delta^2\eta_z^2\right)\eta_{xx}-2\delta^2\eta_x\eta_z\eta_{xz}+(1+\delta^2\eta_x^2)\eta_{zz}}{(1+\delta^2\eta_x^2+\delta^2\eta_z^2)^{3/2}}\Biggr\}_x=0
\end{align}
\\
For the tangential stress boundary conditions, we expand the term $(2\mu\mathbf{D}\cdot\mathbf{n})\cdot\bm{\tau}_1$ as
$$\frac{1}{|n||\tau|}[2h_xh_z(w_z-u_x)+h_z(u_y+v_x)-h_x(v_z+w_y)+(h_x^2-h_z^2)(u_z+w_x)]=0.$$
\\
Non-dimensionalization gives us
\begin{align*}&\Biggl[2\frac{\delta^2U_0}{l_0}h_xh_z(w_z-u_x)+\delta h_z\left(\frac{U_0}{h_0}u_y+\frac{\delta U_0}{l_0}v_x\right)\\&+\delta h_x\left(\frac{\delta U_0}{l_0}v_z+\frac{U_0}{h_0}w_y\right)+\frac{\delta^2U_0}{l_0}(h_x^2+h_z^2)\left(u_z+w_x\right)\Biggr]=0.\end{align*}

\begin{align*}&\Rightarrow[2\delta^2h_xh_z(w_z-u_x)+h_z(u_y+\delta^2v_x)+h_x(\delta^2v_z+w_y)\\&+\delta^2(h_x^2+h_z^2)(u_z+w_x)]=0\hspace{1.5mm}\text{at}\hspace{1.5mm}y=h\end{align*}
\\

The second stress condition can be written in the form

\begin{equation}
0=h_x^2\tau_{11}-(h_x^2+h_z^2)\tau_{22}+h_z^2\tau_{33}+h_x(h_x^2+h_z^2-1)\tau_{12}+2h_xh_z\tau_{13}+h_z(h_x^2+h_z^2-1)\tau_{23},
\end{equation}

which, under the same rescaling as above, the resulting boundary condition is

\begin{align*}
(\delta^2\eta_x^2+\delta^2\eta_z^2-1)&\left[\eta_z(w_y+\delta^2v_z)+\eta_x(\delta^2v_x+u_y+U_y)\right]\\&+2\delta^2\left[\eta_x^2(u_x-\delta^2v_y)+\eta_z^2(w_z-v_y)-h_xh_z(u_z+w_x)\right]=0.
\end{align*}

We begin the asymptotic expansion by considering $\text{Re}=\mathcal{O}(1)$, $\text{Fr}=\mathcal{O}(1),\delta\ll1,$ and $\text{We}=\mathcal{O}(\delta^{-2}),$ where we define the Froude number as $\text{Fr}=gh_0^2\sin\theta/\nu U_0.$ We expand $u,v,w,$ and $\eta$ in a power series of $\delta:$
$$u=\sum_{n=1}^{\infty}\delta^nu^{(n)},$$
$$v=\sum_{n=1}^{\infty}\delta^nv^{(n)},$$
$$w=\sum_{n=1}^{\infty}\delta^nw^{(n)},$$
$$\eta=\sum_{n=1}^{\infty}\delta^n\eta^{(n)}.$$
\\
\\
We also expand time and spatial derivatives as power series:
$$\frac{\partial}{\partial t}=\frac{\partial}{\partial T_0}+\delta\frac{\partial}{\partial T_1}+\delta^2\frac{\partial}{\partial T_2}+...$$
$$\frac{\partial}{\partial x}=\frac{\partial}{\partial X_0}+\delta\frac{\partial}{\partial X_1}+\delta^2\frac{\partial}{\partial X_2}+...$$
$$\frac{\partial}{\partial z}=\frac{\partial}{\partial Z_0}+\delta\frac{\partial}{\partial Z_1}+\delta^2\frac{\partial}{\partial Z_2}+...$$
\\
Up to first order in $\delta,$ we have the following system:
\begin{align}u_{X_0}^{(1)}+v_y^{(1)}+w_{Z_0}^{(1)}=0\label{fo1}\\
v_{yyyy}^{(1)}=0\label{fo2}\\
\left(u_{Z_0}^{(1)}-w_{X_0}^{(1)}\right)_{yy}=0\label{fo3}\\
\eta^{(1)}_{Z_0}U_y=0\text{ at }y=1\label{fo4}\\
\eta_{X_0}^{(1)}U_y=0\text{ at }y=1\label{fo5}\\
u_{yy}^{(1)}=0\text{ at }y=1\label{fo6}\\
\eta_{T_0}^{(1)}+\frac{\text{Fr}}{2}\eta_{X_0}^{(1)}-v^{(1)}=0\text{ at }y=1\label{fo7}
\end{align}

We look for solutions $\eta^{(1)}$ that obey a first-order wave equation, i.e $\eta^{(1)}=\eta^{(1)}(X_0-\text{Fr}T_0).$ Since there is no $Z_0$-dependence of the kinematic boundary condition, $w^{(1)}$ must be zero in order to retain a physical solution. By integrating \ref{fo2} and applying the no-slip boundary conditions, $v^{(1)}$ is of the form $v^{(1)}=C_2(X_0,Z_0,T_0)y^2+C_3(X_0,Z_0,T_0)y^3.$ We can integrate \ref{fo3} and apply no-slip boundary conditions to get a form of $u^{(1)} = D_1(X_0,Z_0,T_0)y.$ Substituting into the continuity equation, we have a relation
$$(D_1)_{X_0}y+2C_2y+3C_3y^2.$$
\\
In order for \ref{fo1} to be satisfied on the domain, it must be the case that $C_3=0.$ In the kinematic boundary condition, seeking a first order wave equation grants us with 
$$C_2=-\frac{\text{Fr}}{2}\eta^{(1)}_{X_0}.$$
Then, using the relation obtained from \ref{fo1} gives us  $u^{(1)}=\text{Fr}\eta^{(1)}y, v^{(1)}=-
\text{Fr}/2\eta^{(1)}_{X_0}y^2, w^{(1)}=0$, which can be easily verified to be a solution of the system. As we sought, the evolution equation becomes
\begin{equation}
\frac{\partial\eta^{(1)}}{\partial T_0}+\text{Fr}\frac{\partial\eta^{(1)}}{\partial X_0}=0,
\end{equation}
implying that $\eta^{(1)}=\eta^{(1)}(X_0-\text{Fr}T_0).$
If we now assume that $\text{We}=\mathcal{O}(\delta^{-2})$ and consider the $\mathcal{O}(\delta^2)$ case, we begin by expanding the kinematic boundary condition. Here it is important to consider the boundary conditions as a power series $y=1+\delta\eta^{(1)}+\delta^2\eta^{(2)}+...$:
\begin{align*}
&\left(\frac{\partial}{\partial T_0}+\delta\frac{\partial}{\partial T_1}+...\right)(\delta\eta^{(1)}+\delta^2\eta^{(2)}+...)\\&
+\text{Fr}\left[(1+\delta\eta^{(1)}+\delta^2\eta^{(2)}+...)-\frac{1}{2}(1+\delta\eta^{(1)}+\delta^2\eta^{(2)}+...)^2+\delta\eta^{(1)}(1+\delta\eta^{(1)}+\delta^2\eta^{(1)}+...)\right]\\&
*\left(\frac{\partial}{\partial X_0}+\delta\frac{\partial}{\partial X_1}+...\right)(\delta\eta^{(1)}+\delta^2\eta^{(2)}+...)+w^{(1)}\eta^{(1)}_{Z_0}-v^{(2)}=0.
\end{align*}
We first notice that $(1+\delta\eta^{(1)}+\delta^2\eta^{(2)}+...)^2=1+2\delta\eta^{(1)}+2\delta^2\eta^{(2)}+\delta^2(\eta^{(1)})^2+\mathcal{O}(\delta^3).$ Using this and substituting our $\mathcal{O}(\delta)$ solution, the expression becomes
\begin{align*}
\frac{\partial\eta^{(2)}}{\partial T_0}+\frac{\partial\eta^{(1)}}{\partial T_1}+\frac{\text{Fr}}{2}\frac{\partial\eta^{(2)}}{\partial X_0}+\frac{\text{Fr}}{2}\frac{\partial\eta^{(1)}}{\partial X_1}+2\text{Fr}\eta^{(1)}\frac{\partial\eta^{(1)}}{\partial X_0}-\text{Fr}\frac{\partial\eta^{(1)}}{\partial X_0}-v^{(2)}=0.
\end{align*}
Our expressions for $\eta^{(2)}$ appear only as derivatives with respect to the fast variables $T_0,X_0.$ We notice that our kinematic boundary condition is now in the form of an inhomogeneous first-order wave equation in $\eta,X_0,T_0$. Since our second order free surface equation then evolves in the same manner as the first order free surface equation aside from a propagation speed of $\text{Fr}/2$, we arrive at a secular term. We may then assume that $\eta^{(2)}\ne\eta^{(2)}(X_0,T_0).$

We also notice that when expanded to $\mathcal{O}(\delta^2)$ that the normal and tangential stress boundary conditions are trivially satisfied, as the expanded expression
\begin{align*}
&\left(\frac{\partial}{\partial Z_0}+\delta\frac{\partial}{\partial Z_1}+...\right)\left(\delta\eta^{(1)}+\delta^2\eta^{(2)}+...\right)
\left(-F\delta^2\eta^{(2)}-...\right)
\end{align*}
has only $\mathcal{O}(\delta^3)$ quantities. A similar result holds for the other stress boundary condition.
We then arrive at the system
\begin{align}
u^{(1)}_{X_1}+w^{(1)}_{Z_1}+u^{(2)}_{X_0}+v^{(2)}_y+w^{(2)}_{Z_0}=0\\
v^{(2)}_{yyyy}+\text{Re}\left[u^{(1)}_{T_0}+Uu^{(1)}_{X_0}+v^{(1)}U_y\right]_{X_0y}=0\\
\left(u^{(2)}_{Z_0}+u^{(1)}_{Z_1}-w^{(2)}_{X_0}-w^{(1)}_{X_1}\right)_{yy}-\text{Re}\left[u^{(1)}_{T_0Z_0}-w^{(1)}_{X_0T_0}+Uu^{(1)}_{X_0Z_0}-Uw^{(1)}_{Z_0Z_0}+U_yv^{(1)}_{Z_0}\right]=0\\
u^{(2)}_{yy}-\text{Re}\left[u^{(1)}_{T_0}+Uu^{(1)}_{X_0}+v^{(1)}U_y\right]-\text{Re}\eta^{(1)}_{X_0}+\text{Re}\tilde{\text{We}}[\eta^{(1)}_{X_0X_0X_0}+\eta^{(1)}_{X_0Z_0Z_0}]=0\text{ at } y=1\\
\eta^{(1)}_{Z_0}u^{(1)}_y+\eta^{(2)}_{Z_0}U_y+\eta^{(1)}_{X_1}U_y-\eta^{(1)}_{X_0}w^{(1)}_y=0\text{ at } y=1\\
\eta^{(1)}_{X_0}u^{(1)}_y+\eta^{(2)}_{X_0}U_y+\eta^{(1)}_{X_1}U_y+\eta^{(1)}_{Z_0}w^{(1)}_y=0\text{ at } y=1\\
\eta^{(1)}_{T_1}+\eta^{(2)}_{T_0}+U\eta^{(1)}_{X_1}+U\eta^{(2)}_{X_0}+u^{(1)}\eta^{(1)}_{X_0}+w^{(1)}\eta^{(1)}_{Z_0}-v^{(2)}=0\text{ at } y=1.
\end{align}

It can be confirmed that solutions for $u^{(2)}$ and $w^{(2)}$ are of the form

\begin{align}
&u^{(2)}=\left(y-\frac{1}{2}y^2\right)\left[\text{ReWe}(\eta^{(1)}_{\xi\xi\xi}+\eta^{(1)}_{\xi\zeta\zeta})-\text{Re}\eta^{(1)}_{\xi}\right]+\frac{\text{ReFr}^2}{3}\eta^{(1)}_{\xi\xi}\left(\frac{y^4}{8}-\frac{y^3}{2}+y\right)
\\&w^{(2)}=\left(y-\frac{1}{2}y^2\right)\left[\text{ReWe}(\eta^{(1)}_{\xi\xi\zeta}+\eta^{(1)}_{\zeta\zeta\zeta})-\text{Re}\eta^{(1)}_\zeta\right].
\end{align}

A simple calculation using the continuity equation then gives us an expression for $v^{(2)}:$

\begin{align}
v^{(2)}&=-\int(u^{(2)}_x+w^{(2)}_z)\hspace{1mm}dy
\\&=-\left(\frac{1}{2}y^2-\frac{1}{6}y^3\right)\left[\text{Re}\tilde{\text{We}}(\eta_{xxxx}+2\eta_{xxzz}+\eta_{zzzz})-\text{R}(\eta_{xx}+\eta_{zz})\right]-\frac{\text{ReFr}^2}{3}\eta^{(1)}_{\xi\xi}\left(\frac{y^5}{40}-\frac{y^4}{8}+\frac{y^2}{2}\right).
\end{align}

Evaluated at the boundary, we have 

\begin{equation}
v^{(2)}(y=1)=-\frac{1}{3}\left[\text{Re}\tilde{\text{We}}\left(\frac{\partial^2}{\partial\xi^2}+\frac{\partial^2}{\partial\zeta^2}\right)^2\eta^{(1)}-\text{Re}(\eta^{(1)}_{\xi\xi}+\eta^{(1)}_{\zeta\zeta})\right]-\frac{2}{15}\text{ReFr}^2\eta^{(1)}_{\xi\xi}
\end{equation}

Applying our expression for $v^{(2)}$ and the non-secularity condition gives us the following evolution equation for the first-order free surface elevation

\begin{equation}
\frac{\partial\eta^{(1)}}{\partial\tau}+2\text{Fr}\eta^{(1)}\frac{\partial\eta^{(1)}}{\partial\xi}-\frac{\text{Re}}{3}\left[\left(1-\frac{2}{5}\text{Fr}^2\right)\frac{\partial^2}{\partial\xi^2}+\frac{\partial^2}{\partial\zeta^2}\right]\eta^{(1)}+\frac{\text{Re}\tilde{\text{We}}}{3}\left(\frac{\partial^2}{\partial\xi^2}+\frac{\partial^2}{\partial\zeta^2}\right)^2\eta^{(1)}=0.
\end{equation}

Ultimately, the $\mathcal{O}(\delta^4)$ solution under the assumption that $Re=\mathcal{O}(\delta)$ will give us the required dispersive waves. We note that (as given in \citet{topper_kawahara}) the requirement that $Re=\mathcal{O}(\delta)$ gives the Nusselt flat film solution for both the $\mathcal{O}(\delta)$ and $\mathcal{O}(\delta^2)$ cases. We proceed with the $\mathcal{O}(\delta^4)$ system. The equations are mostly unaltered, except for the inclusion of two terms. The altered equations are given as

\begin{align}
&v^{(4)}_{yyyy}+2\Delta v^{(2)}_{yy}+\tilde{\text{Re}}\left[u^{(2)}_{T_0}+Uu^{(2)}_{X_0}+v^{(2)}U_y\right]_{X_0y}=0
\\&(u^{(4)}_{Z_0}-w^{(4)}_{X_0}
)_{yy}+\Delta\left(u^{(2)}_{Z_0}-w^{(2)}_{X_0}\right)-\tilde{\text{Re}}\left[\left(u^{(2)}_{Z_0}-w^{(2)}_{X_0}\right)_{T_0}+U\left(u^{(2)}_{Z_0}-w^{(2)}_{X_0}\right)_{X_0}+U_yv^{{(2)}}_{Z_0}\right]=0.
\\& u^{(4)}_{yy}+u^{(2)}_{Z_0Z_0}+u^{(2)}_{X_0X_0}-\tilde{\text{Re}}\left[u^{(2)}_{T_0}+Uu^{(2)}_{X_0}\right]+\tilde{\text{Re}}\tilde{\text{We}}\left[\eta_{X_0X_0X_0}+\eta_{X_0Z_0Z_0}\right]=0 \text{ at } y=1+\eta.
\end{align}

Here we have defined $\tilde{\text{Re}}\equiv\text{Re}\delta.$ The additional terms from the $\mathcal{O}(\delta^4)$ expansion give the following additions to $u^{(4)}$ and $w^{(4)}:$

\begin{align}
&u^{(4)}=\left(y-\frac{1}{2}y^2\right)\left[\tilde{\text{Re}}\tilde{\text{We}}\left(\eta^{(2)}_{\xi\xi\xi}+\eta^{(2)}_{\xi\zeta\zeta}\right)-\tilde{\text{Re}}\eta^{(2)}_{\xi}+3\text{Fr}\left(\eta^{(2)}_{\xi\xi}+\eta^{(2)}_{\zeta\zeta}\right)\right]+\frac{\tilde{\text{Re}}\text{Fr}^2}{3}\eta^{(2)}_{\xi\xi}\left(\frac{y^4}{8}-\frac{y^3}{2}+y\right)
\\&w^{(4)}=\left(y-\frac{1}{2}y^2\right)\left[\tilde{\text{Re}}\tilde{\text{We}}\left(\eta^{(2)}_{\xi\xi\zeta}+\eta^{(2)}_{\zeta\zeta\zeta}\right)-\tilde{\text{Re}}\eta^{(2)}_\zeta\right].
\end{align}

Now, the same procedure with the continuity equation can be done to find the free surface elevation at $\mathcal{O}(\delta^4):$

\begin{equation}
\frac{\partial\eta^{(2)}}{\partial\tilde{\tau}}+2\text{Fr}\eta^{(2)}\frac{\partial\eta^{(2)}}{\partial\xi}-\frac{\tilde{\text{Re}}}{3}\left[\left(1-\frac{2}{5}\text{Fr}^2\right)\frac{\partial^2}{\partial\xi^2}+\frac{\partial^2}{\partial\zeta^2}\right]\eta^{(2)}+\text{Fr}\left(\frac{\partial^2}{\partial\xi^2}+\frac{\partial^2}{\partial\zeta^2}\right)\frac{\partial\eta^{(2)}}{\partial\xi}+\frac{\tilde{\text{Re}}\tilde{\text{We}}}{3}\left(\frac{\partial^2}{\partial\xi^2}+\frac{\partial^2}{\partial\zeta^2}\right)^2\eta^{(2)}=0.
\end{equation}
\renewcommand{\thefootnote}{\fnsymbol{footnote}}
We will now define $\Delta=\frac{\partial^2}{\partial\xi^2}+\frac{\partial}{\partial\zeta^2}$ to recast the equation in the simplified form\footnote{The authors of \citet{topper_kawahara} remark that, upon proper rearrangement of the variables in the equation and removing the $\zeta$-dependence, the general 1-dimensional form $\phi_t+\phi\phi_x+\alpha\phi_{xx} + \beta\phi_{xxx}+\gamma\phi_{xxxx}=0$ is derived, which constitutes \ref{kawahara}.} (removing tildes for brevity)

\begin{equation}
\eta_{\tau}+2\text{Fr}\eta\eta_{\xi}+\frac{2\text{ReFr}}{15}\eta_{\xi\xi}-\frac{\text{Re}}{3}\Delta\eta+\text{Fr}\Delta\eta_{\xi\zeta}+\frac{\text{ReWe}}{3}\Delta^2\eta=0.
\end{equation}

We make substitutions for $\eta,\xi,\zeta,$ and $\tau$, as in \citet{Akrivis2011Linearly}:

\begin{equation}
(\xi,\zeta)=\left(\frac{5\text{We}}{2|\text{Fr}^2-5/2|}\right)^{1/2}(X,Y)
\qquad
\tau=\frac{75\text{We}}{4\text{Re}|F^2-5/2|^2}T
\qquad
\eta=\frac{\sqrt{2}\text{Re}|\text{Fr}^2-5/2|^{3/2}}{15(5\text{We})^{1/2}\text{Fr}}.
\end{equation}

Upon substitution, the expression has the common factor 
\begin{equation}
\frac{4\sqrt{2}\text{Re}|\text{Fr}^2-5/2|^{7/2}}{1125\sqrt{5}\text{We}^{3/2}\text{Fr}},
\end{equation}
such that after some short algebraic steps we arrive at the expression

\begin{equation}
\left(\frac{4\sqrt{2}\text{Re}|\text{Fr}^2-5/2|^{7/2}}{1125\sqrt{5}\text{We}^{3/2}\text{Fr}}\right)\left(H_T+HH_X\pm H_{XX}-\frac{5}{2|\text{Fr}^2-5/2|}H_{YY}+\frac{3\sqrt{5}\text{Fr}}{\text{Re}|\text{Fr}^2-5/2|(2\text{We})^{1/2}}\Delta H_X+\Delta^2H\right)=0,
\end{equation}

with the coefficients in front of $H_{YY}$ and $\Delta H_{X}$ being recast as coefficients $\alpha$ and $\delta,$ respectively. From the analysis in \citet{Akrivis2011Linearly}, the case $\alpha=0$ corresponds to a vertical plate, such that the resulting equation in our analysis is

\begin{equation}
H_T+HH_X+H_{XX}+\delta\Delta_{XY} H_X+\Delta_{XY}^2H=0,
\end{equation}

for $\Delta_{XY}\equiv\frac{\partial^2}{\partial X^2}+\frac{\partial^2}{\partial Y^2}.$ As such, any analysis will be done considering the $X$ and $Y$ coordinates to be the streamwise and spanwise coordinates, respectively, with $H$ being the free surface height evolution.

\section{Neural network details  }

\begin{table}
\caption{Neural networks architectures used for the case with $L=22$ and $\delta=0.002$. The table reports the architectures of the IRMAE-WD and NODE models employed in the main results}
\centering
\begin{tabular}{ccccc}
\hline
Architecture & Shape & Activation  & Learning Rate & Weight Decay\\
\hline
IRMAE-WD    &  $d_h/500/500/500/500/500/d_z$ \quad  & Sigmoid/lin & $[10^{-3},10^{-4},10^{-5}]$ & $10^{-4}$ \\
NODE    & $d_h/500/500/500/d_h$ \quad  & Sigmoid/lin & $[10^{-3},10^{-4},10^{-5}]$ & * \\
\hline
\end{tabular}
\label{table_NN}
\end{table}

\end{appen}

\begin{figure}
\begin{center}
\includegraphics[width=\textwidth]{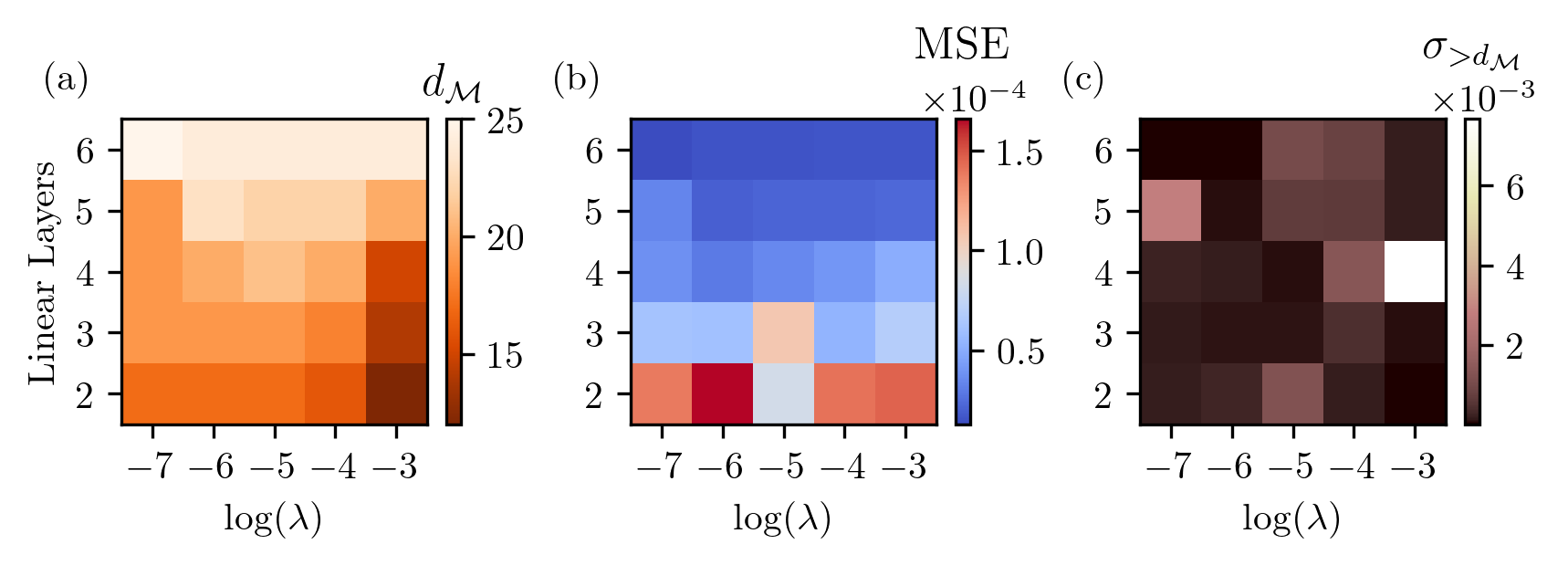}
\caption{Parametric sweep over the number of linear layers and weight decay for models trained on the  $L=22$ and $d=0.002$ dataset with various degrees of implicit and weight regularization. The colors correspond to (a) estimated $d_\mathcal{M}$, (b)  average test MSE measured in POD, (c) fraction of trailing singular values, measured from singular value $\sigma_{d_\mathcal{M}+1}$ to $\sigma_{d_A}$}\label{fig:auto_comparison_plot}
\label{IRMAE_sweep}
\end{center}
\end{figure}

This appendix summarizes the architectures of the neural networks used to train both the autoencoders and the NODE model. Table \ref{table_NN} shows the architecture used for training IRMAE and NODE for  $L=22$ and $\delta=0.002$.  We also include an illustrative example demonstrating the robustness of  IRMAE with different hyperparameters. 
The hyperparameters were selected through systematic trial-and-error refinement, acknowledging the stochastic variability introduced by neural-network optimization during gradient descent.

We assess the robustness of IRMAE using a representative dataset corresponding to $L=22$ and $\delta=0.002$. This regime exhibits sufficiently high-dimensional chaotic dynamics to serve as a rigorous benchmark for reduced-order modelling. To quantify robustness, we examine the sensitivity of the estimated intrinsic dimension $d_\mathcal{M}$, the test MSE, and the singular-value drop as a function of the number of linear layers $n$ and the weight-decay coefficient $\lambda$. Figure~\ref{IRMAE_sweep}a shows the predicted manifold dimension. We observe that the predicted dimension stabilizes for five layers, becoming essentially independent of the weight decay, with values in the range $d_\mathcal{M}\approx 20$--$25$. Figure~\ref{IRMAE_sweep}b presents the corresponding parameter sweep, where the test MSE serves as a proxy for reconstruction accuracy. We observe a progressive deterioration in performance as the number of linear layers decreases, highlighting the importance of network expressivity. Figure~\ref{IRMAE_sweep}c further characterizes the learned embeddings by reporting the ratio $\sigma_{d_\mathcal{M}+1}/\sigma_{d_A}$ of trailing singular values, which provides a quantitative measure of the embedding's ability to compress the data onto a low-dimensional manifold without sacrificing relevant variance.

Hence, the IRMAE architecture is fixed to five linear layers with a weight decay of $10^{-4}$ and kept unchanged across all other $(L,\delta)$ configurations considered in this work.

\vspace{2.0em}
\section*{ Appendix C. Short- and long-time dynamics in manifold coordinates}

This appendix presents additional results assessing both short-time tracking and long-time statistics of the learned models in manifold coordinates. These results demonstrate that the reduced models capture the essential dynamics and can be used in the search for ECS.

\begin{figure}
  \centering
  \begin{tabular}{cc}
    \includegraphics[width=0.52\textwidth]{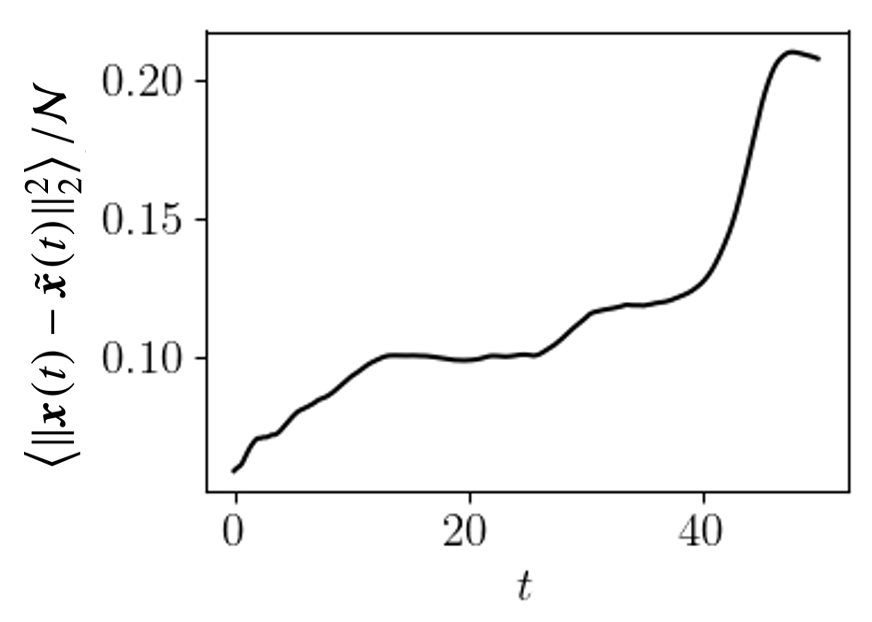} &
    \includegraphics[width=0.47\textwidth]{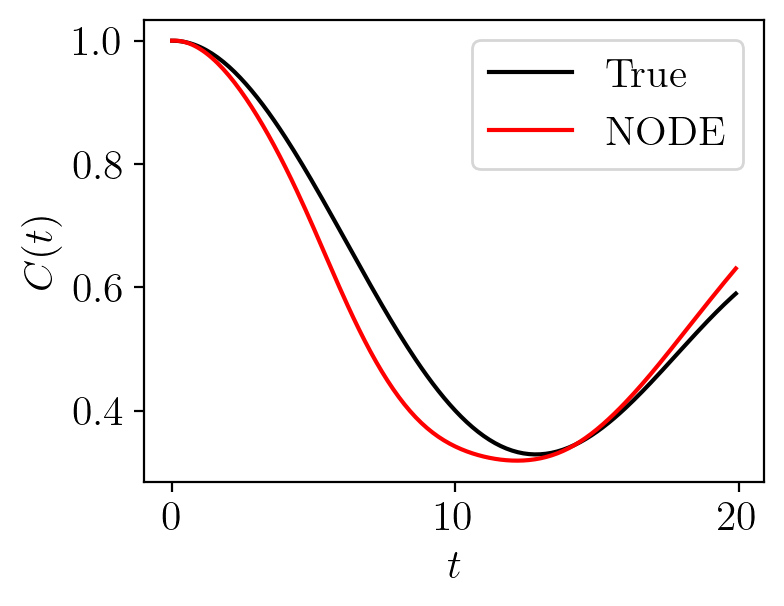} \\
    (a) & (b)
  \end{tabular}
  \caption{Short-time tracking for  $L = 22$ and $\delta = 0.002$. (a) Normalized ensemble-averaged test set reconstruction error for 500 random initial conditions up to $t=20.$
  (b) Temporal autocorrelation of the DNS and NODE trajectories.
  }
  \label{fig:node_autocorrelation}
\end{figure}

Figure~\ref{fig:node_autocorrelation}a presents the short-time tracking error for 500 randomly chosen initial conditions. We plot
$\left\langle \|\boldsymbol{x}(t)-\tilde{\boldsymbol{x}}(t)\|_2^2 \right\rangle/\mathcal{N}$,
where $\mathcal{N}$ denotes the average separation between true solutions at random times $t_i$ and $t_j$, i.e.$
\mathcal{N}=\left\langle \|\boldsymbol{x}(t_i)-\boldsymbol{x}(t_j)\|_2^2 \right\rangle$ .
The error remains below $0.12$ for $t\lesssim 40$ and grows only at later times, as expected for chaotic dynamics. Figure~\ref{fig:node_autocorrelation}b compares the temporal autocorrelation function, $C(t)=
{\langle \boldsymbol{x}(t)\cdot\boldsymbol{x}(t+\tau)\rangle_t}/
{\langle \|\boldsymbol{x}(t)\|_2^2\rangle_t}$,
computed from 500 initial conditions. The NODE reproduces both the decay rate and the location of the minimum with good accuracy, indicating that the dominant temporal coherence of the flow is preserved in the manifold dynamics. Together, these results show that the reduced dynamics capture the short-time  evolution of the DNS and therefore provide a reliable space for convergence to ECS.

\begin{figure}
\centering
\begin{tabular}{c}
\includegraphics[width=0.75\textwidth]{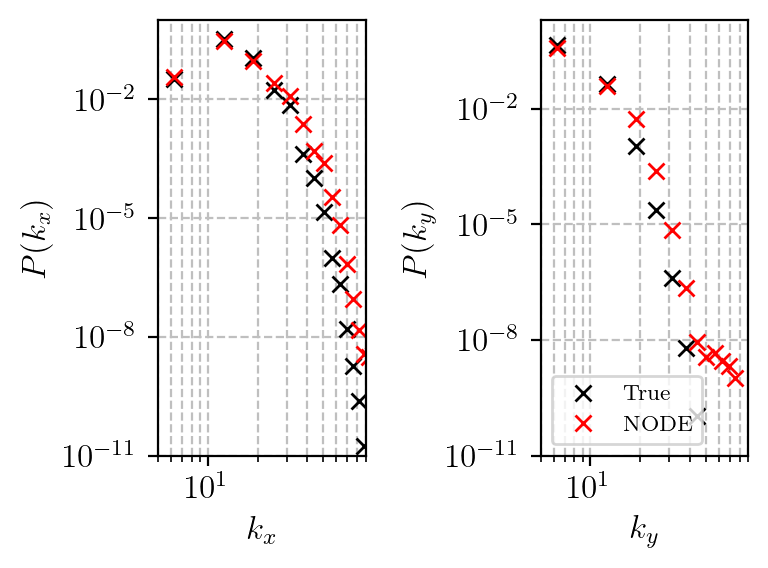}\\
(a)\\
\includegraphics[width=\textwidth]{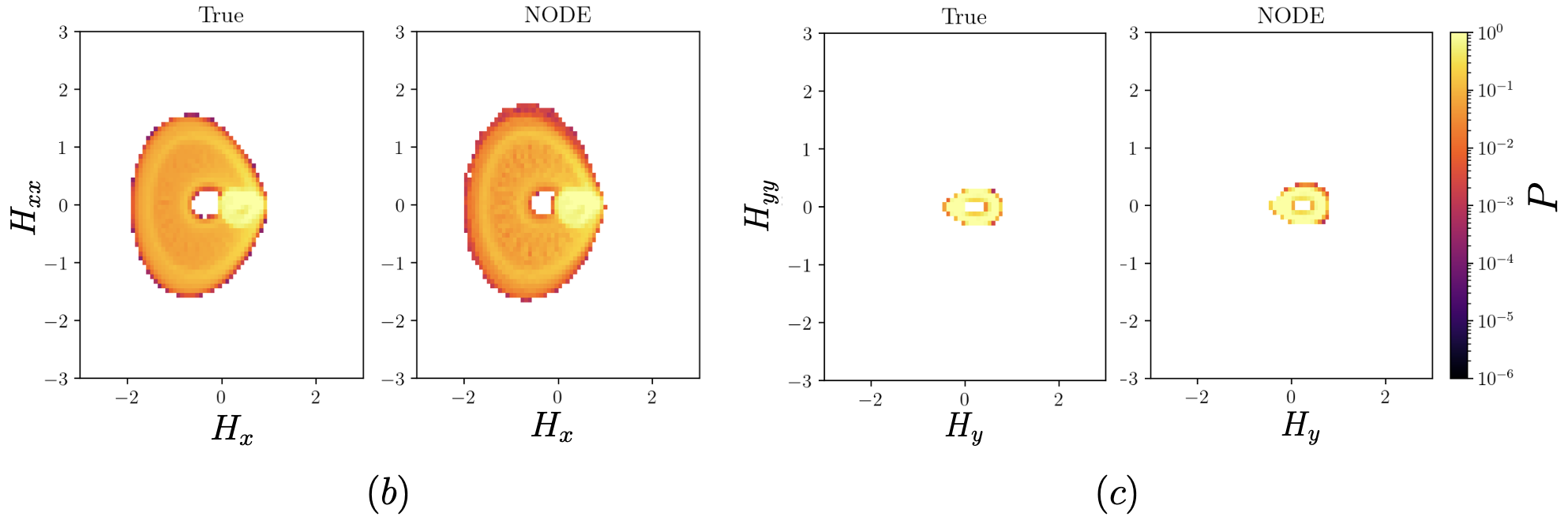}\\  
\end{tabular}
\caption{
Long-time statistics for $L=22$ and $\delta=0.002$. 
Comparison between DNS and NODE-reconstructed dynamics.
(a) Power spectra $P(k_x)$ and $P(k_y)$. 
(b) Joint PDFs in the $(H_x,H_{xx})$ plane along $(x,0)$. 
(c) Joint PDFs in the $(H_y,H_{yy})$ plane along $(0,y)$.}\label{fig:long_time}
\end{figure}

To verify that the model captures the long-time dynamics, we examine both spectral and statistical properties. Figure~\ref{fig:long_time}a compares the time-averaged power spectra of the DNS and reconstructed solutions as functions of $k_x$ and $k_y$, demonstrating good agreement over the energetically dominant range of wavenumbers along a long trajectory. The largest discrepancies appear only at the highest wavenumbers, where the spectral energy is already very small. This is consistent with the limited ability of the autoencoder to reconstruct the weakest short-wavelength content, which contributes little to the overall dynamics.

Figure~\ref{fig:long_time}b,c show the joint probability density functions of the first and second spatial derivatives of the film height. 
Panel (b) shows $(H_x,H_{xx})$ evaluated along the slice $(x,0)$, while panel (c) shows $(H_y,H_{yy})$ along $(0,y)$. 
The PDFs are constructed from all spatial points along the slices and all time samples of a trajectory integrated for $2000$ time units.
This confirm that the model reproduces the long-time statistical behaviour of the system. Although the NODE slightly underestimates some low-probability events in the tails of the distributions, it captures the overall shape of the statistics well for both slices.


\section*{ Appendix D. Calculation of the leading Lyapunov exponent}

In this appendix, we present the framework used to calculate the 
leading Lyapunov exponent, which provides a quantitative measure of the short-time predictability horizon and therefore a reference time scale for assessing the tracking performance of the reduced-order model.
Let $\boldsymbol{\Phi}_t(\boldsymbol{x})$ be the parameterization of the flow map of the dynamical system as above, let $\boldsymbol{x}_0$ and $\boldsymbol{x}_0+\delta\boldsymbol{x}_0$ be two initial conditions where $\|\delta\boldsymbol{x}_0\|_2\ll 1$,
the perturbation evolves as

\begin{equation}
\delta\boldsymbol{x}(t)
=
\boldsymbol{\Phi}_t(\boldsymbol{x}_0+\delta\boldsymbol{x}_0)
-
\boldsymbol{\Phi}_t(\boldsymbol{x}_0),
\end{equation}
and its growth is asymptotically given by
\begin{equation}
\|\delta\boldsymbol{x}(t)\|_2
\approx
\mathrm{e}^{\lambda_{LLE} t}
\|\delta\boldsymbol{x}_0\|_2,
\end{equation}
where $\lambda_{LLE}$ is the leading Lyapunov exponent,
\begin{equation}
\lambda_1
=
\lim_{t\to\infty}
\frac{1}{t}
\ln
\frac{\|\delta\boldsymbol{x}(t)\|_2}
{\|\delta\boldsymbol{x}_0\|_2}.
\end{equation}




The LLE is computed using the standard renormalization algorithm \citep{EDSON_BUNDER_MATTNER_ROBERTS_2019}.
Starting from an initial condition on the attractor, obtained after integrating a random field drawn from $\mathcal{N}(0,10^{-6})$ up to $2000$ time units, we introduce a small perturbation drawn from $\mathcal{N}(0,10^{-10})$. The perturbed and unperturbed trajectories are advanced for a renormalization time $T_r\approx85$. At the end of each cycle the perturbation is rescaled as
\begin{equation}
\delta\boldsymbol{x}
\;\leftarrow\;
\frac{\|\delta\boldsymbol{x}_0\|_2}
{\|\delta\boldsymbol{x}\|_2}
\delta\boldsymbol{x},
\end{equation}
and the finite-time estimate of the exponent is accumulated as
\begin{equation}
\lambda_{LLE}^{(j)}
=
\frac{1}{jT_r}
\sum_{k=1}^j
\ln
\frac{\|\delta\boldsymbol{x}(t_k)\|_2}
{\|\delta\boldsymbol{x}(t_{k-1})\|_2}.
\end{equation}
For the cases $L=22$ and $L=30$ we use $N_r=1000$ renormalization
steps, which yields convergence of $\lambda_{LLE}^{(j)}$.

\bibliographystyle{jfm}
\bibliography{Main}



\end{document}